\documentclass[12pt]{article}

\usepackage[margin=1in]{geometry}

\usepackage{amsmath}
\usepackage{amssymb}
\usepackage{amsfonts}
\usepackage{amsthm}
\usepackage[round,authoryear,sort]{natbib}
\usepackage{times}
\usepackage{bm}
\usepackage[plain,noend]{algorithm2e}
\usepackage{graphicx}
\usepackage{cancel}

\usepackage{times}
\usepackage{enumitem}

\theoremstyle{definition}
\newtheorem{definition}{Definition}[section]
\newtheorem{theorem}[definition]{Theorem}
\newtheorem{proposition}[definition]{Proposition}
\newtheorem{lemma}[definition]{Lemma}
\newtheorem{corollary}[definition]{Corollary}
\newtheorem{example}[definition]{Example}
\newtheorem{remark}[definition]{Remark}
\newtheorem{condition}{Condition}

\newcommand{\n}{n}
\newcommand{\np}{\n '}
\newcommand{\m}{m}
\newcommand{\mpp}{\m '}
\newcommand{\N}{N}
\newcommand{\xx}{x}
\newcommand{\xp}{\xx '}
\newcommand{\yy}{y}
\newcommand{\zz}{z}
\newcommand{\bx}{X}
\newcommand{\bxp}{\bx '}
\newcommand{\by}{Y}
\newcommand{\byp}{\by '}
\newcommand{\bz}{Z}
\newcommand{\bzp}{\bz '}
\newcommand{\bxone}{\bx_1}
\newcommand{\byone}{\by_1}
\newcommand{\bxtwo}{\bx_2}
\newcommand{\bytwo}{\by_2}
\newcommand{\bxthree}{\bx_3}
\newcommand{\bythree}{\by_3}
\newcommand{\lambdap}{\lambda '}
\newcommand{\lambdax}{\lambda_{x}}
\newcommand{\lambday}{\lambda_{y}}
\newcommand{\pOne}{p_{1}}
\newcommand{\mup}{\mu'}
\newcommand{\sigmap}{\sigma'}

\newcommand{\aconst}{a}
\newcommand{\dd}{d}
\newcommand{\kk}{k}
\newcommand{\Tone}{T_1}
\newcommand{\Ttwo}{T_2}

\newcommand{\xstar}{\xx_{\ast}}
\newcommand{\realR}{\mathbb{R}}

\newcommand{\suchthat}{\, : \,}

\newcommand{\absval}[1]{\left \lvert  #1 \right \rvert }

\newcommand{\rankgiven}[2]{R_{#1 \mid #2}}
\newcommand{\rankfun}[2]{\mathrm{rank}_{#1}(#2)}
\newcommand{\subs}[2]{{#1}_{#2}}

\newcommand{\wmwmtext}{W}
\newcommand{\wmwstat}[2]{\wmwmtext(#1, #2)}

\newcommand{\missingI}{I}
\newcommand{\missi}{\iota}
\newcommand{\misss}{s}
\newcommand{\indic}{\mathcal{I}}

\newcommand{\cdfphi}{\Phi}
\newcommand{\cdfwmw}[2]{\Phi_{#1, #2}}
\newcommand{\ptilde}{\tilde{p}}
\newcommand{\pvalfun}{g}
\newcommand{\Wmin}{\wmwmtext_{\min}}
\newcommand{\Wmax}{\wmwmtext_{\max}}
\newcommand{\WminXY}{\Wmin(\bx, \by)}
\newcommand{\WmaxXY}{\Wmax(\bx, \by)}
\newcommand{\WXY}{\wmwstat{\bx}{\by}}
\newcommand{\pset}{\mathcal{P}}
\newcommand{\pmin}{p_{1}}
\newcommand{\pmax}{p_{2}}
\newcommand{\qmin}{q_{\min}}
\newcommand{\qmax}{q_{\max}}
\newcommand{\ptildemin}{\ptilde_{1}}
\newcommand{\ptildemax}{\ptilde_{2}}

\newcommand{\discretespace}{\Omega}
\newcommand{\prob}{\mathrm{pr}}

\newcommand{\dstar}{d_{\max}}

\newcommand{\zzp}{\zz '}
\newcommand{\minrankgiven}[2]{\rankgiven{#1}{#2}^{\min}}

\title{On two-sample testing for data with arbitrarily missing values}

\author{Y. Zeng \and N. M. Adams \and D. A. Bodenham}

\date{{\normalsize Department of Mathematics, Imperial College London,} \\
 {\normalsize South Kensington Campus, London SW7 2AZ, U.K.} \\
{\normalsize yijin.zeng20@imperial.ac.uk, 
\quad n.adams@imperial.ac.uk, 
\quad dean.bodenham@imperial.ac.uk}
}

\begin{document}

\maketitle

\begin{abstract}
    We develop a new rank-based approach for univariate two-sample testing 
    in the presence of missing data which makes no assumptions about 
    the missingness mechanism.
    This approach is a theoretical extension of the Wilcoxon-Mann-Whitney 
    test that controls the Type I error 
    by providing exact bounds for the test statistic after accounting
    for the number of missing values.
    Greater statistical power is shown when the method 
    is extended to account for a bounded domain.
    Furthermore, exact bounds are provided on the proportions of data 
    that can be missing in the two samples 
    while yielding a significant result.
    Simulations demonstrate that our method has good power,
    typically for cases of $10\%$ to $20\%$ missing data, 
    while standard imputation approaches fail to control the Type I 
    error.
    We illustrate our method on complex clinical trial data in which patients' 
    withdrawal
    from the trial lead to missing values.
\end{abstract}

\section{Introduction}

The problem of univariate, nonparameteric two-sample 
hypothesis testing with missing values has not received the
attention that modern data applications require.
Despite the prevalence of missing value problems, no standard approach 
exists to address this issue.
For example, with clinical trials it frequently occurs that participants
drop out before the trial concludes, resulting in missing data
for those participants.
Several analyses of clinical trials data 
\citep{feldmanetal1993, marguliesetal2016} have dealt 
with missing values by assigning
them the worst-possible ranks, or first assigning them the worst-possible 
values and then converting these values to ranks, as
\citet{lachin1999} sought to formalise.
Other analyses have performed several different imputations for
the missing values, and if these different imputation approaches
all lead to a consistent conclusion, this is taken as the result
\citep{bakrisetal2015}.

In this work, we develop a novel extension to the Wilcoxon-Mann-Whitney
rank sum test 
\citep{wilcoxon1945, mannwhitney1947}
designed to accommodate missing data in a manner that is not 
dependent on the values of the missing data.
Broadly, our approach considers all possible 
Wilcoxon-Mann-Whitney test statistic values given the missing data 
in each sample, and if all possible test statistics are significant, 
a significant result is declared.
The theoretical development suggests specific limits to the efficacy of
the enhancement, while empirical results demonstrate the extension provides
good power while controlling the Type I error rate.
Further, empirical results suggest particular resistance to data that is
informatively missing.

An equivalent approach would be to consider all possible imputations
for the missing values and the resulting ranks.
However, for continuous-valued data, 
considering all possible cases could be an intractable problem;
the approach here is to consider all possible imputed rank statistics.
An elegant feature of our proposed approach is that it avoids the 
need to perform combinatorially difficult search to evaluate the
significance level of the new test.

When data is missing completely at random, the proposed method is 
conservative and simply ignoring the missing data will result in greater
power while controlling the Type I error; however, in the absence of such 
information all imputation methods should be used with caution while 
the proposed method can be safely used.

Various authors have suggested modifications and refinements to
the Wilcoxon-Mann-Whitney test to
accommodate missing values, none of which address the problem as
we have. 
The Wilcoxon-Mann-Whitney test with missing data has been considered before 
\citep{lee1997, cheung2005exact}, but only where the missingness mechanism 
is a special case of missing completely at random.
In a sequential testing context, 
calculating the minimum and maximum possible Wilcoxon-Mann-Whitney
statistics has been considered by \citet{alling1963} and 
\citet{halperinware1974}.

A related line of research concerns time-censored data
\citep{gehan1965singly, gehan1965doubly, matsouakabetensky2015},
although this is slightly different to the case of missing values.
An approach for censored data using log-ranks has also been previously 
considered \citep{petopeto1972, cox1972, prentice1978}; 
a discussion comparing this log-rank approach with that of \cite{gehan1965singly}
can be found in \citep{latta1977}.

We start by developing our proposed approach for distinct, real-valued data
in Section~\ref{sec:mainresults}, before extending to the more challenging
case of data with ties or with bounded values in Section~\ref{sec:ties}.


\section{Background}



\subsection{The Wilcoxon-Mann-Whitney rank sum test}
\label{sec:wmwbackground}


Let $\bx = \{ \xx_1, \ldots, \xx_{\n} \}$ and 
let $\by = \{ \yy_1, \ldots, \yy_{\m} \}$ be sets of 
distinct real-valued observations 
and define $\N = \n+\m$ and 
$\bz = \bx \cup \by = \{ \zz_1, \ldots, \zz_{\N} \}$.
The concept of rank is central in this work, so it is carefully defined here.
For a set $\bz \subset \realR$ and a value $\aconst \in \realR$, define the set 
$\subs{\bz}{\aconst} = \{\zz \in \bz \suchthat \zz < \aconst\}$
and let $\absval{\subs{\bz}{\aconst}}$ denote the size of $\subs{\bz}{\aconst}$.
Then, if all values in the set $\bz$ are distinct, the rank of a value
$\zz \in \bz$ is defined as $\rankfun{\bz}{\zz} = \absval{\subs{\bz}{\zz}} + 1$.
In the case that not all values are distinct, this definition is modified 
slightly; see Section~\ref{sec:ties}.
%
    Suppose a sample of observations $\bx$ is a subset of $\bz$. Then
    $\rankgiven{\bx}{\bz}$ denotes the sum of the ranks of the 
    values in $\bx$, where the ranking is considered in the 
    larger set $\bz$, i.e.
    \begin{equation}
        \rankgiven{\bx}{\bz} = \sum_{\xx \in \bx} \rankfun{\bz}{\xx}.
        \nonumber
    \end{equation}
Denoting $\N = \absval{\bz}$, $\rankgiven{\bz}{\bz} = \N(\N+1)/2$.
    Given samples $\bx = \{ \xx_1, \xx_2, \ldots, \xx_{\n} \}$ 
    and $\by=\{ \yy_1, \yy_2, \dots, \yy_{\m} \}$,
    the Wilcoxon-Mann-Whitney statistic $\wmwstat{\bx}{\by}$ is defined as
    \begin{align}
        \wmwstat{\bx}{\by} &= \rankgiven{\bx}{\bx \cup \by} - \n(\n+1)/2.
        \nonumber
    \end{align}
\cite{lehmann1975} show that, when defined in terms 
of random variables, $\wmwstat{\bx}{\by}$ approximately
follows a normal distribution with mean $\n \m/2$ and variance
$\n \m (\n+\m+1) /12$, if all values in $\bx$ and $\by$ are distinct.
This approximation can be shown to be adequate 
for moderate sample sizes; for $\n, \m \geq 50$, 
the \texttt{R} programming language \citep{R:2010} 
uses this normal approximation.
Although
\cite{lehmann1975} define the statistic as $\wmwstat{\by}{\bx}$, 
this results in an equivalent $p$-value.


\subsection{Mechanisms for missing data}
\label{sec:missingdatamech}

The taxonomy in \cite{littlerubin2020} 
describes three categories of an underlying mechanism for missing data.
Following their approach, it is convenient to consider vectors of
data rather than sets in this section.
For univariate data, denote a data
vector as $\bz = (\zz_{1}, \ldots, \zz_{\N})$.
Let  
$\missingI = (\missi_{1}, \ldots, \missi_{\N})$ denote an indicator vector 
where,  
$\missi_{j} = 1$ ($j=1,\ldots, \N$) means $\zz_{j}$ is missing and 
$\missi_{j} = 0$ means $\zz_{j}$ is observed.
The seminal idea in \cite{rubin1976} was to consider the indicator
vector $\missingI$ to itself follow a distribution, or rather to consider
the conditional distribution of $\missi_{j}$ given $\zz_{j}$, denoted
$f_{\missingI \mid \bz}(\missi_{j} \mid \zz_{j}, \theta)$, where $\theta$
denotes any parameters of the distribution.

If $f_{\missingI \mid \bz}(\missi_{j} \mid \zz_{j}, \theta) = 
f_{\missingI \mid \bz}(\missi_{j} \mid \tilde{\zz}_{j}, \theta)$
for every index $j$ and any distinct values $\zz_{j}, \tilde{\zz}_{j}$,
regardless of whether $\missi_{j} = 0$ or $1$, then
the missingness mechanism does not depend on the values of the data, 
whether missing or observed. In such cases, the data are called 
\emph{missing completely at random} and ignoring the missing
values could perhaps be justified.

If this condition is relaxed and 
$f_{\missingI \mid \bz}(\missi_{j} \mid \zz_{j}, \theta) = 
f_{\missingI \mid \bz}(\missi_{j} \mid \tilde{\zz}_{j}, \theta)$
for distinct values $\zz_{j}, \tilde{\zz}_{j}$
for every index $j$ only when $\missi_{j} = 1$, then the missingness
mechanism does not depend on the missing values. In such cases, 
the data are called \emph{missing at random} and imputing the missing
values can be justified \citep{littlerubin2020}.
Two simple approaches for imputing missing values are \emph{mean imputation}, 
where the missing values are replaced with the sample mean of the 
observed values, and \emph{hot deck imputation}, where the missing values
are replaced by randomly-selected observed values.

If the distribution of $\missingI$ depends on the missing values, then
the data are called \emph{missing not at random}. 
There are methods for dealing with this case when the missingness
mechanism is known \citep{littlerubin2020},
but if the missingness mechanism is unknown then imputing the missing
values can lead to unreliable results, 
as shown in Section \ref{sec:numericalexamples}.
A strength of our proposed approach is that it is robust to this 
type of missing data.



\section{Main results}

\label{sec:mainresults}


\subsection{Wilcoxon-Mann-Whitney statistics for missing data}

In this section we focus on the case that the values are distinct and that the 
support is unbounded; e.g. open subsets of the real numbers. 
The former condition allows us to focus on the main ideas of the argument, while
the latter allows us to assume that it is possible for the missing values to
be smaller or larger than those observed.
In Section~\ref{sec:ties} the results are generalised to consider ties and 
the case of compact support.
The following lemma will be used repeatedly to prove 
Proposition~\ref{thm:mainprop}, which is crucial for the main result, 
Theorem~\ref{thm:mainthm}.

\begin{lemma}
    \label{lem:onemissing}
    Suppose $\bxp = \{\xx_1, \ldots, \xx_{\np}\}$ is a sample of  
    observations and $\xstar$ is another observation. 
    Let $\bx=\bxp \cup \{\xstar\}$ and suppose $\bx \subset \bz$. 
    For fixed $\bz \setminus \{ \xstar \}$, the rank sum of $\bx$ in $\bz$,
    \begin{equation}
        \rankgiven{\bx}{\bz} = \rankgiven{\bxp}{\bz} + \rankfun{\bz}{\xstar},
        \nonumber
    \end{equation}
    is minimised when $\rankfun{\bz}{\xstar} < \rankfun{\bz}{\zz}$,
    for all $\zz \in \bz \setminus \{ \xstar \}$, if all values in $\bx$ are distinct.
\end{lemma}
%
\begin{proposition}
    \label{thm:mainprop}
    Suppose that $\bx = \{\xx_1, \ldots, \xx_{\n}\}$ and
    $\by = \{\yy_1, \ldots, \yy_{\m}\}$ are samples of 
    distinct real-valued observations. 
    Suppose that $\bxp \subset \bx$ is a subset of $\np$ distinct 
    values in $\bx$, and suppose that $\byp \subset \by$ is a subset 
    of distinct $\mpp$ values in $\by$.
    Defining $\bz = \bx \cup \by$ and $\bzp = \bxp \cup \byp$
    and supposing only $\bzp$ is known,
    the minimum
    and maximum possible rank sums of $\bx$, over all possible 
    values for the $\dd = \n-\np + \m - \mpp$ observations 
    in $\bz \setminus \bzp$, are
    \begin{align}
        \min_{\bz \setminus \bzp \in \realR^{\dd}} \rankgiven{\bx}{\bz} 
        &= \rankgiven{\bxp}{\bzp} +  (\n - \np)(\n + \np + 1)/2,
        \nonumber \\
        \max_{\bz \setminus \bzp \in \realR^{\dd}} \rankgiven{\bx}{\bz} 
        &= \rankgiven{\bxp}{\bzp}  
           + \{ \n (\n + 2\m + 1) - \np (\np +2 \mpp + 1)\}/2.
           \nonumber 
    \end{align}
\end{proposition}
\begin{theorem}
    \label{thm:mainthm}
    Suppose that $\bx = \{\xx_1, \ldots, \xx_{\n}\}$ and
    $\by = \{\yy_1, \ldots, \yy_{\m}\}$ are samples of 
    distinct real-valued observations and
    $\bxp \subset \bx$ and $\byp \subset \by$ are subsets of
    distinct values with sizes $\np$ and $\mpp$, respectively. Then
    the Wilcoxon-Mann-Whitney statistic $\wmwstat{\bx}{\by}$ is
    bounded as follows:
    \begin{align}
        \wmwstat{\bxp}{\byp}
        \leq
        \wmwstat{\bx}{\by}
        \leq
        \wmwstat{\bxp}{\byp} + (\m \n - \mpp \np).
        \label{eqn:wmwbounds}
    \end{align}
\end{theorem}
While the results above do not explicitly mention missing data, 
one could consider 
$\bxp \cup \byp = \bzp = \{ \zz_{j} \in \bz \suchthat \missi_{j} = 0\}$
to be the observed (non-missing) values of 
$\bz = \bx \cup \by$,
using the indicator vector introduced in Section~\ref{sec:missingdatamech}.
Then, the missing values are $\bz \setminus \bzp$.
The proofs for the above results are in Section 1 of the Supplementary Material.



\subsection{Significant test statistics and $p$-values without imputation}

\label{sec:sigpval}

Although Theorem~\ref{thm:mainthm} gives us bounds for the 
Wilcoxon-Mann-Whitney test statistic, computing a test statistic 
is insufficient for hypothesis testing purposes
without being able to determine its statistical significance.

Supposing as above that $\bx$ and $\by$ are samples of $\n$ and $\m$
observations, respectively, Section~\ref{sec:wmwbackground}
recalls that the statistic $\wmwstat{\bx}{\by}$ approximately
follows a normal distribution with mean $\n\m/2$ and
variance $\n\m(\n+\m+1)/12$
with cumulative distribution function $\cdfwmw{\n}{\m}$.
Considering a two-sided test at significance level $\alpha$, 
given $\wmwstat{\bx}{\by}$ 
one computes a score $\ptilde = \cdfwmw{\n}{\m}(\wmwstat{\bx}{\by})$ 
before computing a $p$-value  
$p = \pvalfun(\ptilde) = 1 - \absval{1 - 2\ptilde}$, 
since extreme $\wmwstat{\bx}{\by}$ statistics
can lead to scores $\ptilde$ close to either $0$ or 
$1$; finally, if $p < \alpha$, the result is declared significant at 
level $\alpha$.

The preceding review is helpful when considering how to proceed 
when data are missing.
The key idea behind our proposed approach is
the following: given the observed data $\bxp \subset \bx$ and 
$\byp \subset \by$, if all potential $\wmwstat{\bx}{\by}$ statistics are 
significant, regardless of the possible values for the missing
observations $(\bx \cup \by) \setminus (\bxp \cup \byp)$, 
then the values of the missing data are irrelevant;
in all possible cases, the $\wmwstat{\bx}{\by}$ 
test statistic is significant, given the observed data. 
Now suppose $\WminXY = \wmwstat{\bxp}{\byp}$ 
and $\WmaxXY = \wmwstat{\bxp}{\byp} + (\m \n - \mpp \np)$ 
are the lower and upper bounds for $\WXY$
given in Theorem~\ref{thm:mainthm}.
Let 
\begin{align}
    \pset = \{\pvalfun \circ \cdfwmw{\n}{\m}( \wmwmtext(\bx, \by) )
    \suchthat \WminXY \leq \wmwmtext (\bx, \by) \leq \WmaxXY,\,
    \textrm{given }\bxp, \byp, \n, \m \},
\nonumber
\end{align}
be the set of all possible $p$-values, given the observed $\bxp$ and $\byp$
of the samples $\bx$ and $\by$, respectively, 
which may contain missing values but the sample sizes are known. 
Then given a significance level $\alpha \in (0, 1)$, if $p < \alpha$ for all 
$p \in \pset$, we declare the result to be statistically significant.

The challenge of checking that all $p \in \pset$ are significant may 
seem as simple as checking that the following condition holds:
\begin{condition}
    \label{cond:A}
$\pmin=\pvalfun \circ \cdfwmw{\n}{\m}(\WminXY) < \alpha$ and
$\pmax=\pvalfun \circ \cdfwmw{\n}{\m}(\WmaxXY) < \alpha$. 
\end{condition}
However, there is a caveat: 
it is possible that $\ptildemin=\cdfwmw{\n}{\m}(\WminXY)$ is close to $0$, 
while $\ptildemax=\cdfwmw{\n}{\m}(\WmaxXY)$ is close to 1, or vice versa,
implying that there may be a $\wmwmtext (\bx, \by) \in [\WminXY, \WmaxXY]$ which
would lead to a non-significant $p$-value.
This is dealt with by 
checking that either $\ptildemin, \ptildemax \leq 0.5$
or $\ptildemin, \ptildemax \geq 0.5$, i.e. the extreme scores are both 
in the lower
or upper tail of the distribution. If at least one of these two conditions does 
not hold, then the result cannot be declared significant.
Equivalently, one can check that
either $\WminXY, \WmaxXY \leq \n\m/2$ or $\WminXY, \WmaxXY \geq \n\m/2$, 
which can be summarised as
\begin{condition}
    \label{cond:B}
    Terms $(\WminXY - \n\m/2)$ and $(\WmaxXY - \n\m/2)$ have the same sign.
\end{condition}
Therefore, to declare a significant result, one needs both
Conditions~\ref{cond:A} and \ref{cond:B} to hold.
Alternatively, 
recalling that $\cdfwmw{\n}{\m}$ is the cumulative distribution function
of a normal distribution with mean $\n\m/2$ and variance $\n\m(\n+\m+1)/12$,
we can define 
\begin{condition}
    \label{cond:C}
Either $\WmaxXY < \cdfwmw{n}{m}^{-1}(\alpha/2)$ or 
    $\WminXY > \cdfwmw{\n}{\m}^{-1}(1-\alpha/2)$.
\end{condition}
Then the following result 
allows us to only check Condition~\ref{cond:C} in order to 
confirm all $p \in \pset$  are significant; 
the proof is in Section 2 of the Supplementary Material.
\begin{lemma}
    Suppose $\bx$ and $\by$ are samples of $\n$ and $\m$ values, 
    respectively, with $\n$ and $\m$ sufficiently large and all values  
    distinct.
    Then Condition~\ref{cond:A} and Condition~\ref{cond:B} are both 
    true if and only if Condition~\ref{cond:C} is true.
\end{lemma}
From the discussion above, the following result summarises how to
determine if the proposed method will declare a significant result.
\begin{lemma}
    The proposed method will declare a significant result 
    if and only if Condition~\ref{cond:C} is true.
\end{lemma}
So far we have only discussed how to declare a result to be significant or not
given a threshold $\alpha$. In certain situations it may be desirable 
to report a $p$-value given the proposed approach, whether or not the $p$-value
is significant.
One way to do this would be to report $p_{\max}=\max\{\pmin, \pmax\}$, defined 
above, provided Condition~\ref{cond:C} holds. If this condition does not hold, 
one could report $p=1$, i.e. not significant. One might consider this 
approach to be computing the \emph{maximum attainable} $p$-value, 
and in the experiments our proposed method will declare a significant
result if $p_{\max} < \alpha$.

In cases where $p_{\max} \geq \alpha$, one could report both
$p_{\max}$ and $p_{\min}=\min \{\pmin, \pmax\}$; 
if also $p_{\min} \geq \alpha$ then there could not be a significant result, 
regardless of the values of the missing data.
However, if $p_{\max} \geq \alpha$ but $p_{\min} < \alpha$, then 
the significance of the result depends on the values missing of the
missing data. In such cases, we would
still fail to reject the null hypothesis since $p_{\max} \geq \alpha$, 
although we may note that the data do not provide conclusive evidence.

A related idea 
is Tarone's concept of computing the minimum attainable $p$-value  
in the context of multiple hypothesis testing for discrete
statistics from contingency tables \citep{tarone1990}.


\subsection{Maximum allowable proportion of missing data}

Since the proposed approach essentially requires all possible
configurations of the missing data to still lead to a significant
statistic, the more data that is missing, the wider the range
of possible statistics. 
This section presents results showing the limits on the amount of
data that can be missing before the proposed method will not be able
to yield a significant result. 

In the following proposition, we consider the case where the samples are large 
enough so that the normal approximation can be used for the distribution of 
the Wilcoxon-Mann-Whitney test statistic.
In the remainder of this section, $\cdfphi$ is the 
cumulative distribution function of the 
standard normal distribution and $\cdfphi^{-1}$ is its inverse.
\begin{proposition}
    \label{thm:missingprop}
    Suppose that $\bx = \{\xx_1, \ldots, \xx_{\n}\}$ and
    $\by = \{\yy_1, \ldots, \yy_{\m}\}$ are samples of 
    real-valued observations which are unknown but assumed to be distinct and
    $\bxp \subset \bx$ and $\byp \subset \by$ are subsets of
    distinct values with sizes $\np$ and $\mpp$, respectively,
    which will be observed. Then for any significance threshold 
    $\alpha \in (0, 1)$ and for sufficiently large $\n,\m$, if
    \begin{align}
        \np \mpp/(\n \m) < 1/2 + 
        \cdfphi^{-1}(1 - \alpha/2) \{(\n + \m + 1)/ (12\n\m) \}^{1/2} ,
        \label{eqn:missingprop}
    \end{align}
    then the proposed method will not yield a significant $p$-value, 
    regardless of the values in $\bxp$ and $\byp$.
\end{proposition}
\begin{example}
    Suppose $\n=\m=100$, and $\alpha=0.05$. Then the right-hand side of 
    Inequality \eqref{eqn:missingprop} is approximately $0.58$.
    If each sample is missing exactly $20\%$ of its values, 
    i.e. $\np/\n= \mpp/\m=0.8$, then 
    the left-hand side of Inequality \eqref{eqn:missingprop} 
    is $0.64$, implying that it may be possible to obtain a significant result.
    However, if $\np/\n=0.8$ but $\mpp/\m=0.7$, then $(\np \mpp)/(\n\m) =0.56$,
    implying that a significant result is not possible.
\end{example}

\begin{remark}
    \label{rem:largenough}
In the statement of Proposition~\ref{thm:missingprop},
$\n$ and $\m$ being ``sufficiently large'' means that the sample sizes
are large enough for the normal approximation to the distribution of the 
Wilcoxon-Mann-Whitney statistic to be considered appropriate. As mentioned in 
Section~\ref{sec:wmwbackground}, $\n,\m \geq 50$ is one rule that is used.
\end{remark}
\begin{remark}
    \label{rem:phi}
    The inequality uses $\cdfphi^{-1}(1 - \alpha/2)$ rather than $\cdfphi^{-1}(\alpha/2)$
    because, for $\alpha \in (0, 1)$, $\cdfphi^{-1}(1 - \alpha/2) > 0$.
\end{remark}
An important case of Proposition~\ref{thm:missingprop} is the following
\begin{corollary}
    \label{cor:missing}
    If samples $\bx$ and $\by$ are sufficiently large and 
    each have at least $30\%$ of their data missing
    then the proposed approach will never yield a significant $p$-value.
\end{corollary}
\begin{proof}
    Since $1-\alpha/2 \geq 0.5$, $\cdfphi^{-1}(1 - \alpha/2) \geq 0$ and
    the right-hand side of Inequality~\eqref{eqn:missingprop} has a lower
    bound of $0.5$. If $\np/\n \leq 0.7$ and $\mpp/\m \leq 0.7$, 
    then $\np \mpp(\n\m)^{-1} \leq 0.49$. 
\end{proof}
While Proposition~\ref{thm:missingprop} gives conditions when a significant
result is not achievable given the proportions of missing samples, it does
not indicate when a significant result \emph{is} achievable.
Fortunately, this is provided by
the following corollary, which essentially
proves that the bound in Inequality~\eqref{eqn:missingprop} in 
Proposition~\ref{thm:missingprop} is sharp.
\begin{corollary}
    \label{supp:thm:missingproptwo}
    Suppose that $\bx = \{\xx_1, \dots, \xx_{\n}\}$ and
    $\by = \{\yy_1, \dots, \yy_{\m}\}$ are samples of 
    real-valued observations which are unknown but assumed to be distinct and
    $\bxp \subset \bx$ and $\byp \subset \by$ are subsets of distinct
    values with sizes $\np$ and $\mpp$, respectively,
    which will be observed. Then for any significance threshold $\alpha \in (0, 1)$
    and for sufficiently large $\n,\m$, if 
    \begin{align}
        \np \mpp/(\n \m) \geq 1/2 + 
        \cdfphi^{-1}(1 - \alpha/2) \{(\n + \m + 1)/ (12\n\m) \}^{1/2} ,
        \label{supp:eqn:corr2}
    \end{align}
    then sets $\bxp$ and $\byp$ exist for which
    the proposed method will yield a significant $p$-value.
\end{corollary}
The proofs of the results in this section are in 
Section 3 of the Supplementary material.


\section{Numerical examples}

\label{sec:numericalexamples}

In this section, we investigate the Type I error and statistical power of  
the proposed method when data are missing, and compare its performance 
with the Wilcoxon-Mann-Whitney test 
when the missing data have been imputed using either mean imputation 
or hot deck imputation, or when the missing data are ignored, which we call 
\emph{ignore missing}, or when the missing values are known, i.e. the data is 
complete.

The first experiment considers the case where data are missing completely at 
random. Observations in $\bx$ are sampled independently from a 
$\mathrm{N}(0,1)$ distribution, while observations in $\by$ are sampled 
independently from a $\mathrm{N}(0,1)$ distribution to evaluate the Type I
error, and from a $\mathrm{N}(1,1)$ distribution to evaluate the statistical 
power. 
A proportion $\misss \in [0, 0.4]$ of the observations
$\bx=\{\xx_1, \dots, \xx_{\n}\}$ are selected completely at random to be marked
as missing. The same proportion $\misss$ of $\by=\{\yy_1, \dots, \yy_{\m}\}$
are selected completely at random to be marked as missing. 

\begin{figure}
    \includegraphics[width=\textwidth]{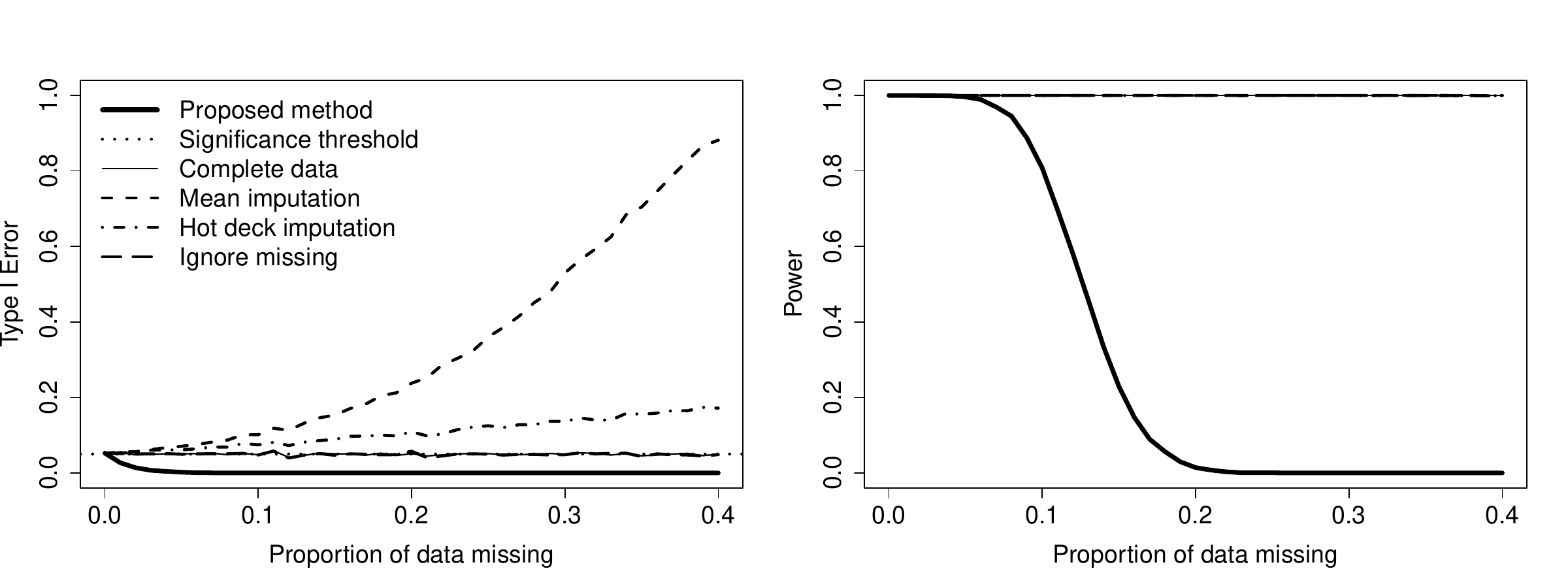}
    \caption{The Type I error and statistical power of the proposed method and 
    the standard Wilcoxon-Mann-Whitney test after the missing data is either known or
    has been imputed or ignored as the proportion of missing data increases.
    The data is missing completely at random.
    (Left) Type I error: $\mathrm{N}(0,1)$ vs $\mathrm{N}(0,1)$; 
    (Right) Power: $\mathrm{N}(0,1)$ vs $\mathrm{N}(1,1)$. For both figures, 
    a significance threshold of $\alpha=0.05$ has been used and the total
    sample sizes are $\n=100$, $\m=100$, and $5000$ trials were used.}
\label{fig:mcar1}
\end{figure}

Figure~\ref{fig:mcar1} shows that the Type I error is not controlled by
either hot deck imputation or mean imputation, although it is controlled
for the proposed method and for the case deletion case when the missing data 
are ignored. On the other hand, all methods have good power, except that
the power for the proposed method decreases significantly when more
than $10\%$ of the data is missing.

For this experiment,
$\n=\m=100$, but the Supplementary Material contains experiments with different
values for $\n$ and $\m$, although the results are similar.
Furthermore, if there is a larger difference between the two samples for the
power experiment, for example if the second sample follows a $\mathrm{N}(3, 1)$
distribution rather than a $\mathrm{N}(1, 1)$ distribution, then the proposed
method can still have good power for over $20\%$ of the data is missing; see
Figure~S1 of the Supplementary Material.


\begin{figure}
    \includegraphics[width=\textwidth]{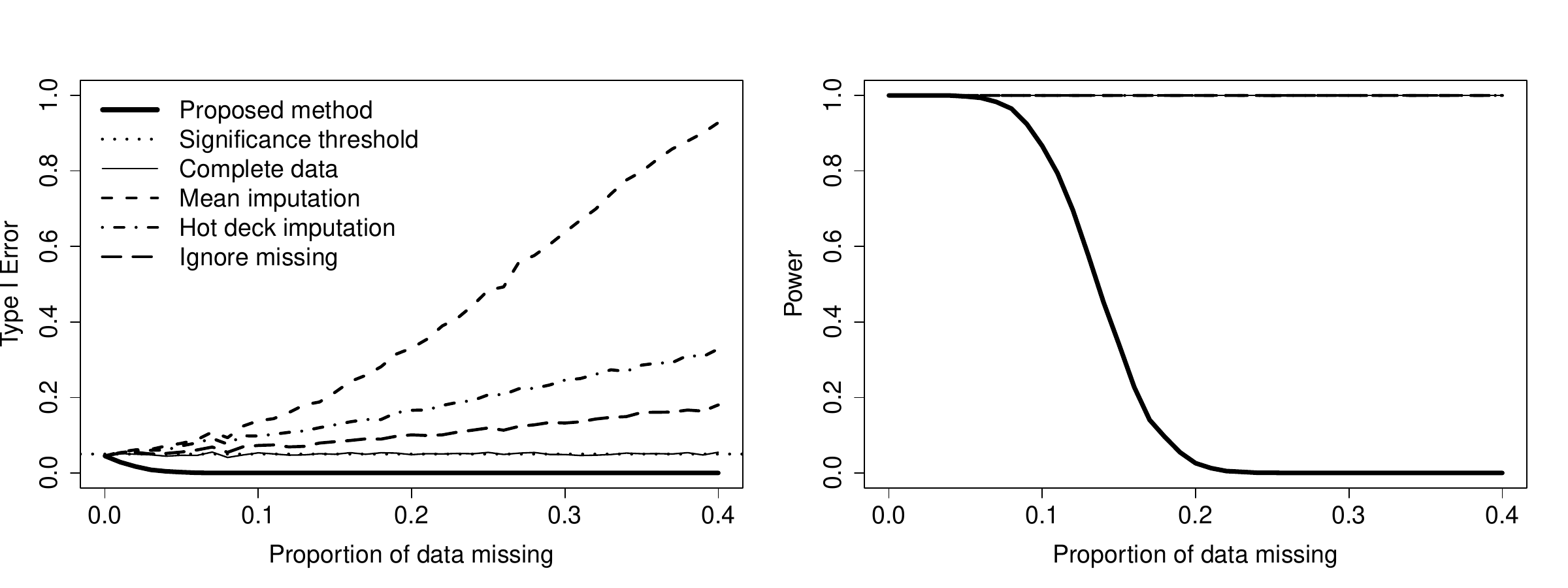}
    \caption{The Type I error and statistical power of the proposed method and 
    the standard Wilcoxon-Mann-Whitney test after the missing data is either known or
    has been imputed or ignored as the proportion of missing data increases.
    The data is missing not at random, according to the mechanism in 
    Equation~\eqref{eqn:missmechone}, where only observations greater than $0$
    are possibly missing from both samples.
    (Left) Type I error: $\mathrm{N}(0,1)$ vs $\mathrm{N}(0,1)$; 
    (Right) Power: $\mathrm{N}(0,1)$ vs $\mathrm{N}(1,1)$. For both figures, 
    a significance threshold of $\alpha=0.05$ has been used and the total
    sample sizes are $\n=100$, $\m=100$, and $5000$ trials were used.}
\label{fig:mnar1}
\end{figure}

The second experiment is the same as the first, except in this case the data 
are missing not at random. The missingness mechanism is as follows: if $\misss$
is the proportion of observations to be missing, then for any observation
$\xx_i \in \{\xx_1, \dots, \xx_{\n}\}$, the probability of $\xx_i$ being 
missing is 
\begin{align}
    \prob(\textrm{$\xx_i$ is missing}) = 
    \begin{cases}
       q, &\quad\text{if $\xx_i > 0$}, \\
       0, &\quad\text{otherwise}, \\
     \end{cases}
     \label{eqn:missmechone}
\end{align}
where $q = \min(1, \misss \n / \sum_{j=1}^{\n} \indic_{ \{\xx_j > 0 \} })$ 
and $\indic_{A}$ is the indicator function for the event $A$.
In other words, only values greater than $0$ can be missing and $q$ is
specified so that the total proportion of missing data for 
$\{\xx_1, \dots, \xx_{\n}\}$ will be approximately $\misss$.
For $\yy_i \in \{\yy_1, \dots, \yy_{\n}\}$, the same missingness mechanism
is used.
Figure~\ref{fig:mnar1} shows that when the data from both samples
are missing not at random and
follow the above missingness mechanism, then the Type I error is not controlled 
for either of the imputation methods or when the missing data are ignored.
This relatively simple example illustrates the potential peril of not taking 
missing data into account.
On the other hand, the proposed method controls the Type I error rate for this case.
The statistical power of these methods appears similar to that for the missing 
completely at random case in Figure~\ref{fig:mcar1}; all methods have good power, 
although the proposed method's power decreases as the proportion of missing
data increases beyond $10\%$ of the total.
Additional figures in Section~7.2 of the Supplementary Material shows how the 
power increases when the sample sizes are increased to $n=m=1000$ and/or the 
data in the second sample are observations from a $\mathrm{N}(3, 1)$ distribution. 


\section{Additional results}


\subsection{Theoretical results for data missing completely at random}

\begin{table}
    \def~{\hphantom{0}}
    \caption{Estimated power of the proposed method given by $1000$ Monte Carlo 
    replications and theoretical estimates (in parentheses) from 
    Proposition~\ref{prop:mcar1} when data is missing completely at random. 
    $s$, proportion of data missing; $\n,\m$, sample sizes for $X$ and $Y$
    respectively; 
        $\Delta$, value of mean shift such that all samples in $X$ generated 
        from $\mathrm{N}(0,1)$ and $Y$ from $\mathrm{N}(\Delta,1)$.
    }{%
        \scalebox{1.0}{
\begin{tabular}{cccccccc}
\\
& &  \multicolumn{6}{c}{Proportion of data missing} \\
&  & $s = 0$ & $s=5\%$ & $s =10\%$ & $s = 15\%$ & $s=20\%$ & $s=30\%$ \\
$n,m$ & $\Delta$ &  \multicolumn{6}{c}{}\\
\hline
20 & 0 & 0.06 (0.05)  & 0.01 (0.01)  & 0.00 (0.00) & 0.00 (0.00) & 0.00 (0.00) & 0.00 (0.00) \\
& 0.5 & 0.30 (0.31) & 0.10 (0.10) & 0.01 (0.01) & 0.00 (0.00) & 0.00 (0.00) & 0.00 (0.00) \\
& 1 & 0.85 (0.85) & 0.54 (0.53) & 0.17 (0.16) & 0.00 (0.01) & 0.00 (0.00) & 0.00 (0.00) \\
& 2 & 1.00 (1.00) & 1.00 (1.00) & 0.95 (0.96) & 0.34 (0.32) & 0.00 (0.00) & 0.00 (0.00) \\
50 & 0 & 0.06 (0.05) & 0.00 (0.00) & 0.00 (0.00) & 0.00 (0.00) & 0.00 (0.00) & 0.00 (0.00) \\
& 0.5 & 0.68 (0.67) & 0.33 (0.23) & 0.02 (0.02) & 0.00 (0.00) & 0.00 (0.00) & 0.00 (0.00) \\
& 1 & 1.00 (1.00) & 0.96 (0.95) & 0.54 (0.52) & 0.08 (0.04) & 0.00 (0.00) & 0.00 (0.00) \\
& 2 & 1.00 (1.00) & 1.00 (1.00) & 1.00 (1.00) & 1.00 (0.99) & 0.08 (0.10) & 0.00 (0.00) \\
100 & 0 & 0.06 (0.05) & 0.00 (0.00) & 0.00 (0.00) & 0.00 (0.00) & 0.00 (0.00) & 0.00 (0.00) \\
& 0.5 & 0.93 (0.93) & 0.46 (0.45) & 0.03 (0.03) & 0.00 (0.00) & 0.00 (0.00) & 0.00 (0.00) \\
& 1 & 1.00 (1.00) & 1.00 (1.00) & 0.90 (0.89) & 0.12 (0.12) & 0.00 (0.00) & 0.00 (0.00) \\
& 2 & 1.00 (1.00) & 1.00 (1.00) & 1.00 (1.00) & 1.00 (1.00) & 0.78 (0.76) & 0.00 (0.00)\\
200 & 0  & 0.05 (0.05) & 0.00 (0.00) & 0.00 (0.00) & 0.00 (0.00) & 0.00 (0.00) & 0.00 (0.00) \\
& 0.5 & 1.00 (1.00) & 0.77 (0.78)& 0.04 (0.05) & 0.00 (0.00) & 0.00 (0.00)& 0.00 (0.00) \\
& 1 & 1.00 (1.00) & 1.00 (0.95) & 1.00 (1.00) & 0.36 (0.35) & 0.00 (0.00) & 0.00 (0.00) \\
& 2 & 1.00 (1.00) & 1.00 (1.00) & 1.00 (1.00) & 1.00 (1.00) & 1.00 (1.00) & 0.00 (0.00) \\
300 & 0  & 0.05 (0.05) & 0.00 (0.00) & 0.00 (0.00) & 0.00 (0.00) & 0.00 (0.00) & 0.00 (0.00) \\
& 0.5 & 1.00 (1.00) & 0.92 (0.92) & 0.06 (0.06) & 0.00 (0.00) & 0.00 (0.00) & 0.00 (0.00) \\
& 1 & 1.00 (1.00) & 1.00 (1.00) & 1.00 (1.00) & 0.59 (0.58) & 0.00 (0.00) & 0.00 (0.00) \\
& 2 & 1.00 (1.00) & 1.00 (1.00) & 1.00 (1.00) & 1.00 (1.00) & 1.00 (1.00) & 0.00 (0.00) \\
500 & 0  & 0.05 (0.05) & 0.00 (0.00) & 0.00 (0.00) & 0.00 (0.00) & 0.00 (0.00) & 0.00 (0.00) \\
& 0.5 & 1.00 (1.00) & 0.99 (0.99) & 0.09 (0.10) & 0.00 (0.00) & 0.00 (0.00) & 0.00 (0.00) \\
& 1 & 1.00 (1.00) & 1.00 (1.00) & 1.00 (1.00) & 0.89 (0.88) & 0.00 (0.00) & 0.00 (0.00) \\
& 2 & 1.00 (1.00) & 1.00 (1.00) & 1.00 (1.00) & 1.00 (1.00) & 1.00 (1.00) & 0.00 (0.00) \\
1000 & 0  & 0.05 (0.05) & 0.00 (0.00) & 0.00 (0.00) & 0.00 (0.00) & 0.00 (0.00) & 0.00 (0.00) \\
& 0.5 & 1.00 (1.00) & 1.00 (1.00) & 0.20 (0.21) & 0.00 (0.00) & 0.00 (0.00) & 0.00 (0.00) \\
& 1& 1.00 (1.00) & 1.00 (1.00) & 1.00 (1.00) & 1.00 (1.00) & 0.00 (0.00) & 0.00 (0.00) \\
& 2 & 1.00 (1.00) & 1.00 (1.00) & 1.00 (1.00) & 1.00 (1.00) & 1.00 (1.00) & 0.00 (0.00) \\
5000 & 0  & 0.05 (0.05) & 0.00 (0.00) & 0.00 (0.00) & 0.00 (0.00) & 0.00 (0.00) & 0.00 (0.00) \\
& 0.5 & 1.00 (1.00) & 1.00 (1.00) & 0.88 (0.88) & 0.00 (0.00) & 0.00 (0.00) & 0.00 (0.00) \\
& 1& 1.00 (1.00) & 1.00 (1.00) & 1.00 (1.00) & 1.00 (1.00) & 0.00 (0.00) & 0.00 (0.00) \\
& 2 & 1.00 (1.00) & 1.00 (1.00) & 1.00 (1.00) & 1.00 (1.00) & 1.00 (1.00) & 0.00 (0.00) \\
10000 & 0  & 0.06 (0.05) & 0.00 (0.00) & 0.00 (0.00) & 0.00 (0.00) & 0.00 (0.00) & 0.00 (0.00) \\
& 0.5 & 1.00 (1.00) & 1.00 (1.00) & 1.00 (1.00) & 0.00 (0.00) & 0.00 (0.00) & 0.00 (0.00) \\
& 1& 1.00 (1.00) & 1.00 (1.00) & 1.00 (1.00) & 1.00 (1.00) & 0.00 (0.00) & 0.00 (0.00) \\
& 2 & 1.00 (1.00) & 1.00 (1.00) & 1.00 (1.00) & 1.00 (1.00) & 1.00 (1.00) & 0.00 (0.00) \\
\hline
\end{tabular}}
}
\label{tab:additionaltheorytable}
\end{table}

The experiments in the 
preceding section show that 
the proposed method performs well even when the data is missing not at random.
This section provides results showing we can estimate the power of the proposed
method when data can be assumed to be missing completely at random.
Consequently, we change notation slightly and consider sets of
\emph{random variables} rather than observations.

\begin{proposition} 
    \label{prop:mcar1}
    Let $\bx = \{ \bx_1, \cdots, \bx_{\n} \}$ and 
    $\by = \{\by_1, \cdots, \by_{\m} \}$ be sets of random variables 
    identically and independently 
    distributed according to continuous distributions $F$ and $G$, respectively. 
    Let $\bxp \subset \bx$ and $\byp \subset \by$ denote the subsets of distinct
    random variables with sizes $\np$ and $\mpp$, respectively, which can be 
    observed. 
    Assume that the observations of the 
    random variables in the complements $\bx \setminus \bxp$ 
    and $\by \setminus \byp$ are missing completely at random,
    and $0 < \prob(\bx_1 < \by_1) < 1$. Then, for any given significance level 
    $\alpha \in (0,1)$, with probability approximately equal to 
    $\cdfphi (\{L - \mu'\}/\sigmap) + 1 - \cdfphi( \{R - \mup\}/\sigmap)$
    the proposed method yields a significant $p$-value, 
	where $\cdfphi$ is the cumulative distribution function of the standard 
    normal distribution and $\mu = \n \m/2$ and $\sigma^2 = \n \m(\n+\m+1)/12$ 
    and
	\begin{align}
		&L = \sigma \cdfphi^{-1}(\alpha/2) + \mu - \n\m +\np\mpp , \quad
		R = \sigma \cdfphi^{-1}(1-\alpha/2) + \mu ,
        \nonumber\\ 
        &p_1 = \prob(X_1 < Y_1), \,\,
        p_2 = \prob(X_1 < Y_1 \textrm{ and } X_1 < Y_2),  \,\,
        p_3 = \prob(X_1 < Y_1 \textrm{ and } X_2 < Y_1).
        \nonumber\\ 
		&\mup  = \mpp \np p_1, \quad
        (\sigmap)^2 = \mpp \np p_1(1-p_1) + \mpp \np(\np-1)(p_2-p_1^2) + \np\mpp (\mpp-1)(p_3-p_1^2).
        \nonumber 
	\end{align}
\end{proposition}
The proof, based on similar results in \citep[Section 2.3]{lehmann1975} 
for the case when data is not missing, is in Section~4 of the Supplementary Material.

Table~\ref{tab:additionaltheorytable} shows the estimated power obtained
from Monte Carlo simulations compared to the theoretically-estimated power
from Proposition~\ref{prop:mcar1}, when data are missing completely at random,
as the proportion of data missing, $s$, varies. The table shows that  
the theoretical results closely match those obtained in simulation, at least
for this case where the data are sampled from normal distributions.
\begin{remark}
    \label{rem:propxy01}
    The condition $0 < \prob(\bx_1 < \by_1) < 1$ required in the 
    above proposition is necessary for the rescaled Wilcoxon-Mann-Whitney
    statistic to tend to the standard normal distribution asymptotically
    \citep[Appendix, Example 20]{lehmann1975}. If this condition does
    not hold, i.e. $\prob(\bx_1 < \by_1) \in \{0,1\}$, then one
    of the distributions $F$ or $G$ lies entirely to the left of the 
    other distribution \citep[Section 2.3]{lehmann1975}.
\end{remark}

This section ends with the following result which shows that, 
in the case data are missing completely at random, the probability
of the proposed method yielding a significant $p$-value tends to either
$0$ or $1$, depending on the proportion of missing data,
as the total sample sizes increase. The assumptions 
are the same as for Proposition~\ref{prop:mcar1}.
\begin{proposition}
\label{prop:mcar2lim}
Let $\bx = \{ \bx_1, \cdots, \bx_{\n} \}$ and 
$\by = \{\by_1, \cdots, \by_{\m} \}$ be sets of random variables 
identically and independently 
distributed according to continuous distributions $F$ and $G$, respectively. 
Let $\bxp \subset \bx$ and $\byp \subset \by$ denote the subsets of distinct
random variables with sizes $\np$ and $\mpp$, respectively, which can be 
observed. 
Assuming that the random variables in the complements $\bx \setminus \bxp$ 
and $\by \setminus \byp$ are missing completely at random,
and $0 < \prob(\bx_1 < \by_1) < 1$, and additionally assuming 
$\mpp \leq \np$ and
%
\begin{align}
& \mpp/\np \to \lambdap \text{ when } 
    \n,\m \to \infty, \text{ where } 0 < \lambdap \leq 1, 
    \nonumber \\
&\np/\n \to \lambdax, \text{ when } \n,\m \to \infty, 
    \text{ where } 0 < \lambdax \leq 1, 
    \nonumber \\
&\mpp/\m \to \lambday, \text{ when } \n,\m \to \infty, 
    \text{ where } 0 < \lambday \leq 1 
    \nonumber \\
&\lambda_{x} \lambday (\pOne - 1) + 1/2  \neq 0 \text{ and } 
    \lambda_{x} \lambday \pOne - 1/2  \neq 0, \text{ where }
    \pOne=\prob(\bx_1 < \by_1). \nonumber
\end{align}
As $\n,\m \to \infty$, with probability 
\begin{align}
    p=
\left\{ \begin{array}{cl}
    0, &  \text{ if }\lambdax \lambday (\pOne - 1)+1/2 > 0 \text{ and }  
    \lambdax \lambday \pOne -1/2  < 0,\\
    1, &\text{ otherwise,} 
\nonumber
\end{array}\right.
\end{align}
the proposed method will yield a significant $p$-value.
\end{proposition}

The next section investigates the case when we drop the assumption
that the data are distinct and allow ties.


\subsection{Bounding the statistics in the presence of ties and closed support}

\label{sec:ties}

For discrete-valued data, it is possible that two 
or more observations can be equal. 
In such cases, these values 
would all be assigned the same rank, e.g. if $\zz_{1}, \zz_{2} \in \bz$ 
are such that $\zz_{1} = \zz_{2}$, 
then $\rankfun{\bz}{\zz_{1}} = \rankfun{\bz}{\zz_{2}}$; 
in this section, we shall
consider $\bx, \by$ and $\bz=\bx \cup \by$ to be \emph{multisets}, rather 
than sets. We follow the practice of 
using \emph{midranks}; if there were exactly $\kk-1$ smaller values than 
$\zz_{1}$ and $\zz_{2}$, and these two values were distinct,
then $\zz_{1}$ and $\zz_{2}$ would occupy ranks $\kk$ and $\kk+1$, but if 
$\zz_{1}=\zz_{2}$, then these two values are each assigned the 
arithmetic mean rank $\{\kk + (\kk+1) \}/2 = \kk+ 1/2$.
The same argument holds for cases when more than two values are equal;
when there
is an odd number of equal values then their midranks will be an integer, and 
for an even number of equal values then their midranks will be a half-integer.
Recalling the earlier definition of the set
$\subs{\bz}{\aconst} = \{\zz \in \bz \suchthat \zz < \aconst\}$, we now define
$\partial \subs{\bz}{\aconst} = \{\zz \in \bz \suchthat \zz = \aconst\}$; 
the notation implying that $\partial \subs{\bz}{\aconst}$ is the boundary of the 
set $\subs{\bz}{\aconst}$.
Again using $\absval{S}$ to denote the size of the set $S$, 
the general definition of the rank of $\zz \in \bz$, 
including in the presence of ties, is
$\rankfun{\bz}{\zz} = \absval{\subs{\bz}{\zz}} + (\absval{\partial \subs{\bz}{\zz}} + 1)/2.$
The appendix contains 
a short proof showing this is equivalent to the definition in the 
above discussion.
This is consistent with the definition of rank when
all the elements are distinct, i.e. when 
$\absval{\partial \subs{\bz}{\zz}} = 1$ then
$\rankfun{\bz}{\zz} = \absval{\subs{\bz}{\zz}} + 1$, as before.
A nice property of using midranks is that the identity
$\rankgiven{\bz}{\bz} = \N(\N+1)/2$ still holds, where $\absval{\bz} = \N$. 

When there are ties, the Wilcoxon-Mann-Whitney statistic for
multisets $\bx = \{ \bx_1, \cdots, \bx_{\n} \}$ and 
$\by = \{\by_1, \cdots, \by_{\m} \}$ still
asymptotically follows a normal distribution with mean $\mu=\n\m/2$, 
but the variance $\sigma^2$ takes the ties into account; 
if $\bx \cup \by$ contains $e$ distinct values, with
multiplicities $d_1, \dots, d_{e}$, 
where $d_i$ is the $i$th smallest distinct value
(so $d_1 + \dots + d_{e} = \n+\m$), then 
\citet{lehmann1975} gives the variance as
\begin{align}
    \sigma^2 &=
    \sigma^2 (\bx, \by) =
    \n\m(\n+\m+1)/12 
    - \n\m \{ 12(\n+\m)(\n+\m-1) \}^{-1} \sum_{i=1}^{e} (d_i^3 - d_i).
    \label{eqn:varianceties}
\end{align}
Comparing this with the variance $\n\m(\n+\m+1)/12$ when there are no ties, 
the second term in the above equation essentially corrects for ties and
reduces the variance.

When considering ties, we also consider the case of \emph{closed support}, 
or when there is a defined upper and/or lower limit to the space from 
which the data are observed. We prefer this terminology since it includes
cases where data may only be bounded on one side, such as 
data with support $[0, \infty)$.
\begin{remark}
    As an example of why it may be necessary to consider the case of closed 
    support, suppose $\bx$ consists of $5$ values in the set 
    $\mathbb{Z}^{+}=\{1, 2, \dots\}$, and $\bxp \subset \bx$ is observed as
    $\{2, 3, 5\}$. Although $\bxp$ consists of distinct values, the two missing
    values in $\bx \setminus \bxp$ could both be $1$, to potentially minimise
    any sum of ranks of $\bx$ within some larger multiset. 
    The space need not be discrete; for example, in the space $[0, \infty)$
    there could be multiple missing values which could potentially have had 
    the value $0$.
\end{remark}
Following the above discussions, one can extend 
Theorem~\ref{thm:mainthm} to the case where ties are allowed.

\begin{proposition}
    \label{prop:mainthmties}
    Suppose that $\bx = \{\xx_1, \ldots, \xx_{\n}\}$ and
    $\by = \{\yy_1, \ldots, \yy_{\m}\}$ are samples of 
    observations, which need not necessarily be distinct, 
    from a space $\discretespace \subset \realR$.
    Suppose that $\bxp \subset \bx$ and $\byp \subset \by$ are sub-multisets
    with sizes $\np$ and $\mpp$, respectively.
    Let $a = \min \discretespace$ if the minimum exists, otherwise define
    $\absval{ \partial \bxp_{a} } = \absval{ \partial \byp_{a} } =  0$.
    Let $b = \max \discretespace$ if the maximum exists, otherwise define
    $\absval{ \partial \bxp_{b} } = \absval{ \partial \byp_{b} } =  0$.
    Then
    the Wilcoxon-Mann-Whitney statistic $\wmwstat{\bx}{\by}$ is
    bounded as follows:
    \begin{align}
        \wmwstat{\bxp}{\byp} + \Tone/2
        \leq
        \wmwstat{\bx}{\by}
        \leq
        \wmwstat{\bxp}{\byp} + (\m \n - \mpp \np) - \Ttwo/2,
        \label{eqn:wmwboundsties}
    \end{align}
    where
    $\Tone = \absval{ \partial \byp_{a} } (\n - \np) 
    + \absval{ \partial \bxp_{b} } (\m - \mpp)$ and 
    $\Ttwo = \absval{ \partial \bxp_{a} } (\m - \mpp) 
    + \absval{ \partial \byp_{b} } (\n - \np)$
    are non-negative values
    depending on $a, b, \m, \mpp, \n, \np, \bxp$ and $\byp$.
\end{proposition}
\begin{remark}
    The result above covers cases where the space $\discretespace$ is 
    discrete and bounded, e.g. $\{0, 1, 2\}$, or continuous and bounded, e.g. 
    $[0, 1]$, or bounded only on one side, e.g. $\mathbb{Z}^{+}=\{1, 2,\dots\}$,
    or $[0, \infty)$. See 
    Section~5 in the Supplementary Material for 
    the proof and examples. If 
    $a, b \not \in \bxp \cup \byp$, then 
    $\Tone = \Ttwo = 0$ and the result reduces to Theorem~\ref{thm:mainprop}.
\end{remark}

Now that the bounds for statistic $\wmwstat{\bx}{\by}$ are available, the next
step is to compute the bounds for its $p$-value. 
Care must be taken, however, not to simply compute the $p$-values for the 
minimum and maximum statistics from Proposition~\ref{prop:mainthmties};
the reason is that now the variance can take on different values depending on
ties, as shown in Equation~\eqref{eqn:varianceties}. The following result
illustrates the issue.

\begin{proposition}
    \label{prop:tiescandidates}
    Suppose that $\bx = \{\xx_1, \ldots, \xx_{\n}\}$ and
    $\by = \{\yy_1, \ldots, \yy_{\m}\}$ are samples of real-valued 
    observations, which need not necessarily be distinct. 
    Suppose that $\bxp \subset \bx$ and $\byp \subset \by$ are sub-multisets
    with sizes $\np$ and $\mpp$, respectively, which are known.
    Define $\bz = \bx \cup \by$ and $\bzp=\bxp \cup \byp$
    and suppose that $\bz \setminus \bzp$ is unknown.
    Suppose $\bxone, \byone$ and $\bxtwo, \bytwo$ are such that
    $\bxp \subset \bx_{i}$ and $\byp \subset \by_{i}$ for $i = 1,2$ and
    \begin{align}
        \wmwstat{\bxone}{\byone} = \min_{\bz \setminus \bzp} 
        \wmwstat{\bx}{\by}, \qquad
        \wmwstat{\bxtwo}{\bytwo} = \max_{\bz \setminus \bzp} 
        \wmwstat{\bx}{\by}.
        \nonumber
    \end{align}
    Then, among candidates for $\bx, \by$,
    it is possible $\bxthree, \bythree$ exist such that 
    $\bxp \subset \bx_{3}$ and $\byp \subset \by_{3}$ and
    \begin{align}
        \wmwstat{\bxone}{\byone} \leq \wmwstat{\bxthree}{\bythree} 
        \leq \wmwstat{\bxtwo}{\bytwo},
        \nonumber
    \end{align}
    but moreover, after defining $\mu=\n\m/2$, the standardised statistics are such that
    \begin{align}
        (\wmwstat{\bxthree}{\bythree} -\mu)/\sigma(\bxthree, \bythree) \not \in
        [ (\wmwstat{\bxone}{\byone} -\mu)/\sigma(\bxone, \byone),
        (\wmwstat{\bxtwo}{\bytwo} -\mu)/\sigma(\bxtwo, \bytwo)],
        \nonumber
    \end{align}
    and consequently $p_3 \not \in [p_{\min}, p_{\max}]$, where
    $p_{\min} = \min\{p_1, p_2\}$ and $p_{\max} = \max\{p_1, p_2\}$ if 
    Condition~\ref{cond:B} holds otherwise $p_{\max} = 1$, and
    $p_1$, $p_2$ and $p_3$ are the $p$-values for 
    $\wmwstat{\bxone}{\byone}$, $\wmwstat{\bxtwo}{\bytwo}$ and $\wmwstat{\bxthree}{\bythree}$, 
    respectively.
\end{proposition}
In order to deal with the issue raised by Proposition~\ref{prop:tiescandidates}, 
we start by computing the minimum and maximum
possible values for the variance $\sigma^2(\bx, \by)$ in the presence of ties
and closed support.
\begin{proposition}
    \label{prop:sigmaties}
    Suppose that $\bx = \{\xx_1, \ldots, \xx_{\n}\}$ and
    $\by = \{\yy_1, \ldots, \yy_{\m}\}$ are samples of 
    observations, which need not necessarily be distinct, from a space
    $\discretespace \subset \realR$. 
    Suppose that $\bxp \subset \bx$ and $\byp \subset \by$ are sub-multisets
    with sizes $\np$ and $\mpp$, respectively, which are known, and that
    $\bxp \cup \byp$ contains $e'$ distinct values with   
    multiplicities $d'_1, \dots, d'_{e'}$ and $d'_1 \leq \dots \leq d'_{e'}$.
    Define
    \begin{align}
        \sigma_{\max}^2(\bx, \by) &=
    \n\m(\n+\m+1)/12 
        - \n\m \{ 12(\n+\m)(\n+\m-1) \}^{-1}\sum_{i=1}^{e'} \{(d'_i)^3 - d'_i\},
        \nonumber \\
        \sigma_{\min}^2(\bx, \by) &= \sigma_{\max}^2(\bx, \by)
        - \n\m \{ 12(\n+\m)(\n+\m-1) \}^{-1} 
        \{ \dstar^3 - \dstar - (d'_{e'})^3 + d'_{e'} \},
        \nonumber
    \end{align}
    where $\dstar = d'_{e'} + \n+\m-\np-\mpp$,
    then $\sigma^2(\bx, \by)$, the variance of Wilcoxon-Mann-Whitney statistic
    $\wmwstat{\bx}{\by}$, 
    is bounded by
    $\sigma_{\min}^2(\bx, \by) \leq \sigma^2(\bx, \by) 
    \leq \sigma_{\max}^2(\bx, \by).$
\end{proposition}



\subsection{The proposed method in the presence of ties and closed support}

Extending results from the previous section, we can apply the proposed method
to cases where there may be tied observations and the sample space may have
closed support, by using the following result.
\begin{corollary}
    \label{cor:tiesresult}
    Suppose that $\bx = \{\xx_1, \ldots, \xx_{\n}\}$ and
    $\by = \{\yy_1, \ldots, \yy_{\m}\}$ are samples of 
    observations, which need not necessarily be distinct,
    from a space $\discretespace \subset \realR$.
    Suppose that $\bxp \subset \bx$ and $\byp \subset \by$ are sub-multisets
    with sizes $\np$ and $\mpp$, respectively, which are known, 
    and compute $\partial \bxp_{a}$, $\partial \bxp_{b}$,
    $\partial \byp_{a}$ and $\partial \byp_{b}$ from 
    Proposition~\ref{prop:mainthmties} and 
    $\sigma_{\max}^2(\bx, \by)$ from Proposition~\ref{prop:sigmaties},
    and define
    \begin{align}
        &\WminXY =
        \wmwstat{\bxp}{\byp} + \absval{ \partial \byp_{a} } (\n - \np)/2 
        + \absval{ \partial \bxp_{b} } (\m - \mpp)/2,
        \quad \mu = \n\m/2,
        \nonumber \\
        &\WmaxXY =
        \wmwstat{\bxp}{\byp} + (\m \n - \mpp \np) - 
        \absval{ \partial \bxp_{a} } (\m - \mpp) 
        + \absval{ \partial \byp_{b} } (\n - \np),
        \nonumber \\
        &p_1 = 2\cdfphi( -\absval{\WminXY - \mu}/\sigma_{\max}(\bx, \by) ),
        \quad
        p_2 = 2\cdfphi( -\absval{\WmaxXY - \mu}/\sigma_{\max}(\bx, \by) ).
        \nonumber
    \end{align}
    Then given a significance threshold $\alpha \in (0, 1)$ and 
    defining $p = \max\{p_1, p_2\}$, if $p < \alpha$
    and the two terms $(\WminXY - \mu)$ and $(\WminXY - \mu)$ 
    have the same sign, then the data 
    will yield a significant result, regardless of the values in 
    $(\bx \cup \by) \setminus (\bxp \cup \byp)$.
\end{corollary}

\begin{remark}
    It may seem counterintuitive that Corollary~\ref{cor:tiesresult} only 
    relies on $\sigma_{\max}(\bx, \by)$ and not $\sigma_{\min}(\bx, \by)$,
    but reasons for this are made clear in the proof, in the 
    Section 6 of the Supplementary Material.
\end{remark}

\begin{figure}
    \includegraphics[width=\textwidth]{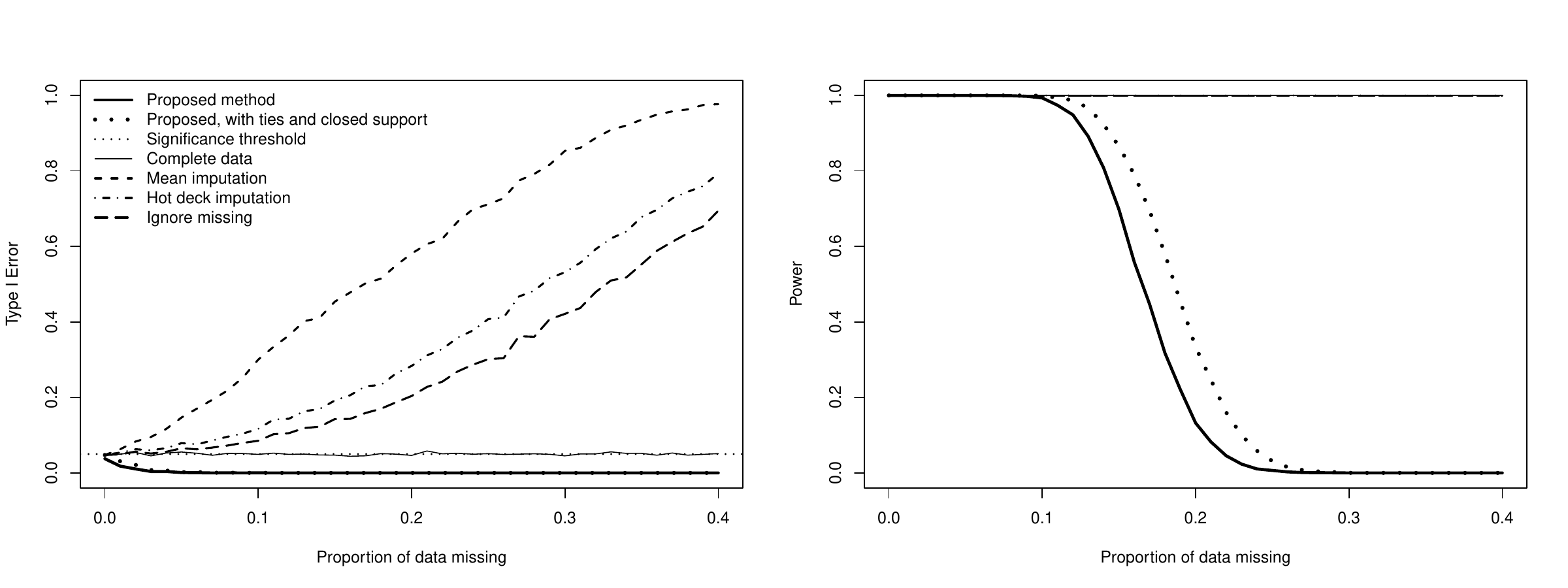}
    \caption{The Type I error and statistical power of the proposed method and 
    the standard Wilcoxon-Mann-Whitney test after the missing data is either known or
    has been imputed or ignored as the proportion of missing data increases.
    The data for the first sample $\bx$ is missing completely at random but the
    data for the second sample $\by$ is missing not at random, according to the mechanism in 
    Equation~\eqref{eqn:missmechone}, where only observations greater than $0$
    are possibly missing.
    (Left) Type I error: $\mathrm{Pois}(1)$ vs $\mathrm{Pois}(1)$; 
    (Right) Power: $\mathrm{Pois}(1)$ vs $\mathrm{Pois}(3)$. For both figures, 
    a significance threshold of $\alpha=0.05$ has been used and the total
    sample sizes are $\n=100$, $\m=100$, and $5000$ trials were used.}
\label{fig:mnar1ties}
\end{figure}

We ran a simulation to demonstrate the advantage of accounting for ties
when using the proposed method instead of using the original version from
Section~\ref{sec:mainresults}. The experiment is similar to the one shown in 
Figure~\ref{fig:mnar1}, but uses Poisson distributions rather than
normal distributions.
The data in first sample $\bx$ are independent observations of
a $\mathrm{Pois}(1)$ random variable with data missing completely at random.
The data in the second sample $\by$ are independent observations of 
a $\mathrm{Pois}(\Delta)$ random variable with data missing not at random;
only values greater than $0$ can be missing. For the Type I error plot
$\Delta=1$ and for the plot of statistical power, $\Delta=3$.
The results in Figure~\ref{fig:mnar1ties} show the same methods as in 
previous plot, but with the proposed method accounting for ties as
the thickly-dotted line. While still controlling the Type I error, 
the modified methods shows an improvement in statistical power over the 
original proposed method which doesn't account for ties or closed support.
Section 7.3 in the Supplementary Material shows additional figures for 
this case, such as when the sample sizes increase.


\section{Application to clinical trial data}

In this section we demonstrate the proposed method to analyze
clinical trial data that contains missing values.

Patients with type 2 diabetes may suffer from kidney disease, or 
\emph{diabetic nephropathy}. 
Albumin is a protein in the blood that a healthy kidney will 
not allow to pass into the urine \citep{NIDDK}. However, a damaged kidney 
may allow albumin to pass into the urine, and a high level of 
albumin in the urine may indicate kidney disease. 
Persistent \emph{albuminuria} is defined as a 
urinary albumin-creatinine ratio (UACR) that is greater than or equal to
$30$ mg/g \citep{bakrisetal2015}.

The clinical trial NCT01874431 \citep{bayervivli} 
assessed the efficacy of different doses
of the drug finerenone (BAY 94-8862) for reducing the 
urinary albumin-creatinine ratios of patients 
with type 2 diabetes and diabetic nephropathy.
This dataset is available from Bayer via the Vivli 
platform. 
There were 1501 patients enrolled in the trial, of whom 821 were 
randomly allocated to eight different treatment groups while the remainder
were excluded. 
One treatment group received a placebo, while the other seven groups
each received finerenone at a different dosage level.
The number of patients in each treatment group is
shown in Table~\ref{tab:UACRcohort}.

\begin{table}
    \def~{\hphantom{0}}
    \caption{Treatment groups in clinical trial NCT01874431.
        The number of patients in each treatment group, by dosage level,
        that start and complete the trial, with the percentage missing shown.}{%
\begin{tabular}{lccc}
\\
    Finerenone dosage    & Started & Completed & $\%$ missing\\
\hline
    $0$~mg (Placebo)       & ~94     & ~90  & ~4.3 \\
    $1.25$~mg              & ~96     & ~90  & ~6.3 \\
    $2.5$~mg              & ~92     & ~87  & ~5.4 \\
    $5$~mg                 & 100     & ~89  & 11.0 \\
    $7.5$~mg               & ~97     & ~89  & ~8.2 \\
    $10$~mg                & ~98     & ~89  & ~9.2 \\
    $15$~mg                & 125     & 114  & ~8.8 \\
    $20$~mg                & 119     & 112  & ~5.8 \\
    \hline
\end{tabular}}
\label{tab:UACRcohort}
\end{table}

The patients' urinary albumin-creatinine ratios were measured 
at the start of the trial (baseline value) as well as on a further five visits, 
the fifth and final visit being on day 90 of the trial. 
The difference in patients' urinary albumin-creatinine ratios at the start and 
end of the trial was compared across the different treatment groups 
and the placebo by \citet{bakrisetal2015} using an ANCOVA model and an $F$-test.
They found a significant improvement for the treatment groups with 
finerenone dosage levels of $7.5$~mg, $10$~mg, $15$~mg and $20$~mg, which in 
this case means that the obtained $p$-values are $0.004$ or less.
However, not all patients who 
start the trial completed the trial; there is
a drop-out rate of between $4\%$ and $11\%$ for each treatment group; 
see Table~\ref{tab:UACRcohort} for details.
In other words, there are \emph{missing data} in the final 
urinary albumin-creatinine ratio measurements.
In the original analysis, these missing data were imputed by 
using the last recorded value from an earlier visit as the imputed value. 
A sensitivity analysis was then performed by repeating the primary analysis 
using different imputation methods; see \citet{bakrisetal2015} for details.

We now analyze these data, avoiding any imputation, using our proposed method.
Let $\bx$ be the set of placebo group measurements which consists of 
$\n=94$ patients. For $i=1, \dots, 94$, let $\xx_i = (c_i - s_i)/s_i$, 
where $s_i$ is patient $i$'s baseline (starting) urinary albumin-creatinine 
ratio measurement and $c_i$ is patient $i$'s day 90 (completion) 
corresponding
measurement. The $\xx_i$ values measure the proportional increase
or decrease in urinary albumin-creatinine ratio.
Four of the patients in the placebo group are missing $c_i$ values, and so four
$\xx_i$ values are missing values.
Next, let $\by_{1}$ be the similarly transformed values for the first treatment
group with finerenone dosage $1.25$~mg, and let $\by_{2}, \dots, \by_{7}$
be the data for the remaining treatment groups. Again, several values in 
$\by_{1}, \dots, \by_{7}$ will be missing data. We now use the proposed method
to compare the $\bx$ values with the $\by_{j}$ values for $j=1, \dots, 7$.

If each element in $\bx$ is an observation of a random variable $U$ and 
each element in $\by_{j}$ is an observation of a random variable $V$, then the
null hypothesis is that $U = V$. Since the alternative hypothesis is that
the treatment lowers the patients' urinary albumin-creatinine ratios, 
the alternative 
hypothesis is $U > V$, and so the $p$-values are one-sided.
The obtained $p$-values are shown in 
the second row of Table~\ref{tab:UACRpval}, with the 
$p$-values for the $15$~mg and $20$~mg treatment groups computed as $0.011$
and $1.4 \times 10^{-4}$, respectively.

To correct for multiple hypothesis
testing, we use the Holm method \citep{holm1979} to compute the adjusted 
$p$-values; this results in the $p$-values $0.064$ and $9.4 \times 10^{-4}$
for the $15$~mg and $20$~mg treatment groups, respectively. 
The Holm method is preferred since it is both more powerful than the Bonferroni
method and does not impose any independence assumptions on the $p$-values, 
unlike other approaches. 

\begin{table}
    \def~{\hphantom{0}}
    \caption{Results from statistical analysis, with $p$-values}{%
        \scalebox{0.9}{
\begin{tabular}{lccccccc}
\\
                                     &   1.25 mg  & 2.5 mg    & 5 mg      & 7.5 mg       & 10 mg                 & 15 mg                 & 20 mg                 \\ 
\hline
    Ignoring missing                 & 0.196      & 0.208     & 0.264     & 0.002        & $3.9 \times 10^{-4}$  & $7.4 \times 10^{-6}$  & $6.5 \times 10^{-8}$  \\
    Proposed method                  & 1.000      & 1.000     & 1.000     & 0.112        & 0.072                 & 0.011                 & $1.4 \times 10^{-4}$  \\
    Proposed, Holm-corrected  & 1.000      & 1.000     & 1.000     & 0.450        & 0.358                 & 0.064                 & $9.4 \times 10^{-4}$  \\
    \hline
\end{tabular}}
}
\label{tab:UACRpval}
\end{table}

Table~\ref{tab:UACRpval} shows all the computed $p$-values for our proposed
method and, for comparison, the $p$-values computed when ignoring missing data
and our proposed method reduces to the Wilcoxon-Mann-Whitney test.
This table shows two interesting features. First, our proposed
method yields a $p$-value for the $20$~mg treatment group that is significant
at the $0.001$ significance level, after correction for multiple hypothesis 
testing. However, we do not obtain significant (corrected) $p$-values for the 
other treatment groups. The second interesting feature is that the 
Wilcoxon-Mann-Whitney test on the data, ignoring missing values, 
gives significant $p$-values for the $7.5$~mg, $10$~mg, $15$~mg and $20$~mg
treatment groups, as in the analysis in \citet{bakrisetal2015}.
After correction, the Wilcoxon-Mann-Whitney $p$-values are all less than 
$0.007$; see the Supplementary Material for a version of 
Table~\ref{tab:UACRpval} with these details included.

To summarize these results, our proposed method obtains a significant result 
only for the $20$~mg treatment group and notably this result 
is robust to any form of missingness.
However, our purpose in providing the analysis of this dataset is to illustrate 
the impact of missing data in a complex study, rather than to provide
clinical recommendations.


\section{Acknowledgements}

The authors wish to thank H. Battey for feedback on an earlier draft of the 
manuscript.
This publication is based on research using data from data contributor 
Bayer that has been made available through Vivli, Inc. 
Vivli has not contributed to or approved, and is not in any way responsible 
for, the contents of this publication.
The Vivli platform can be accessed at \texttt{https://vivli.org/}.


\clearpage


\renewcommand{\thepage}{S\arabic{page}}
\renewcommand{\thesection}{S\arabic{section}}
\renewcommand{\thetable}{S\arabic{table}}
\renewcommand{\thefigure}{S\arabic{figure}}
\renewcommand{\theequation}{S\arabic{equation}}
\setcounter{section}{0}
\setcounter{equation}{0}



\begin{center}
\textbf{\LARGE Supplementary Material}
\end{center}

\section{Proofs of results in Section~3.1}

\subsection{Proof of Lemma~1}
We restate the result before providing its proof.

\begin{lemma}
    \label{lem:onemissing}
    Suppose $\bxp = \{\xx_1, \ldots, \xx_{\np}\}$ is a sample of  
    observations and $\xstar$ is another observation. 
    Let $\bx=\bxp \cup \{\xstar\}$ and suppose $\bx \subset \bz$. 
    For fixed $\bz \setminus \{ \xstar \}$, the rank sum of $\bx$ in $\bz$,
    \begin{equation}
        \rankgiven{\bx}{\bz} = \rankgiven{\bxp}{\bz} + \rankfun{\bz}{\xstar},
        \nonumber
    \end{equation}
    is minimised when $\rankfun{\bz}{\xstar} < \rankfun{\bz}{\zz}$,
    for all $\zz \in \bz \setminus \{ \xstar \}$, if all values in $\bx$ are distinct.
\end{lemma}

\begin{proof}
    Let us start by defining 
    $\bzp = \bz  \setminus \{ \xstar \}$ and 
    $\by = \bzp \setminus \bxp$. Then
    $\bzp = \bxp \cup \by$
    and
    \begin{equation}
        \rankgiven{\bxp}{\bzp} = \sum_{i=1}^{\np} \rankfun{\bzp}{ \xx_i }.
        \nonumber
    \end{equation}
    The statement of the result assumes
    $\bzp = \bz  \setminus \{ \xstar \}$ is fixed.
    Since all the values in $\bz$ are assumed to be distinct, 
    they have distinct ranks, and there are two main cases to consider:
    either
    \begin{enumerate}
        \item for all $\zzp \in \bzp$,
            $\rankfun{\bz}{\xstar} < \rankfun{\bz}{\zzp}$, or 
        \item there is at least one $\zzp \in \bzp$ such that 
            $\rankfun{\bz}{\zzp} < \rankfun{\bz}{\xstar}$.
    \end{enumerate}
    We will show that the first case yields 
    $\rankgiven{\bx}{\bz}^{\min}$,
    the smallest possible 
    $\rankgiven{\bx}{\bz}$, while in the second case, which will be further 
    partitioned into subcases, yields 
    $\rankgiven{\bx}{\bz} \geq \minrankgiven{\bx}{\bz}$
    or
    $\rankgiven{\bx}{\bz} > \minrankgiven{\bx}{\bz}$.

    In the first case, $\rankfun{\bz}{\xstar} < \rankfun{\bz}{\zzp}$
    implies, for $i=1, 2, \dots, \np$, that $\xstar < \xx_i$ 
    (where $\xx_i \in \bxp \subset \bzp$, for $i=1, 2, \ldots, \np$) which in 
    turn implies $\rankfun{\bz}{\xx_{i}} = \rankfun{\bzp}{\xx_{i}} + 1$ and so
    \begin{align}
        \rankgiven{\bxp}{\bz} 
        = \sum_{i=1}^{\np} \rankfun{\bz}{ \xx_i } 
        = \sum_{i=1}^{\np} (\rankfun{\bzp}{ \xx_i } + 1 ) 
        = \rankgiven{\bxp}{\bzp} + \np .
        \nonumber
    \end{align}
    Now by definition, 
    \begin{align}
        \rankgiven{\bx}{\bz} &= \rankgiven{\bxp}{\bz} + \rankfun{\bz}{\xstar}.
        \nonumber
    \end{align}
    Since $\rankfun{\bz}{\xstar} < \rankfun{\bz}{\zz}$ 
    for all $\zz \in \bzp = \bz  \setminus \{ \xstar \}$,
    this implies $\rankfun{\bz}{\xstar} = 1$ and so
    \begin{align}
        \rankgiven{\bx}{\bz} &= \rankgiven{\bxp}{\bz} + \rankfun{\bz}{\xstar}
                            = \rankgiven{\bxp}{\bzp}+\np + \rankfun{\bz}{\xstar}
                            = \rankgiven{\bxp}{\bzp}+\np + 1.
        \label{eqn:lemmaproof:case1:rankxz}
    \end{align}
    Equation~\eqref{eqn:lemmaproof:case1:rankxz} always holds for 
    the first case
    (that $\rankfun{\bz}{\xstar} < \rankfun{\bz}{\zz}$ for all $\zz \in \bz$), 
    regardless of the ranks of the observations in $\by$.
    We shall now define $\minrankgiven{\bx}{\bz}=\rankgiven{\bxp}{\bzp}+\np+1$
    and later show that this is the minimum possible value for
    $\rankgiven{\bx}{\bz}$.

    In the second case, there is at least one $\zzp \in \bzp$ such that 
    $\rankfun{\bz}{\zzp} < \rankfun{\bz}{\xstar}$. This can be further broken
    down into two subcases:
    \begin{enumerate}[label=2\alph*.]
        \item there is at least one $\xp \in \bxp$ such that 
            $\rankfun{\bz}{\xp} < \rankfun{\bz}{\xstar}$, or
        \item there is at least one $\yy \in \by = \bz \setminus \bxp$ 
            such that $\rankfun{\bz}{\yy} < \rankfun{\bz}{\xstar}$.
    \end{enumerate}
    Cases 2a and 2b are not mutually exclusive, and we consider each
    in turn.
    For case 2a, there is at least one 
    $\xx \in \bxp$ with $\xx < \xstar$.
    Let $\xp = \max \{\xx \in \bxp \suchthat \xx < \xstar\}$;
    this is well-defined, since $\bxp$ is finite.
    Suppose further that $\xp$ is the $\kk$th smallest value in $\bxp$, 
    where $\kk \in \{1, 2, \dots, \np \}$.
    Then there are exactly $\np - \kk$ values, which we label 
    $\xx_1, \dots, \xx_{\np-\kk} \in \bxp$,
    that are larger than
    $\xstar$ and for each of these values 
    $\rankfun{\bz}{\xx_{i}} = \rankfun{\bzp}{\xx_{i}} + 1$
    (for $i=1, \ldots, \np-\kk$).
    Therefore, similar to the first case,
    \begin{align}
        \rankgiven{\bxp}{\bz} = \rankgiven{\bxp}{\bzp} + (\np - \kk).
        \label{eqn:lemmaproof:case3:rankxpz}
    \end{align}
    Since $\xp$ is the $\kk$th smallest value in $\bxp$, this implies
    $\rankfun{\bz}{\xp} \geq \kk$, since there could be a value 
    $\yy \in \bzp \setminus \bxp$ such that $\yy < \xp$.
    Therefore
    $\rankfun{\bz}{\xstar} \geq \rankfun{\bz}{\xp} + 1 \geq \kk + 1$. 
    Therefore, for case 2a,
    \begin{align}
        \rankgiven{\bx}{\bz} &= \rankgiven{\bxp}{\bz} + \rankfun{\bz}{\xstar}
                 =\rankgiven{\bxp}{\bzp} + (\np - \kk)
                 + \rankfun{\bz}{\xstar}
                 \nonumber \\
                  &\geq \rankgiven{\bxp}{\bzp} + (\np - \kk)
                            + \kk + 1
                  = \rankgiven{\bxp}{\bzp} + \np + 1,
        \nonumber
    \end{align}
    where the second equality is from Equation~\eqref{eqn:lemmaproof:case3:rankxpz}.
    So, for case 2a, $\rankgiven{\bx}{\bz} \geq \minrankgiven{\bx}{\bz}$, 
    from case 1.

    Finally, for case 2b, suppose there is a $\yy \in \by=\bzp \setminus \bxp$
    such that
    $\rankfun{\bz}{\yy} < \rankfun{\bz}{\xstar}$. 
    If there is no $\xp \in \bxp$ such that $\xp < \xstar$, then 
    $\xstar < \xp$ for all $\xp \in \bxp$ 
    (recall that all values in $\bz$ are distinct), and similar to case 1, 
    $\rankgiven{\bxp}{\bz} = \rankgiven{\bxp}{\bzp} + \np$.
    However, since there is a $\yy \in \by$ with 
    $\rankfun{\bz}{\yy} < \rankfun{\bz}{\xstar}$, now
    $\rankfun{\bz}{\xstar} > 1$, and 
    \begin{align}
        \rankgiven{\bx}{\bz} &= \rankgiven{\bxp}{\bz} + \rankfun{\bz}{\xstar}
                            > \rankgiven{\bxp}{\bzp}+\np + 1, 
        \nonumber
    \end{align}
    and so $\rankgiven{\bx}{\bz} > \minrankgiven{\bx}{\bz}$.

    If, on the other hand, 
    there is at least one $\xp \in \bxp$ such that $\xp < \xstar$, 
    then as in case 2a, define
    $\xp = \max \{\xx \in \bxp \suchthat \xx < \xstar\}$
    and suppose $\xp$ is the $\kk$th smallest value in $\bxp$.
    Then, again,
    $\rankgiven{\bxp}{\bz} = \rankgiven{\bxp}{\bzp} + (\np - \kk)$
    and
    $\rankfun{\bz}{\xp} \geq \kk$,
    but
    $\rankfun{\bz}{\xstar} > \kk + 1$, since $\xstar > \xp$ and 
    $\xstar > \yy$ (and so $\xstar$ is strictly larger than $\kk+1$ values 
    in $\bz$). Then
    \begin{align}
        \rankgiven{\bx}{\bz} &= \rankgiven{\bxp}{\bz} + \rankfun{\bz}{\xstar}
                 =
                 \rankgiven{\bxp}{\bzp} + (\np - \kk)
                 + \rankfun{\bz}{\xstar}
                 \nonumber \\
                  &> \rankgiven{\bxp}{\bzp} + (\np - \kk)
                            + \kk + 1
                 > \rankgiven{\bxp}{\bzp} + \np + 1, 
        \nonumber
    \end{align}
    which implies $\rankgiven{\bx}{\bz} > \minrankgiven{\bx}{\bz}$ for case 2b, 
    regardless of whether or not there is an $\xp \in \bxp$ such that 
    $\xp < \xstar$.

    Since in case 2a $\rankgiven{\bx}{\bz} \geq \minrankgiven{\bx}{\bz}$, and 
    in case 2b $\rankgiven{\bx}{\bz} > \minrankgiven{\bx}{\bz}$, together case 2
    (which is cases 2a and 2b together), implies  
    that $\rankgiven{\bx}{\bz} \geq \minrankgiven{\bx}{\bz}$.
    Therefore $\minrankgiven{\bx}{\bz} = \rankgiven{\bxp}{\bzp} + \np + 1$,
    which is the value of $\rankgiven{\bx}{\bz}$ in
    case 1, is the minimum possible value for 
    $\rankgiven{\bx}{\bz}$.

    Therefore, if $\rankfun{\bz}{\xstar} < \rankfun{\bz}{\zz}$ 
    for all $\zz \in \bz$, the value of $\rankgiven{\bx}{\bz}$ is minimized.

\end{proof}



\subsection{Proof for Proposition~1}

\begin{proposition}
    \label{prop:mainprop}
    Suppose that $\bx = \{\xx_1, \ldots, \xx_{\n}\}$ and
    $\by = \{\yy_1, \ldots, \yy_{\m}\}$ are samples of 
    distinct real-valued observations. 
    Suppose that $\bxp \subset \bx$ is a subset of $\np$ distinct 
    values in $\bx$, and suppose that $\byp \subset \by$ is a subset 
    of distinct $\mpp$ values in $\by$.
    Defining $\bz = \bx \cup \by$ and $\bzp = \bxp \cup \byp$
    and supposing only $\bzp$ is known,
    the minimum
    and maximum possible rank sums of $\bx$, over all possible 
    values for the $\dd = \n-\np + \m - \mpp$ observations 
    in $\bz \setminus \bzp$, are
    \begin{align}
        \min_{\bz \setminus \bzp \in \realR^{\dd}} \rankgiven{\bx}{\bz} 
        &= \rankgiven{\bxp}{\bzp} +  (\n - \np)(\n + \np + 1)/2,
        \label{eqn:prop:minR} \\
        \max_{\bz \setminus \bzp \in \realR^{\dd}} \rankgiven{\bx}{\bz} 
        &= \rankgiven{\bxp}{\bzp}  
           + \{ \n (\n + 2\m + 1) - \np (\np +2 \mpp + 1)\}/2.
           \label{eqn:prop:maxR}
    \end{align}
\end{proposition}

\begin{proof}
    Let $\dd = \n-\np + \m - \mpp$ denote the number of missing values 
    in $\bz$. We can think of the set of missing values, $\bz \setminus \bzp$,
    as an element of $\realR^{\dd}$. We first consider the minimum possible
    sum of ranks over all possible values for $\bz \setminus \bzp$.

    For convenience, relabel the elements of 
    $\bx = \{ \xx_1, \ldots, \xx_{\np}, \xx_{\np+1}, \ldots, \xx_{\n} \}$ so 
    that $\bxp = \{ \xx_1, \ldots, \xx_{\np} \}$ are the observed values and 
    $\bx \setminus \bxp = \{ \xx_{\np+1}, \ldots, \xx_{\n} \}$ are the 
    $\n-\np$ 
    missing values. Then
    \begin{align}
        \rankgiven{\bx}{\bz} 
        &= \sum_{i=1}^{\n} \rankfun{\bz}{ \xx_i } 
        \nonumber \\
        &= \sum_{i=1}^{\np} \rankfun{\bz}{ \xx_i } 
            + \sum_{j=1}^{\n-\np} \rankfun{\bz}{\xx_{\np + j} }
        \nonumber \\
        \Rightarrow 
        \rankgiven{\bx}{\bz} 
        &= \rankgiven{\bxp}{\bz}
            + \sum_{j=1}^{\n-\np} \rankfun{\bz}{\xx_{\np + j} }.
        \nonumber 
    \end{align}
    Using Lemma~\ref{lem:onemissing} repeatedly for each of these $\n - \np$ 
    missing values, 
    \begin{align}
        \min_{\bz \setminus \bzp \in \realR^{\dd}} \rankgiven{\bx}{\bz} 
        &= \min_{\bz \setminus \bzp \in \realR^{\dd}} \rankgiven{\bxp}{\bz}
            + (\n-\np)(\n-\np+1) /2.
            \label{eqn:mainprop:firsteq}
    \end{align}
    The last line follows since by taking the minimum over
    all $\bz \setminus \bzp \in \realR^{\dd}$, the
    values $\{\xx_{\np+1}, \ldots, \xx_{\n}\}$
    must be ranked $\{1, 2, \ldots, \n-\np\}$ in $\bz$ following 
    Lemma~\ref{lem:onemissing}, and
    \begin{align}
        1 + 2 + \cdots + (\n-\np) = (\n-\np)(\n-\np+1)/2. 
        \nonumber
    \end{align}
    Moreover, since the values $\{\xx_{\np+1}, \ldots, \xx_{\n}\}$ are
    ranked $\{1, 2, \ldots, \n-\np\}$, this implies they are smaller
    than all other values in $\bz$; in particular, they are smaller
    than all $\np$ observations in $\bxp$.
    Therefore for each $\xx \in \bxp$, 
    $\rankfun{\bz}{\xx} = \rankfun{\bzp}{\xx} + (\n-\np)$, and so
    \begin{align}
        \min_{\bz \setminus \bzp \in \realR^{\dd}} \rankgiven{\bxp}{\bz}
        &= \rankgiven{\bxp}{\bzp} + \np (\n - \np).
        \label{eqn:mainprop:secondeq}
    \end{align}
    Combining Equations~\eqref{eqn:mainprop:firsteq} 
    and \eqref{eqn:mainprop:secondeq},
    one obtains 
    \begin{align}
        \min_{\bz \setminus \bzp \in \realR^{\dd}} \rankgiven{\bx}{\bz} 
        &= \rankgiven{\bxp}{\bzp} + \np (\n - \np)
            + (\n-\np)(\n-\np+1)/2 
        \nonumber \\
        &= \rankgiven{\bxp}{\bzp} 
            + (\n-\np)(\n-\np+1 +2\np)  /2
        \nonumber \\
        \Rightarrow
        \min_{\bz \setminus \bzp \in \realR^{\dd}} \rankgiven{\bx}{\bz} 
        &= \rankgiven{\bxp}{\bzp} 
            + (\n-\np)(\n+\np+1) /2. 
        \nonumber
    \end{align}
    which is Equation~\eqref{eqn:prop:minR}.

    Next, to maximize $\rankgiven{\bx}{\bz}$, one starts as before 
    with the decomposition
    \begin{align}
        \rankgiven{\bx}{\bz} 
        &= \rankgiven{\bxp}{\bz}
            + \sum_{j=1}^{\n-\np} \rankfun{\bz}{\xx_{\np + j} }.
        \label{supp:eqn:sumranks}
    \end{align}
    One calculates that 
    \begin{align}
        \max_{\yy \in \realR} \rankgiven{\bxp}{ \bzp \cup \{ \yy \} }
        &= \rankgiven{\bxp}{\bzp} + \np,
        \label{eqn:prop:thirdeq}
    \end{align}
    because the maximum would occur when $\yy < \xp$ for all $\xp \in \bxp$, 
    which would imply $\rankfun{\bz}{\xp} = \rankfun{\bzp}{\xp} + 1$, 
    for all $\xp \in \bxp$, and there are $\np$ observations in $\bxp$. 
    Similarly to the previous case, one writes 
    $\by = \{ \yy_1, \ldots, \yy_{\mpp}, \yy_{\mpp+1}, \ldots, \yy_{\m} \}$ so 
    that $\byp = \{ \yy_1, \ldots, \yy_{\mpp} \}$ are the observed values and 
    $\by \setminus \byp = \{ \yy_{\mpp+1}, \ldots, \yy_{\m} \} \subset \bz \setminus \bzp$ 
    are the $\m - \mpp$ unknown values (the lemma assumes only $\bzp$ is known).
    Therefore, repeatedly applying Equation~\eqref{eqn:prop:thirdeq},
    \begin{align}
        \max_{\bz \setminus \bzp \in \realR^{\dd}} \rankgiven{\bxp}{\bz}
        &= \rankgiven{\bxp}{\bzp} + \np(\m - \mpp),
        \label{eqn:thm:main:max1} 
    \end{align}
    by considering all $\m-\mpp$ values in $\by \setminus \byp $ to be 
    less than all $\np$ values in $\bxp$.
    Next, 
    by considering the values in 
    $\bx \setminus \bxp = \{ \xx_{\np+1}, \ldots, \xx_{\n} \}$ to have maximal
    rank in $\bz$, and so be larger than all $\m$ values in $\by$ and 
    all $\np$ values in $\bxp$,
    \begin{align}
        \max_{\bz \setminus \bzp \in \realR^{\dd}} 
        \sum_{j=1}^{\n-\np} \rankfun{\bz}{\xx_{\np + j} }
        &= (\m + \np + 1) + (\m + \np + 2) + \cdots + (\m + \n) 
        \nonumber \\
        &= (\n - \np) (2\m + \n + \np + 1)/2,
        \nonumber 
    \end{align}
    where the second line follows by the formula for the sum
    of an arithmetic progression and some simplification (see the proof of
    Proposition~\ref{prop:simplifone}).
    Combining this expression with Equations~\eqref{supp:eqn:sumranks} and 
    \eqref{eqn:thm:main:max1},
    one obtains
    \begin{align}
        \max_{\bz \setminus \bzp \in \realR^{\dd}} \rankgiven{\bx}{\bz}
        &=  \rankgiven{\bxp}{\bzp} + \np(\m - \mpp) 
        + (\n - \np) (2\m + \n + \np + 1) /2
        \nonumber \\
        \Rightarrow
        \max_{\bz \setminus \bzp \in \realR^{\dd}} \rankgiven{\bx}{\bz}
        &=  \rankgiven{\bxp}{\bzp} 
           + \{ \n (\n + 2\m + 1) - \np (\np +2 \mpp + 1) \} /2,
    \end{align}
    after some simplification, which is Equation~\eqref{eqn:prop:maxR} and
    finally proves the result.
\end{proof}

\begin{proposition}
    \label{prop:simplifone}
    The following simplification holds:
    \begin{equation}
        \np(\m - \mpp) + \sum_{j=1}^{\n - \np} (\m + \np + j)
        = \{ \n (\n + 2\m + 1) - \np (\np +2 \mpp + 1) \} /2.
        \nonumber
    \end{equation}
\end{proposition}

\begin{proof}
    First, using the formula for an arithmetic progression with $q$ terms,
    \begin{align}
        a + (a+d) + (a+2d) + \dots + (a+ (q-1)d)
        &= q \{ 2a + (q-1)d \} /2,
        \nonumber
    \end{align}
    one obtains
    \begin{align}
        \sum_{j=1}^{\n-\np} (\m + \np + j)
        &= (\n - \np) \{ 2(\m + \np + 1) + (\n - \np - 1)(1)\} /2 
        \label{supp:eqn:subap} \\
        &= (\n - \np)(2\m + 2\np + 2 + \n - \np - 1  ) /2
        \nonumber \\
        &= (\n - \np)(2\m + \n + \np + 1) /2,
        \nonumber
    \end{align}
    where Equation~\eqref{supp:eqn:subap} 
    uses the formula for the arithmetic progression with 
    $a=\m + \np +1$, $d=1$ and $q= \n - \np$
    (since $q$ is the number of terms). 
    Then
    \begin{align}
        &\np(\m - \mpp) +  (\n - \np) (2\m + \n + \np + 1) / 2
        \nonumber \\
        &=  \m \np -  \mpp \np 
        + (2\m\n + \n^2 + \cancel{\n\np} + \n 
        - 2\m\np - \cancel{\n\np} - \np^2 - \np) /2
        \nonumber \\
        &= ( \cancel{2\m \np} - 2 \mpp \np  + 2 \m \n 
                        - \cancel{2 \m \np} + \n^2 + \n - \np^2 - \np) / 2
        \nonumber \\
        &= \{ (\n^2 + 2\m\n + \n) - (\np^2 +2 \mpp \np + \np ) \} /2
        \nonumber \\
        &= \{ \n (\n + 2\m + 1) - \np (\np +2 \mpp + 1) \} /2,
        \nonumber 
    \end{align}
    proving the result.
\end{proof}

\begin{remark}
    Another, and perhaps more elegant, approach to computing 
    $\max \rankgiven{\bx}{\bz}$ would be to start by
    to computing $\min \rankgiven{\by}{\bz}$ in the same way 
    as $\min \rankgiven{\bx}{\bz}$. Then one can use the
    fact that $\rankgiven{\bx}{\bz} + \rankgiven{\by}{\bz} 
    = \tfrac{1}{2}(\n+\m)(\n+\m+1)$, and that
    maximising $\rankgiven{\bx}{\bz}$ is the same as minimising
    $\rankgiven{\by}{\bz}$, so 
    $\max \rankgiven{\bx}{\bz} 
    = \tfrac{1}{2}(\n+\m)(\n+\m+1) - \min \rankgiven{\by}{\bz}$. 
    This yields the same expression as
    Equation~\eqref{eqn:prop:maxR} , via a more involved computation; 
    details are provided below. 
\end{remark}

\subsection{Alternative proof for Proposition~\ref{prop:mainprop}}

In the proof of Proposition~\ref{prop:mainprop}, 
suppose we wish to find an expression for $\max \rankgiven{\bx}{\bz}$. We start 
by using the first part of Proposition~\ref{prop:mainprop} which, 
computes $\min \rankgiven{\bx}{\bz}$. 
Applying the same argument to
$\by$ instead of $\bx$, by symmetry would give
\begin{align}
        \min_{\bz \setminus \bzp \in \realR^{\dd}} \rankgiven{\by}{\bz} 
        &= \rankgiven{\byp}{\bzp} + (\m - \mpp)(\m + \mpp + 1) /2.
        \label{eqn:altmaxproof:1}
\end{align}
Considering the set $\bzp = \bxp \cup \byp$ of $\np + \mpp$ values which have ranks
$\{1, 2, \ldots, \np + \mpp\}$, the following
equation must hold:
\begin{align}
    &\rankgiven{\bxp}{\bzp} + \rankgiven{\byp}{\bzp} = (\np+\mpp)(\np+\mpp+1) /2,
    \nonumber 
\end{align}
which in turn implies
\begin{align}
    \Rightarrow
    \rankgiven{\byp}{\bzp} &= - \rankgiven{\bxp}{\bzp} + (\np+\mpp)(\np+\mpp+1) /2
    \nonumber \\
    \Rightarrow
    -\rankgiven{\byp}{\bzp} &= \rankgiven{\bxp}{\bzp} - (\np+\mpp)(\np+\mpp+1) /2.
    \label{eqn:altmaxproof:2}
\end{align}
Similarly to above, since $\bz = \bx \cup \by$ consists of of $\n + \m$ values,
\begin{align}
    \rankgiven{\bx}{\bz} + \rankgiven{\by}{\bz} = (\n+\m)(\n+\m+1) /2 ,
    \nonumber
\end{align}
must also hold. Therefore, when $\rankgiven{\by}{\bz}$ is minimised, this means 
$\rankgiven{\bx}{\bz}$ is maximized, since their sum is constant. Therefore,
\begin{align*}
    &\max_{\bz \setminus \bzp \in \realR^{\dd}} \rankgiven{\bx}{\bz} 
    + \min_{\bz \setminus \bzp \in \realR^{\dd}}\rankgiven{\by}{\bz} 
    = (\n+\m)(\n+\m+1)/2
    \nonumber \\
    \Rightarrow
    &\max_{\bz \setminus \bzp \in \realR^{\dd}} \rankgiven{\bx}{\bz} 
    = - \min_{\bz \setminus \bzp \in \realR^{\dd}}\rankgiven{\by}{\bz} 
    + (\n+\m)(\n+\m+1)/2. 
\end{align*}
Applying equation \eqref{eqn:altmaxproof:1}, we further have
\begin{align*}
	\max_{\bz \setminus \bzp \in \realR^{\dd}} \rankgiven{\bx}{\bz} =     \left [-\rankgiven{\byp}{\bzp} - (\m - \mpp)(\m + \mpp + 1) /2 \right ]
	+ (\n+\m)(\n+\m+1) /2. 
\end{align*}    
Replacing $-\rankgiven{\byp}{\bzp}$ in above using the equation \eqref{eqn:altmaxproof:2} gives us    
\begin{align} 	\label{eqn:altmaxproof:3}
	\max_{\bz \setminus \bzp \in \realR^{\dd}} \rankgiven{\bx}{\bz} &=
	\left[ \left\{  \rankgiven{\bxp}{\bzp} - (\np+\mpp)(\np+\mpp+1) /2 \right \}
	- (\m - \mpp)(\m + \mpp + 1) /2 \right ]
	\nonumber \\
	& \qquad \qquad + (\n+\m)(\n+\m+1) /2 .
\end{align}
We now show that the following term simplifies as follows:
\begin{align}
\begin{split}
	    &- (\np+\mpp)(\np+\mpp+1) - (\m - \mpp)(\m + \mpp + 1)
	+ (\n+\m)(\n+\m+1) 
	\\
	\\
	=& - (\np^2 + 2 \np \mpp + \cancel{\mpp^2} + \np + \xcancel{\mpp})
	- ( \bcancel{\m^2} - \cancel{\mpp^2} + \cancel{\m} - \xcancel{\mpp})
	\\
	& \qquad+ (\n^2 + 2 \n \m + \bcancel{\m^2} + \n + \cancel{\m})
	\\
	\\
	=& - (\np^2 + 2 \np \mpp + \np  ) + (\n^2 + 2 \n \m +  \n )
	\\
	\\
	=&  \n(\n + 2 \m + 1 ) -\np (\np + 2 \mpp + 1  ) . 
\end{split} \label{supp:eqn:combine}
\end{align}
which combined with the Equation~\eqref{eqn:altmaxproof:3} above gives,
\begin{align}
    \max_{\bz \setminus \bzp \in \realR^{\dd}} \rankgiven{\bx}{\bz} 
    &=  \rankgiven{\bxp}{\bzp} 
    + \left \{ \n(\n + 2 \m + 1 ) -\np (\np + 2 \mpp + 1)  \right \} /2,
    \nonumber
\end{align}
which is equivalent to the expression for the maximum in 
Proposition~\ref{prop:mainprop}.


\subsection{Proof of Theorem~1}

\begin{theorem}
    \label{thm:mainthm}
    Suppose that $\bx = \{\xx_1, \ldots, \xx_{\n}\}$ and
    $\by = \{\yy_1, \ldots, \yy_{\m}\}$ are samples of 
    distinct real-valued observations and
    $\bxp \subset \bx$ and $\byp \subset \by$ are subsets of
    distinct values with sizes $\np$ and $\mpp$, respectively. Then
    the Wilcoxon-Mann-Whitney statistic $\wmwstat{\bx}{\by}$ is
    bounded as follows:
    \begin{align}
        \wmwstat{\bxp}{\byp}
        \leq
        \wmwstat{\bx}{\by}
        \leq
        \wmwstat{\bxp}{\byp} + (\m \n - \mpp \np).
        \label{eqn:wmwbounds}
    \end{align}
\end{theorem}

\begin{proof}
    Writing $\bz = \bx \cup \by$, 
    Definition~1 for the Wilcoxon-Mann-Whitney statistic gives us
    \begin{align*}
    	\wmwstat{\bx}{\by} =
    	\rankgiven{\bx}{\bz} - \n(\n+1)/2.
    \end{align*}
    Applying Proposition~\ref{thm:mainthm}, we further have
    \begin{alignat}{2}
        \wmwstat{\bx}{\by}
        &{\geq}
        \rankgiven{\bxp}{\bzp} + (\n - \np)(\n + \np + 1) /2
        - \n(\n+1)/2
        \nonumber \\
        &= \rankgiven{\bxp}{\bzp} 
        + (\n^2 - \np^2 + \n - \np) /2
        - (\n^2 + \n) /2
        \nonumber \\
        &= \rankgiven{\bxp}{\bzp} 
        - \np(\np + 1) /2
        \nonumber \\
        &{=}
        \wmwstat{\bxp}{\byp}
        \nonumber \\
        \Rightarrow
        \wmwstat{\bx}{\by}
        &\geq
        \wmwstat{\bxp}{\byp} .
        \nonumber
    \end{alignat}
    Similarly, using
    Definition~1 and Equation~\eqref{eqn:prop:maxR} 
    of Proposition~\ref{prop:mainprop}, this gives us
    \begin{alignat}{2}
        \wmwstat{\bx}{\by}
        &{=} 
        \rankgiven{\bx}{\bz} - \n(\n+1) /2
        \nonumber \\
        &{\leq}
        \rankgiven{\bx}{\bz} 
        + \left \{\n (\n + 2\m + 1) - \np (\np +2 \mpp + 1) \right \} /2
        - \n(\n+1) /2
        \nonumber \\
        &= \rankgiven{\bx}{\bz} 
        + \left \{ 2\m\n + \n (\n + 1) - \np (\np +2 \mpp + 1) \right \} /2
        - \n(\n+1) /2
        \nonumber \\
        &= \rankgiven{\bx}{\bz} 
        + \left\{ 2\m\n - 2\mpp\np  - \np (\np + 1) \right \} /2
        \nonumber \\
        &= \rankgiven{\bx}{\bz} - \np (\np + 1) /2
        + (\m\n - \mpp\np)
        \nonumber \\
        &{=} 
        \wmwstat{\bxp}{\byp} + (\m\n - \mpp\np)
        \nonumber \\
        \Rightarrow
        \wmwstat{\bx}{\by} &\leq \wmwstat{\bxp}{\byp} + (\m\n - \mpp\np) ,
        \nonumber
    \end{alignat}
    which proves the theorem.
\end{proof}


\section{Proof of Lemma~2}

Recalling that $\cdfwmw{\n}{\m}$ is the cumulative distribution function of
the normal distribution with mean $\n\m/2$ and variance $\n\m(\n+\m+1)/12$,
and $\pvalfun$ is the function $\pvalfun(x) = 1 - \absval{1 - 2 x}$,
it will be helpful to restate the three conditions:
\begin{condition}
    \label{cond:SA}
$\pvalfun \circ \cdfwmw{\n}{\m}(\WminXY) < \alpha$ and
$\pvalfun \circ \cdfwmw{\n}{\m}(\WmaxXY) < \alpha$.
\end{condition}
\begin{condition}
    \label{cond:SB}
    Terms $(\WminXY - \n\m/2)$ and $(\WmaxXY - \n\m/2)$ have the same sign.
\end{condition}
\begin{condition}
    \label{cond:SC}
Either $\WmaxXY < \cdfwmw{n}{m}^{-1}(\alpha/2)$ or 
    $\WminXY > \cdfwmw{\n}{\m}^{-1}(1-\alpha/2)$.
\end{condition}
It will be useful to prove the following lemma and proposition,
before proving Proposition 1 and Corollary 2. 
\begin{lemma}
    \label{lem:con12equiv3}
    Suppose $\bx$ and $\by$ are samples of $\n$ and $\m$ values, 
    respectively, with $\n$ and $\m$ sufficiently large and all values  
    distinct.
    Then Condition~\ref{cond:SA} and Condition~\ref{cond:SB} are both 
    true if and only if Condition~\ref{cond:SC} is true.
\end{lemma}

\begin{proof}
    Assume Condition~\ref{cond:SC} is true, i.e.
    Either $\WmaxXY < \cdfwmw{n}{m}^{-1}(\alpha/2)$ or 
    $\WminXY > \cdfwmw{\n}{\m}^{-1}(1-\alpha/2)$ is true.
    Assume $\WmaxXY < \cdfwmw{n}{m}^{-1}(\alpha/2)$ is true; then
    \begin{align}
        &\WmaxXY < \cdfwmw{n}{m}^{-1}(\alpha/2)
        \nonumber \\
        \Leftrightarrow
        & \cdfwmw{n}{m}(\WmaxXY) < \alpha/2
        \nonumber \\
        \Leftrightarrow
        & \pvalfun \circ \cdfwmw{n}{m}(\WmaxXY) < \alpha,
        \nonumber 
    \end{align}
    which implies Condition~\ref{cond:SA} is true. Also, 
    since $\WminXY \leq \WmaxXY$, and since 
    $\cdfwmw{n}{m}^{-1}(\alpha/2) < \cdfwmw{n}{m}^{-1}(0.5) = \n\m/2$,
    \begin{align}
        &\WmaxXY < \cdfwmw{n}{m}^{-1}(\alpha/2)
        \nonumber \\
        \Leftrightarrow
        &\WminXY \leq \WmaxXY < \cdfwmw{n}{m}^{-1}(\alpha/2) < \n\m/2,
        \nonumber 
    \end{align}
    which implies that Condition~\ref{cond:SB} is true. If instead of 
    $\WmaxXY < \cdfwmw{n}{m}^{-1}(\alpha/2)$ is true but rather
    $\WminXY > \cdfwmw{\n}{\m}^{-1}(1-\alpha/2)$ is true (at least one of
    these inequalities is true if we assume Condition~\ref{cond:SC} is true),
    we similarly conclude that Conditions~\ref{cond:SA} and \ref{cond:SB}
    are true. Therefore, if Condition~\ref{cond:SC} is true, then
    Conditions~\ref{cond:SA} and \ref{cond:SB} are true.

    For the other direction, assume Conditions~\ref{cond:SA} and \ref{cond:SB}
    are both true. Each condition states that one of two inequalities is true, 
    so there are four cases to consider. 
    For Condition~\ref{cond:SA}, suppose 
    $\pvalfun \circ \cdfwmw{\n}{\m}(\WmaxXY) < \alpha$ is true, and for 
    Condition~\ref{cond:SB} suppose that $\WminXY, \WmaxXY \leq \n\m/2$
    is true. Then
    \begin{align}
        &\pvalfun \circ \cdfwmw{\n}{\m}(\WmaxXY) < \alpha
        \nonumber \\
        \Leftrightarrow &\cdfwmw{\n}{\m}(\WmaxXY) < \alpha/2
        \;\;\textrm{or} \;\;
        \cdfwmw{\n}{\m}(\WmaxXY) > 1- \alpha/2
        \nonumber \\
        \Leftrightarrow &\WmaxXY < \cdfwmw{\n}{\m}^{-1}(\alpha/2)
        \;\;\textrm{or} \;\;
        \WmaxXY > \cdfwmw{\n}{\m}^{-1}(1-\alpha/2)
        \nonumber \\
        \Leftrightarrow &\WmaxXY < \cdfwmw{\n}{\m}^{-1}(\alpha/2)
        \;\;(\textrm{since} \;\;\WminXY, \WmaxXY \leq \n\m/2),
        \nonumber
    \end{align}
    which implies that Condition~\ref{cond:SC} is true. Considering
    the other three cases also leads to the conclusion that 
    Condition~\ref{cond:SC} is true. 

    Therefore, Condition~\ref{cond:SC} is true if and only if
    Conditions~\ref{cond:SA} and \ref{cond:SB} are both true.
   \end{proof}

   The following lemma summarises previous results.
\begin{lemma}
    The proposed method will declare a significant result 
    if and only if Condition~\ref{cond:SC} is true.
\end{lemma}
\begin{proof}
    \begin{align}
        \textrm{Condition~\ref{cond:SC} is true} 
        &\Leftrightarrow \textrm{Conditions~\ref{cond:SA} and \ref{cond:SA} are true}
        \qquad \textrm{(by Lemma~\ref{lem:con12equiv3})}
        \nonumber \\
        &\Leftrightarrow \textrm{the proposed method will declare a significant result}
        \qquad \textrm{(see Section 3.2)}
        \nonumber
    \end{align}
\end{proof}


\section{Proofs for Section 3.3}

The following proposition, while not stated in the main paper,
is needed to prove the results in Section~3.3.

\begin{proposition}
    \label{supp:prop:S3}
    Suppose $\bx = \{ \xx_{1}, \xx_{2}, \cdots, \xx_{\n} \}$ and 
    $\by = \{\yy_{1}, \yy_{2},\cdots, \yy_{\m} \}$ are samples of distinct
    real-valued observations and $\bxp \subset \bx$, $\byp \subset \by$ are 
    subsets of distinct values with size $\np$ and $\mpp$ respectively, which 
    are observed. 
    Then, for any significance level $\alpha \in (0,1)$ and sufficiently 
    large $\n, \m$, Condition~\ref{cond:SC} is true if and only if 
    either of the following two inequalities are true
  \begin{align}
  	\wmwstat{\bxp}{\byp} & < \cdfphi^{-1}(\alpha/2) \{\n\m(\n+\m+1)/12\}^{1/2} 
      - \n \m/2 + \np \mpp,
      \nonumber \\ 
    \wmwstat{\bxp}{\byp} & > \cdfphi^{-1}(1-\alpha/2) \{\n\m(\n+\m+1)/12\}^{1/2}  
      + \n\m/2,
    \nonumber
  \end{align}
    where $\cdfphi(x)$ is cumulative distribution function of the standard normal 
    distribution. 
\end{proposition}	
\begin{proof}
    First, since $\cdfwmw{n}{m}$ is the cumulative distribution function with 
    mean $\mu=\n\m/2$ and variance $\sigma^2=\n\m(\n+\m+1)/12$, and 
    $\cdfphi$ is the cumulative distribution function for the standard normal
    distribution with mean $0$ and variance $1$, then for any value 
    $\xx \in (0, 1)$, $\cdfwmw{\n}{\m}^{-1}(\xx)=\cdfphi^{-1}(\xx)\sigma+\mu$.

    By Lemma~\ref{lem:con12equiv3}, Conditions~\ref{cond:SA} and \ref{cond:SB}
    are true if and only if $\WmaxXY < \cdfwmw{n}{m}^{-1}(\alpha/2)$
    or $\WminXY > \cdfwmw{\n}{\m}^{-1}(1-\alpha/2)$. Then, 
    since $\WmaxXY = \wmwstat{\bxp}{\byp} + \n\m - \np\mpp$
    and $\WminXY = \wmwstat{\bxp}{\byp}$,
    \begin{align}
        &\WmaxXY < \cdfwmw{n}{m}^{-1}(\alpha/2)
        \nonumber \\
        \Leftrightarrow &\wmwstat{\bxp}{\byp} + \n\m - \np\mpp
        < \cdfwmw{n}{m}^{-1}(\alpha/2)
        \nonumber \\
        \Leftrightarrow &\wmwstat{\bxp}{\byp} + \n\m - \np\mpp
        < \cdfphi^{-1}(\alpha/2) \{\n\m(\n+\m+1)/12\}^{1/2} + \n\m/2
        \nonumber \\
        \Leftrightarrow &\wmwstat{\bxp}{\byp} 
        < \cdfphi^{-1}(\alpha/2) \{\n\m(\n+\m+1)/12\}^{1/2} - \n\m/2 - \np\mpp .
        \nonumber
    \end{align}
    Similarly, 
    \begin{align}
        &\WminXY > \cdfwmw{\n}{\m}^{-1}(1-\alpha/2)
        \nonumber \\
        \Leftrightarrow &\wmwstat{\bxp}{\byp} 
        > \cdfwmw{n}{m}^{-1}(1-\alpha/2)
        \nonumber \\
        \Leftrightarrow &\wmwstat{\bxp}{\byp} 
        > \cdfphi^{-1}(1-\alpha/2) \{\n\m(\n+\m+1)/12\}^{1/2} + \n\m/2,
        \nonumber
    \end{align}
    which proves the result.
\end{proof}
The following well-known result regarding the Wilcoxon-Mann-Whitney statistic 
will be useful to prove the results in Section~3.3.
\begin{lemma}
    \label{supp:lem:S3}
    If $\bx$ and $\by$ are samples of $\n$ and $\m$ values, respectively, with 
    all values distinct, then the Wilcoxon-Mann-Whitney test statistic 
    $\wmwstat{\bx}{\by}$ is bounded by $0 \leq \wmwstat{\bx}{\by} \leq \n \m$.
\end{lemma}
\begin{proof}
    By definition,
    $\wmwstat{\bx}{\by} = \rankgiven{\bx}{\bx \cup \by} - \n(\n+1)/2$, and
    \begin{align}
        \rankgiven{\bx}{\bx \cup \by} 
        = \sum_{\xx \in \bx} \rankfun{\bx \cup \by}{\xx}.
        \nonumber
    \end{align}
    The rank sum $\rankgiven{\bx}{\bx \cup \by}$ is minimised when 
    the values in $\bx$ occupy ranks $1, \dots, \n$ in $\bx \cup \by$, and so 
    $\rankgiven{\bx}{\bx \cup \by} = \n(\n+1)/2$, which implies
    $\wmwstat{\bx}{\by} = 0$ is the minimum possible value 
    for $\wmwstat{\bx}{\by}$.
    On the other hand, $\rankgiven{\bx}{\bx \cup \by}$ is maxmised when
    the values in $\bx$ occupy ranks $\n+1, \dots, \n+\m$ in $\bx \cup \by$,
    which implies
    $\rankgiven{\bx}{\bx \cup \by} = \n\m + \n(\n+1)/2$, which implies 
    $\wmwstat{\bx}{\by} = \n\m$ is the maximum possible value for 
    $\wmwstat{\bx}{\by}$.
\end{proof}
We are finally ready to provide Proposition~2 in the main paper which is
restated here:
%
\begin{proposition}
    \label{supp:thm:missingprop}
    Suppose that $\bx = \{\xx_1, \ldots, \xx_{\n}\}$ and
    $\by = \{\yy_1, \ldots, \yy_{\m}\}$ are samples of 
    real-valued observations which are unknown but assumed to be distinct and
    $\bxp \subset \bx$ and $\byp \subset \by$ are subsets of
    distinct values with sizes $\np$ and $\mpp$, respectively,
    which will be observed. Then for any significance threshold 
    $\alpha \in (0, 1)$ and for sufficiently large $\n,\m$, if
    \begin{align}
        \np \mpp/(\n \m) < 1/2 + 
        \cdfphi^{-1}(1 - \alpha/2)\{(\n + \m + 1)/ (12\n\m) \}^{1/2} ,
        \label{eqn:missingprop}
    \end{align}
    where $\cdfphi$ is the cumulative distribution function of the standard 
    normal distribution,
    then the proposed method will not yield a significant $p$-value, 
    regardless of the values in $\bxp$ and $\byp$.
\end{proposition}

\begin{proof}
    Suppose the conditions of the proposition are satisfied, and 
    \begin{align}
        \np \mpp/(\n \m) < 1/2 + 
        \cdfphi^{-1}(1 - \alpha/2)\{(\n + \m + 1)/ (12\n\m) \}^{1/2}.
        \label{supp:ineq:prop30}
    \end{align}
    Then
    \begin{align}
        &\np \mpp/(\n \m) < 1/2 + 
        \cdfphi^{-1}(1 - \alpha/2)\{(\n + \m + 1)/ (12\n\m) \}^{1/2}
        \nonumber \\
        \Leftrightarrow
        &\np \mpp< \n \m /2 + 
        \cdfphi^{-1}(1 - \alpha/2)\{\n\m(\n + \m + 1)/12 \}^{1/2}
        \nonumber \\
        \Leftrightarrow
        &0 < - \np \mpp  + \n \m /2 
        + \cdfphi^{-1}(1 - \alpha/2)\{\n\m(\n + \m + 1)/12 \}^{1/2}
        \nonumber \\
        \Leftrightarrow
        &0 >  \np \mpp  - \n \m /2  
        -\cdfphi^{-1}(1 - \alpha/2)\{\n\m(\n + \m + 1)/12 \}^{1/2}
        \nonumber \\
        \Leftrightarrow
        & \wmwstat{\bxp}{\byp} \geq 0 >  \np \mpp  - \n \m /2 
        -\cdfphi^{-1}(1 - \alpha/2)\{\n\m(\n + \m + 1)/12 \}^{1/2}
        \nonumber \\
        \Leftrightarrow
        & \wmwstat{\bxp}{\byp} > 
        \cdfphi^{-1}(\alpha/2)\{\n\m(\n + \m + 1)/12 \}^{1/2} 
          - \n \m /2 + \np \mpp ,
          \label{supp:ineq:prop3a}
    \end{align}
    since $\cdfphi^{-1}(1-\xx) = -\cdfphi^{-1}(\xx)$, for $\xx \in (0, 1)$,
    and using Lemma~\ref{supp:lem:S3} applied to $\bxp$ and $\byp$, i.e.
    $0 \leq \wmwstat{\bxp}{\byp} \leq \np \mpp$.
    Similarly, 
    \begin{align}
        &\np \mpp/(\n \m) < 1/2 + 
        \cdfphi^{-1}(1 - \alpha/2)\{(\n + \m + 1)/ (12\n\m) \}^{1/2}
        \nonumber \\
        \Leftrightarrow
        &\np \mpp< \n \m /2 + 
        \cdfphi^{-1}(1 - \alpha/2)\{\n\m(\n + \m + 1)/12 \}^{1/2}
        \nonumber \\
        \Leftrightarrow
        & \wmwstat{\bxp}{\byp} \leq \np \mpp < \n \m /2 + 
        \cdfphi^{-1}(1 - \alpha/2)\{\n\m(\n + \m + 1)/12 \}^{1/2}
        \nonumber \\
        \Leftrightarrow
        & \wmwstat{\bxp}{\byp}  < 
        \cdfphi^{-1}(1 - \alpha/2)\{\n\m(\n + \m + 1)/12 \}^{1/2}
         + \n \m /2. 
          \label{supp:ineq:prop3b}
    \end{align}
    Given that if Inequality~\eqref{supp:ineq:prop30} is true it implies both 
    Inequalities~\eqref{supp:ineq:prop3a} and \eqref{supp:ineq:prop3b} are 
    true, this means that neither of the inequalities in 
    Proposition~\ref{supp:prop:S3} are true, 
    which implies that Conditions~\ref{cond:SC} is not true, which by 
    Lemma~3 implies that the proposed method will not yield a significant 
    $p$-value.
\end{proof}

\begin{corollary}
    \label{supp:thm:missingproptwo}
    Suppose that $\bx = \{\xx_1, \dots, \xx_{\n}\}$ and
    $\by = \{\yy_1, \dots, \yy_{\m}\}$ are samples of 
    real-valued observations which are unknown but assumed to be distinct and
    $\bxp \subset \bx$ and $\byp \subset \by$ are subsets of
    values with sizes $\np$ and $\mpp$, respectively,
    which will be observed. Then for any significance threshold $\alpha \in (0, 1)$
    and for sufficiently large $\n,\m$, if 
    \begin{align}
        \np \mpp/(\n \m) \geq 1/2 + 
        \cdfphi^{-1}(1 - \alpha/2)\{(\n + \m + 1)/ (12\n\m) \}^{1/2} ,
        \label{supp:eqn:corr2}
    \end{align}
    where $\cdfphi$ is the cumulative distribution function of the standard 
    normal distribution,
    then sets $\bxp$ and $\byp$ exist for which
    the proposed method will yield a significant $p$-value.
\end{corollary}

\begin{proof}
    Given $\alpha \in (0, 1)$, suppose we have $\n, \m$ with $\np \leq \n$ 
    and $\mpp \leq \m$ satisfying Inequality~\eqref{supp:eqn:corr2}. 
    If we
    consider the range of possible values for sum of ranks of $\bxp$ in 
    $\bxp \cup \byp$, following the same logic as in the proof of 
    Lemma~\ref{supp:lem:S3},
    $\rankgiven{\bxp}{\bxp \cup \byp}$ 
    is minimised when the elements of
    $\bxp$ occupy ranks $1, \dots, \np$ in $\bxp \cup \byp$, and so 
    $\rankgiven{\bxp}{\bxp \cup \byp} = \np(\np+1)/2$. 
    This occurs when all values in $\bxp$ are smaller than all values
    in $\byp$, and all values are distinct.
    Recalling that 
    $\wmwstat{\bxp}{\byp} = \rankgiven{\bxp}{\bxp \cup \byp} - \np(\np+1)/2$,
    this implies that
    $\wmwstat{\bxp}{\byp} = 0$ is the minimum possible value 
    for $\wmwstat{\bxp}{\byp}$.

    The goal is to prove that
    if Inequality~\eqref{supp:eqn:corr2} holds then 
    sets $\bxp$ and $\byp$ exist for which the proposed method will yield
    a significant $p$-value. Let us therefore choose sets $\bxp$ and $\byp$
    where all values in $\bxp$ are smaller than all values
    in $\byp$, and all values are distinct, and so $\wmwstat{\bxp}{\byp} = 0$.

    We proceed by rearranging Inequality~\eqref{supp:eqn:corr2}, starting
    by multiplying both sides by $\n \m$:
    \begin{align}
        &\np \mpp/(\n \m) \geq 
        1/2 + \cdfphi^{-1}(1 - \alpha/2)\{(\n + \m + 1)/ (12\n\m) \}^{1/2} 
        \nonumber \\
        \Leftrightarrow
        &\np \mpp \geq 
        \n \m/2 + \cdfphi^{-1}(1 - \alpha/2)\{\n\m(\n + \m + 1)/ 12 \}^{1/2} 
        \nonumber \\
        \Leftrightarrow
        &\np \mpp - \n \m/2  \geq 
        \cdfphi^{-1}(1 - \alpha/2)\{\n\m(\n + \m + 1)/ 12 \}^{1/2} 
        \nonumber \\
        \Leftrightarrow
        &(\np \mpp - \n \m/2) / \{\n\m(\n + \m + 1)/ 12 \}^{1/2}  \geq 
        \cdfphi^{-1}(1 - \alpha/2) 
        \nonumber \\
        \Leftrightarrow
        &(\n \m/2 - \np \mpp) / \{\n\m(\n + \m + 1)/ 12 \}^{1/2}  \leq
        -\cdfphi^{-1}(1 - \alpha/2) .
        \label{supp:eqn:corr2proof}
    \end{align}

    Recall that the statistic $\wmwstat{\bx}{\by}$ follows
    a normal distribution with mean $\mu=\n\m/2$ and variance
    $\sigma^2 = \n\m(\n+\m+1)/12$ and cumulative distribution function
    $\cdfwmw{\n}{\m}$. If $\WmaxXY < \cdfwmw{\n}{\m}^{-1}(\alpha/2)$, 
    then Condition~3 is true, and if Condition~3 is true then by Lemma~3
    the proposed method will declare a significant result. We note
    $\WmaxXY < \cdfwmw{\n}{\m}^{-1}(\alpha/2)$ is equivalent to
    $(\WmaxXY - \mu)/\sigma < \cdfphi^{-1}(\alpha/2)$, where
    $\cdfphi$ is the cumulative distribution function of the standard normal
    distribution. Noting $\cdfphi^{-1}(\alpha/2) = -\cdfphi^{-1}(1-\alpha/2)$,
    if 
    $$(\WmaxXY - \mu)/\sigma < -\cdfphi^{-1}(1-\alpha/2),$$ 
    then the proposed
    method will yield a significant result.

    We recall from Theorem 1 that the $\WmaxXY$ is given by
    $$\WmaxXY = \wmwstat{\bxp}{\byp} + \n\m - \np\mpp , $$
    but since we earlier assumed $\wmwstat{\bxp}{\byp} = 0$, then
    $\WmaxXY = \n\m - \np\mpp$. 
    But then
    \begin{align}
        (\WmaxXY - \mu)/\sigma 
        &= (\n\m - \np\mpp - \n\m/2)/ \{\n\m(\n + \m + 1)/ 12\}^{1/2}
        \nonumber \\
        &= (\n\m/2 - \np\mpp)/ \{\n\m(\n + \m + 1)/ 12\}^{1/2}
        \nonumber \\
        \Rightarrow
        (\WmaxXY - \mu)/\sigma 
        &\leq -\cdfphi^{-1}(1 - \alpha/2),
        \label{supp:eqn:corr2proof2}
    \end{align}
    where the last line follows from Inequality~\eqref{supp:eqn:corr2proof}.
    But Inequality~\eqref{supp:eqn:corr2proof2} implies that the proposed 
    method will yield a significant $p$-value, which completes the proof.
\end{proof}


\section{The proof of results in Section 5.1}



\begin{proposition} 
    \label{supp:prop:mcar1}
    Let $\bx = \{ \bx_1, \cdots, \bx_{\n} \}$ and 
    $\by = \{\by_1, \cdots, \by_{\m} \}$ be sets of random variables 
    identically and independently 
    distributed according to continuous distributions $F$ and $G$, respectively. 
    Let $\bxp \subset \bx$ and $\byp \subset \by$ denote the subsets of distinct
    random variables with sizes $\np$ and $\mpp$, respectively, which can be 
    observed. 
    Assume that the observations of the 
    random variables in the complements $\bx \setminus \bxp$ 
    and $\by \setminus \byp$ are missing completely at random,
    and $0 < \prob(\bx_1 < \by_1) < 1$. Then, for any given significance level 
    $\alpha \in (0,1)$, with probability approximately equal to 
    $\cdfphi (\{L - \mu'\}/\sigmap) + 1 - \cdfphi( \{R - \mup\}/\sigmap)$
    the proposed method yields a significant $p$-value, 
	where $\cdfphi$ is the cumulative distribution function of the standard 
    normal distribution and $\mu = \n \m/2$ and $\sigma^2 = \n \m(\n+\m+1)/12$ 
    and
	\begin{align}
		&L = \sigma \cdfphi^{-1}(\alpha/2) + \mu - \n\m +\np\mpp , \quad
		R = \sigma \cdfphi^{-1}(1-\alpha/2) + \mu ,
        \nonumber\\ 
        &p_1 = \prob(X_1 < Y_1), \,\,
        p_2 = \prob(X_1 < Y_1 \textrm{ and } X_1 < Y_2),  \,\,
        p_3 = \prob(X_1 < Y_1 \textrm{ and } X_2 < Y_1).
        \nonumber\\ 
		&\mup  = \mpp \np p_1,
        \nonumber \\
        &(\sigmap)^2 = \mpp \np p_1(1-p_1) + \mpp \np(\np-1)(p_2-p_1^2) + \np\mpp (\mpp-1)(p_3-p_1^2).
        \label{supp:eqn:755} 
	\end{align}
\end{proposition}

\begin{proof}
	The proof is based on the results in Section 3, Chapter 2 in \cite{lehmann1975}. 
	
	Since the missingness mechanism is assumed missing completely at random for both $X$ and $Y$, for all observed variables in $X$ and $Y$, we must have $X'$ and $Y'$ are also identical and independently distributed according to $F$ and $G$, respectively.
	
	Then, from (2.14) in \cite{lehmann1975}, since $F,G$ are both continuous cumulative density function and $0 \le P(X_1 < Y_1) \le 1$, $W(X',Y')$ approximately follows a normal distribution, i.e.
	\begin{align}
		(W(X',Y') - \mu')/\sigma' \sim  N(0,1)
        \label{supp:eqn:757}
	\end{align}
	approximately, where $\mu' = E(W(X',Y'))$ and $\sigma' = (\text{var}(W(X',Y')))^{1/2}$, and from the equations (2.17) and (2.21) in  \cite{lehmann1975} respectively,  we have
	\begin{align*}
		&\mu'  = m'n'p_1, \\
		&\sigma'^2 = m'n'p_1(1-p_1) + m'n'(n'-1)(p_2-p_1^2) + n'm'(m'-1)(p_3-p_1^2),
	\end{align*}
	where
	\begin{align*}
		&p_1 = \prob(X_1 < Y_1),
		p_2 = \prob(X_1 < Y_1 \text{ and } X_1 < Y_2),
		p_3 = \prob(X_1 < Y_1 \text{ and } X_2 < Y_1).
	\end{align*}
	
	In Proposition \ref{supp:prop:S3}, we showed that the proposed method yields a significant $p$-value if and only if either of the following two inequalities are true
	\begin{align}
		  	\wmwstat{\bxp}{\byp} & < \cdfphi^{-1}(\alpha/2) \{\n\m(\n+\m+1)/12\}^{1/2} 
		- \n \m/2 + \np \mpp,
		\nonumber \\ 
		\wmwstat{\bxp}{\byp} & > \cdfphi^{-1}(1-\alpha/2) \{\n\m(\n+\m+1)/12\}^{1/2}  
		+ \n\m/2, \nonumber
	\end{align}
	Recall that $\sigma = \{\n\m(\n+\m+1)/12\}^{1/2}$ and $\mu = nm/2$. We can denote the right hand sides of the above two inequalities as $L =  \sigma \Phi^{-1}(\alpha/2) + \mu -nm +n'm'$, and $R =  \sigma \Phi^{-1}(1-\alpha/2) + \mu$, respectively. Subsequently, we have
	\begin{align*}
		&\prob(\text{Proposed method yields a significant $p$-value}) \\	
		&= \prob\left(W(X',Y') < L \text{ or } W(X',Y') > R \right) \nonumber\\
		&=   \prob\left(W(X',Y') < L\right) + \prob\left(W(X',Y') > R\right) \nonumber.
	\end{align*}
	Notice that, according to the approximation result in {\eqref{supp:eqn:757}}, we have
	\begin{align*}
		\begin{split}
			&\prob\left(W(X',Y') < L\right) 
			= \prob\left\{ (W(X',Y')-\mu')/\sigma' < (L - \mu')/\sigma'\right\} 
            \approx \Phi\left\{ (L - \mu')/\sigma'\right\},\\
			&\prob\left(W(X',Y') > R \right) 
			= \prob\left\{(W(X',Y')-\mu')/\sigma'> (R - \mu')/\sigma'  \right\} 
			\approx 
            1- \Phi\left\{(R -\mu')/\sigma'\right\} .
		\end{split}
	\end{align*} 
	Hence, we have
	\begin{align*}
		\prob(\text{Proposed method yields a significant $p$-value}) \approx \Phi\left((L - \mu')/\sigma'\right) + 1 - \Phi\left((R-\mu')/\sigma'\right),
	\end{align*}
	which completes our proof.
\end{proof}


\begin{proposition}
\label{supp:prop:mcar2lim}
Let $\bx = \{ \bx_1, \cdots, \bx_{\n} \}$ and 
$\by = \{\by_1, \cdots, \by_{\m} \}$ be sets of random variables 
identically and independently 
distributed according to continuous distributions $F$ and $G$, respectively. 
Let $\bxp \subset \bx$ and $\byp \subset \by$ denote the subsets of distinct
random variables with sizes $\np$ and $\mpp$, respectively, which can be 
observed. 
Assuming that the random variables in the complements $\bx \setminus \bxp$ 
and $\by \setminus \byp$ are missing completely at random,
and $0 < \prob(\bx_1 < \by_1) < 1$, and additionally assuming 
$\mpp \leq \np$ and
\begin{align}
& \mpp/\np \to \lambdap \text{ when } 
    \n,\m \to \infty, \text{ where } 0 < \lambdap \leq 1, 
    \label{supp:eqn:761} \\
&\np/\n \to \lambdax, \text{ when } \n,\m \to \infty, 
    \text{ where } 0 < \lambdax \leq 1, 
    \label{supp:eqn:762} \\
&\mpp/\m \to \lambday, \text{ when } \n,\m \to \infty, 
    \text{ where } 0 < \lambday \leq 1 
    \label{supp:eqn:763} \\
&\lambda_{x} \lambday (\pOne - 1) + 1/2  \neq 0 \text{ and } 
    \lambda_{x} \lambday \pOne - 1/2  \neq 0, \text{ where }
    \pOne=\prob(\bx_1 < \by_1). 
    \label{supp:eqn:764} 
\end{align}
As $\n,\m \to \infty$, with probability 
\begin{align}
    p=
\left\{ \begin{array}{cl}
    0, &  \text{ if }\lambdax \lambday (\pOne - 1)+1/2 > 0 \text{ and }  
    \lambdax \lambday \pOne -1/2  < 0,\\
    1, &\text{ otherwise,} 
    \label{supp:eqn:765} 
\end{array}\right.
\end{align}
the proposed method will yield a significant $p$-value.
\end{proposition}

\begin{proof}
	For any given $n,m,n',m'$, since the missingness mechanism is assumed missing completely at random for both $X$ and $Y$, we have for all observed variables in $X$ and $Y$, $X'$ and $Y'$ are also identical and independently distributed according to $F$ and $G$ respectively.
	
	Since we assume $0 < \lambda' \le 1$ and $0 < \prob(X_1 < Y_1) < 1$, we can use the Corollary 5 in  \cite{lehmann1975} (see Appendix: U-Statistic Corollary 5 in \cite{lehmann1975}) and get the result that
	\begin{align}
		(W(X',Y') - \mu')/\sigma' \to^{d} N(0,1), \text{ when } n,m \to \infty,
    \label{supp:eqn:766} 
	\end{align}
	where $\mu' = m'n'p_1$ and $\sigma' = \{m'n'p_1(1-p_1) + m'n'(n'-1)(p_2-p_1^2) + n'm'(m'-1)(p_3-p_1^2)\}^{1/2}$ as defined in Proposition~\ref{supp:prop:mcar1}.
	
	Notice that we can rewrite $m/n$ as
	\begin{align*}
		{m}/{n} =  \frac{m}{m'}\frac{m'}{n'}\frac{n'}{n}. 
	\end{align*}
    According to equations \eqref{supp:eqn:763}, \eqref{supp:eqn:761} and \eqref{supp:eqn:762}, we have
    \begin{align*}
    	&\lim_{n,m \to \infty} m /m' \to 1/\lambda_y, \text{where } 0 < \lambda_y \le 1, \\
    	&\lim_{n,m \to \infty} m' /n' \to \lambda', \text{where } 0 < \lambda' \le 1, \\
        &\lim_{n,m \to \infty} n' /n \to \lambda_x, \text{where } 0 < \lambda_x \le 1,
    \end{align*}
    separately. All limitations of the three terms exist and well defined. Thus, the limitation of $m/n$ exists and
	\begin{align*}
		\lim_{n,m \to \infty} {m}/{n}		
		= \left(\lim_{n,m \to \infty}{m}/{m'}\right)\left(\lim_{n,m \to \infty}{m'}/{n'}\right)\left(\lim_{n,m \to \infty}{n'}/{n}\right)
		=\lambda'\lambda_x/\lambda_y.
	\end{align*}
	We shall denote 
	\begin{align*}
		\lambda := \lambda'\lambda_x/\lambda_y.
	\end{align*}
	
	Let us now consider the probability of the proposed method yielding a significant $p$-value. According to \text{Proposition } \ref{supp:prop:S3}, we have
	\begin{align*}
		&\prob(\text{Proposed method yields a significant $p$-value})  \\
		&= \prob\left( W(X',Y') \le \sigma \Phi^{-1}(\alpha/2) + \mu - nm + n'm'\right) 
        + \prob\left(W(X',Y')  \ge \sigma\Phi^{-1}(1-\alpha/2)+\mu\right)
    \end{align*}
    This equation can be further written as 
    \begin{align*}
    	&\prob(\text{Proposed method yields a significant $p$-value})  \\
	    & = \prob\left\{W(X',Y') - \mu + nm -n'm' \le \sigma\Phi^{-1}(\alpha/2)\right\} 
        + \prob\left\{{W(X',Y') - \mu} \ge \sigma\Phi^{-1}(1-\alpha/2)\right\} \\
		& = \prob\left\{ \frac{(W(X',Y') - \mu + nm -n'm')}{\sigma} \le \Phi^{-1}(\alpha/2)\right\} + 
        \prob\left\{\frac{W(X',Y') - \mu}{\sigma} \ge \Phi^{-1}(1-\alpha/2)\right\} \\
		&= {\prob\left\{ \frac{\sigma'}{\sigma}\frac{W(X',Y') - \mu'}{\sigma'} + \frac{\mu' - \mu + nm -n'm'}{\sigma} \le \Phi^{-1}(\alpha/2)\right\}} \\
		& + {\prob\left\{\frac{\sigma'}{\sigma}\frac{W(X',Y')-\mu'}{\sigma'} + \frac{\mu' - \mu}{\sigma} \ge \Phi^{-1}(1-\alpha/2) \right\}}.
	\end{align*}
   Denote
   \begin{align*}
   a_1(n,m,n'm') = {(\mu' - \mu + nm -n'm')/{\sigma}},~
   a_2(n,m,n',m') = {({\mu' - \mu})/{\sigma}}.
   \end{align*}
   We have
   \begin{align}
   	\begin{split}
   	   &\prob(\text{Proposed method yields a significant $p$-value}) \\
    	& = {\prob\left\{ ({\sigma'}/{\sigma})(W(X',Y') - \mu')/{\sigma'} + a_1(n,m,n'm') \le \Phi^{-1}(\alpha/2)\right\}} \\
    	& + {\prob\left\{ ({\sigma'}/{\sigma})({W(X',Y')-\mu'})/{\sigma'} + a_2(n,m,n'm') \ge \Phi^{-1}(1-\alpha/2) \right\}}.
   	\end{split}
   	\label{supp:eqn:33}
   \end{align}
  We are ultimately interested in the limitation of the above equation \eqref{supp:eqn:33} when $n,m \to \infty$. To start, let us first dealing with the limitation of the term  ${\sigma'}/{\sigma}$. Recall that $\sigma^2 = nm(n+m+1)/12$, and $\sigma'$ is defined in \eqref{supp:eqn:755}. We expand the term $(\sigma'/\sigma)^2$ as
	\begin{align*}
		\left(\frac{\sigma'}{\sigma}\right)^2 &= 12 \frac{m'n'p_1(1-p_1) + m'n'(n'-1)\left(p_2-p_1^2\right) + n'm'(m'-1)\left(p_3-p_1^2\right)}{nm(n+m+1)}.
	\end{align*}
    Divide the numerator and denominator of the above equation both by $n^2m$, we have
    \begin{align*}
       \left(\frac{\sigma'}{\sigma}\right)^2  &= {12}  \frac{\frac{m'}{m}\frac{n'}{n}\frac{1}{n}p_1(1-p_1) + \frac{m'}{m}\frac{n'}{n}(\frac{n'}{n}-\frac{1}{n})(p_2-p_1^2) \frac{n'}{n}\frac{m'}{m}(\frac{m'}{n}-\frac{1}{n})(p_3-p_1^2)} {1+({m}/{n})+({1}/{n})}. 
    \end{align*}
    By replacing $m'/n$ in the above equation as $(m'/m)(m/n)$, it follows
    \begin{align*}
       \left(\frac{\sigma'}{\sigma}\right)^2  &= {12}  \frac{\frac{m'}{m}\frac{n'}{n}\frac{1}{n}p_1(1-p_1) + \frac{m'}{m}\frac{n'}{n}(\frac{n'}{n}-\frac{1}{n})(p_2-p_1^2) \frac{n'}{n}\frac{m'}{m}(\frac{m'}{m}\frac{m}{n}-\frac{1}{n})(p_3-p_1^2)} {1+({m}/{n})+({1}/{n})}. 
    \end{align*} 
    According to equations \eqref{supp:eqn:763}, \eqref{supp:eqn:761}, \eqref{supp:eqn:762}, we have 
    \begin{align*}
	&\lim_{n,m \to \infty} m /m' \to 1/\lambda_y, \text{where } 0 < \lambda_y \le 1,\\
	&\lim_{n,m \to \infty} m' /n' \to \lambda', \text{where } 0 < \lambda' \le 1, \\
	&\lim_{n,m \to \infty} n' /n \to \lambda_x, \text{where } 0 < \lambda_x \le 1,
	\end{align*}
    separately. Also, recall that we have shown $\lim_{n,m \to \infty} m /n \to \lambda$ and $\lim_{n,m \to \infty} 1 /n \to 0$. We have 
    \begin{align*}
    	\lim_{n,m \to \infty} \left(\frac{\sigma'}{\sigma}\right)^2	&\to {12} \frac{0+\lambda_x^2 \lambda_y(p_2-p_1^2) + \lambda_x^2\lambda_y\lambda' (p_3 - p_1^2)}{1 + \lambda},
    \end{align*}
    which leads to
    \begin{align*}
    	\lim_{n,m \to \infty} \left(\frac{\sigma'}{\sigma}\right)	&\to {12} \left(\frac{0+\lambda_x^2 \lambda_y(p_2-p_1^2) + \lambda_x^2\lambda_y\lambda' (p_3 - p_1^2)}{1 + \lambda}\right)^{1/2} := C,
    \end{align*}
    From \eqref{supp:eqn:766}, we have $(W(X',Y') - \mu')/{\sigma'}$ convergences in distribution to $N(0,1)$. Since $\sigma'/\sigma$ convergences to a constant $C$, we can proceed by using Slutsky's theorem and get 
	\begin{align}
		\lim_{n,m \to \infty} {({\sigma'}/{\sigma})(W(X',Y') - \mu')}/{\sigma'} \to^{d} CN(0,1)
        \label{supp:eqn:7613}
	\end{align}
	
	Let us now deal with other terms in equation \eqref{supp:eqn:33}. Recall that $\mu', \mu$ and $\sigma$ are defined in Proposition~\ref{supp:prop:mcar1} as
	\begin{align*}
		\mu' = m'n'p_1,~
		\mu = nm/2,~
		\sigma = \{{nm(n+m+1)}/12\}^{1/2}.
	\end{align*}
	We can therefore rewrite $a_1(n,m,n',m')$ and $a_2(n,m,n',m')$ in equation \eqref{supp:eqn:33} as
	\begin{align}
		a_1(n,m,n',m') &= (\mu' - \mu +nm - n'm')/{\sigma} \nonumber\\
		&= (m'n'p_1 - nm/2 + nm -n'm')/{\{{nm(n+m+1)}/{12}\}^{1/2}} \nonumber\\
		& = \{({n'}/{n})({m'}/{m})(p_1-1) + {1}/{2} \}/{\{{(n+m+1)}/{(12nm)}\}^{1/2}},
        \label{supp:eqn:7614}
	\end{align}
	and
	\begin{align}
		a_2(n,m,n',m') &= (\mu' - \mu)/\sigma  \nonumber \\
		&= (m'n'p_1 - nm/2 )/{\{{nm(n+m+1)}/{12}\}^{1/2}} \nonumber \\
		& = \{({m'}/{m})({n'}/{n})p_1 - {1}/{2} \}/{\{{(n+m+1)}/(12nm)\}^{1/2}},
        \label{supp:eqn:7615}
	\end{align}
     respectively. Using \eqref{supp:eqn:762} and \eqref{supp:eqn:763} again saying that $\lim\limits_{n,m \to \infty} n'/n \to \lambda_x$ and $\lim\limits_{n,m \to \infty} m'/m \to \lambda_y$, we can see that for the numerator of \eqref{supp:eqn:7614},
	\begin{align}
		\lim_{n,m \to \infty} \{({n'}/{n})({m'}/{m})(p_1-1) + {1}/{2}\} \to \lambda_x\lambda_y(p_1 - 1) + 1/2; 
        \label{supp:eqn:7616}
	\end{align}
    for the numerator of \eqref{supp:eqn:7615}
    \begin{align}
    	\lim_{n,m \to \infty}  \{({m'}/{m})({n'}/{n})p_1 - {1}/{2}\} \to \lambda_x\lambda_yp_1 - 1/2.
        \label{supp:eqn:7617}
    \end{align}
    Meanwhile, we can see that the denominators of both \eqref{supp:eqn:7614} and \eqref{supp:eqn:7615} follow
    \begin{align}
    	\lim_{n,m \to \infty} \{{(n+m+1)}/(12nm)\}^{1/2} \to 0^+.
        \label{supp:eqn:7618}
    \end{align}	
    Put \eqref{supp:eqn:7616} and \eqref{supp:eqn:7618} together, we can see that
    \begin{align}
    \lim_{n,m \to \infty} a_1(n,m,n'm') \to + \infty \text{ if } \lambda_x\lambda_y(p_1 - 1) + 1/2 > 0, 
        \label{supp:eqn:7619} \\
    \lim_{n,m \to \infty} a_1(n,m,n'm') \to - \infty \text{ if } \lambda_x\lambda_y(p_1 - 1) + 1/2 < 0.
        \label{supp:eqn:7620}
    \end{align}
    Similarly, put \eqref{supp:eqn:7617} and \eqref{supp:eqn:7618} together, we can see that
    \begin{align}
     \lim_{n,m \to \infty} a_2(n,m,n'm') \to + \infty \text{ if }  \lambda_x\lambda_yp_1 - 1/2 > 0, 
        \label{supp:eqn:7621} \\
     \lim_{n,m \to \infty} a_2(n,m,n'm') \to - \infty \text{ if } \lambda_x\lambda_yp_1 - 1/2 < 0.
        \label{supp:eqn:7622}
    \end{align}
     Recall that we assume in \eqref{supp:eqn:764} that $\lambda_x\lambda_y(p_1-1)+1/2 \neq 0$ and $\lambda_x\lambda_yp_1 -1/2  \neq 0$. Consequently, this allows us to categorize all possible interactions among $\lambda_x, \lambda_y, p_1$ into the following 4 cases:
	\begin{align*}
		\left\{ \begin{array}{lcl}
		\text{ Case I}: & \lambda_x\lambda_y(p_1-1)+1/2 < 0 \text{ and }\lambda_x\lambda_yp_1 -1/2  < 0\\
		\text{ Case II}: & \lambda_x\lambda_y(p_1-1)+1/2 > 0 \text{ and }\lambda_x\lambda_yp_1 -1/2  < 0\\
		\text{ Case III}: & \lambda_x\lambda_y(p_1-1)+1/2 < 0 \text{ and }\lambda_x\lambda_yp_1 -1/2  > 0\\
		\text{ Case IV}: & \lambda_x\lambda_y(p_1-1)+1/2 > 0 \text{ and }\lambda_x\lambda_yp_1 -1/2  > 0
		\end{array}\right.
	\end{align*}
	
	{Case I.} Let us consider Case I first, i.e.
	\begin{align*}
		\lambda_x\lambda_y(p_1 - 1) + 1/2 < 0 \text{ and } \lambda_x\lambda_yp_1 < 1/2.
	\end{align*}
	Notice that this case is possible. For example, we can consider $\lambda_x = \lambda_y = 0.9$ and  $p_1 = 0.1$, then $\lambda_x\lambda_y(p_1 - 1) = -0.9^3 = -0.729 < -0.5$, and $\lambda_x\lambda_yp_1 = 0.081 < 1/2$.
	
	If Case I holds, we can see from \eqref{supp:eqn:7620} and \eqref{supp:eqn:7622} that
	\begin{align*}
		\lim_{n,m \to \infty} a_1(n,m,n'm') \to - \infty, \lim_{n,m \to \infty} a_2(n,m,n'm') \to - \infty.
	\end{align*}
    Recall that we have shown in \eqref{supp:eqn:7613} that 
    \begin{align*}
    	\lim_{n,m \to \infty} {(W(X',Y') - \mu')}/{\sigma'}\cdot({\sigma'}/{\sigma}) \to^{d} C N(0,1),
    \end{align*}
	where $C$ is a constant.
	We then have
	\begin{align*}
		&\lim_{n,m \to \infty} {\prob\left\{ ({\sigma'}/{\sigma})(W(X',Y') - \mu')/{\sigma'} 
        + a_1(n,m,n'm') \le \Phi^{-1}(\alpha/2)\right\}} \to 1 \\
		&\lim_{n,m \to \infty} {\prob\left\{({\sigma'}/{\sigma}) ({W(X',Y')-\mu'})/{\sigma'}
        + a_2(n,m,n'm') \ge \Phi^{-1}(1-\alpha/2) \right\}} \to 0.
	\end{align*}
    According to equation \eqref{supp:eqn:33}, it follows
	\begin{align*}
	\begin{split}
		&\lim_{n,m \to \infty} \prob(\text{Proposed method yields a significant $p$-value}) \\
		& = \lim_{n,m \to \infty} {\prob\left\{ ({\sigma'}/{\sigma})(W(X',Y') - \mu')/{\sigma'}  + a_1(n,m,n'm') \le \Phi^{-1}(\alpha/2)\right\}} \\
		& + \lim_{n,m \to \infty} {\prob\left\{({\sigma'}/{\sigma})({W(X',Y')-\mu'})/{\sigma'} + a_2(n,m,n'm') \ge \Phi^{-1}(1-\alpha/2) \right\}} = 1.
	\end{split}
	\end{align*}

	{Case II.} Let us now consider Case II, i.e.
	\begin{align*}
		\lambda_x\lambda_y(p_1 - 1)+1/2 > 0 \text{ and } \lambda_x\lambda_yp_1 < 1/2.
	\end{align*}
	Notice that this case is possible. For example, we can consider $\lambda_x = \lambda_y = 0.5$ and  $p_1 = 0.5$, then $\lambda_x\lambda_y(p_1-1) = -0.5^3 = -0.125 > -0.5$, and $\lambda_x\lambda_yp_1 = 0.0125 < 1/2$.
	
	If Case II holds, we can see from \eqref{supp:eqn:7619} and \eqref{supp:eqn:7622} that
	\begin{align*}
		\lim_{n,m \to \infty} a_1(n,m,n'm') \to + \infty, \lim_{n,m \to \infty} a_2(n,m,n'm') \to - \infty.
	\end{align*}
	Recall that we have shown in \eqref{supp:eqn:7613} that 
	\begin{align*}
		\lim_{n,m \to \infty} {({\sigma'}/{\sigma})(W(X',Y') - \mu')}/{\sigma'} \to^{d} C N(0,1),
	\end{align*}
	where $C$ is a constant.
	We then have
	\begin{align*}
		&\lim_{n,m \to \infty} {\prob\left\{ ({\sigma'}/{\sigma})(W(X',Y') - \mu')/{\sigma'} + a_1(n,m,n'm') \le \Phi^{-1}(\alpha/2)\right\}} \to 0 \\
		&\lim_{n,m \to \infty} {\prob\left\{({\sigma'}/{\sigma})({W(X',Y')-\mu'})/{\sigma'} + a_2(n,m,n'm') \ge \Phi^{-1}(1-\alpha/2) \right\}} \to 0,
	\end{align*}
    According to equation \eqref{supp:eqn:33}, it follows
	\begin{align*}
		\begin{split}
			&\lim_{n,m \to \infty} \prob(\text{Proposed method yields a significant $p$-value}) \\
			& = \lim_{n,m \to \infty} {\prob\left\{ ({\sigma'}/{\sigma})(W(X',Y') - \mu')/{\sigma'} + a_1(n,m,n'm') \le \Phi^{-1}(\alpha/2)\right\}} \\
			& + \lim_{n,m \to \infty} {\prob\left\{({\sigma'}/{\sigma})({W(X',Y')-\mu'})/{\sigma'} + a_2(n,m,n'm') \ge \Phi^{-1}(1-\alpha/2) \right\}} = 0.
		\end{split}
	\end{align*}
	
	{Case III.} Let us now consider case III, i.e.
	\begin{align*}
		\lambda_x\lambda_y(p_1 - 1) +1/2 < 0 \text{ and } \lambda_x\lambda_yp_1 > 1/2.
	\end{align*}
	Notice that
	\begin{align*}
		\lambda_x\lambda_y(p_1 - 1) +1/2 < 0 &\Leftrightarrow \lambda_x\lambda_y(1 - p_1) -1/2 > 0 \\
		& \Leftrightarrow \lambda_x\lambda_y - \lambda_x\lambda_yp_1 > 1/2 \\
		&\Leftrightarrow  \lambda_x\lambda_y > \lambda_x\lambda_yp_1 + 1/2 \\
		&\Leftrightarrow \lambda_x\lambda_y > 1.
	\end{align*}
	It means the Case III is impossible provided $0 < \lambda_x, \lambda_y \le 1$.

	{Case IV.} Let us consider Case IV, i.e.
	\begin{align*}
		\lambda_x\lambda_y(p_1 - 1) + 1/2 > 0 \text{ and } \lambda_x\lambda_yp_1 > 1/2.
	\end{align*}
	Notice that this case is possible. For example, we can consider $p_1 = 0.9$ and $\lambda_x =\lambda_y =  0.9$. Then $\lambda_x\lambda_y(p_1-1) = -0.081 > -1/2$, and $\lambda_x\lambda_yp_1 = 0.9^3 = 0.729 > 1/2$.
	
	If Case IV holds, we can see from \eqref{supp:eqn:7619} and \eqref{supp:eqn:7621} that
	\begin{align*}
		\lim_{n,m \to \infty} a_1(n,m,n'm') \to + \infty, \lim_{n,m \to \infty} a_2(n,m,n'm') \to + \infty.
	\end{align*}
	Recall that we have shown in \eqref{supp:eqn:7613} that 
	\begin{align*}
		\lim_{n,m \to \infty} {({\sigma'}/{\sigma})(W(X',Y') - \mu')}/{\sigma'} \to^{d} C N(0,1),
	\end{align*}
	where $C$ is a constant.
	We then have
	\begin{align*}
		&\lim_{n,m \to \infty} {\prob\left\{({\sigma'}/{\sigma})(W(X',Y') - \mu')/{\sigma'} + a_1(n,m,n'm') \le \Phi^{-1}(\alpha/2)\right\}} \to 0 \\
		&\lim_{n,m \to \infty} {\prob\left\{({\sigma'}/{\sigma})({W(X',Y')-\mu'})/{\sigma'} + a_2(n,m,n'm') \ge \Phi^{-1}(1-\alpha/2) \right\}} \to 1,
	\end{align*}
    According to equation \eqref{supp:eqn:33}, it follows
	\begin{align*}
		\begin{split}
			&\lim_{n,m \to \infty} \prob(\text{Proposed method yields a significant $p$-value}) \\
			& = \lim_{n,m \to \infty} {\prob\left\{ ({\sigma'}/{\sigma})(W(X',Y') - \mu')/{\sigma'} + a_1(n,m,n'm') \le \Phi^{-1}(\alpha/2)\right\}} 
             \\
			& + \lim_{n,m \to \infty} {\prob\left\{({\sigma'}/{\sigma})({W(X',Y')-\mu'})/{\sigma'} + a_2(n,m,n'm') \ge \Phi^{-1}(1-\alpha/2) \right\}} = 1.
		\end{split}
	\end{align*}
	
	Overall, among all possible 4 cases, we have shown Case III is impossible, Case II leads to the rejection probability to $0$ when $n,m \to \infty$, and Case I, Case IV both leads to the rejection probability to $1$ when $n,m \to \infty$. Thus, if Case II holds, the rejection probability goes to $0$ when $n,m \to \infty$, otherwise the probability goes to 1. Notice that the Case II can be written as 
	\begin{align*}
		\lambda_x\lambda_y(1-p_1) < 1/2, \lambda_x \lambda_yp_1 < 1/2. 
	\end{align*}
	Hence, we conclude our results.
\end{proof}	


\section{The proof of results in Section 5.2}

Before we can prove Proposition~5, we need the following result:

\begin{proposition} 
	\label{prop:ranksumties}
	Suppose that $\bx = \{\xx_1, \ldots, \xx_{\n}\}$ and
	$\by = \{\yy_1, \ldots, \yy_{\m}\}$ are samples of 
	observations, which need not necessarily be distinct, 
	from a space $\discretespace \subset \realR$.
	Suppose that $\bxp \subset \bx$ and $\byp \subset \by$ are sub-multisets
	with sizes $\np$ and $\mpp$, respectively. Define
	$\bz = \bx \cup \by$, $\bzp = \bxp \cup \byp$ and suppose only $\bzp$ is known. Let $a = \min \discretespace$ if the minimum exists, otherwise define
	$\absval{ \partial \bxp_{a} } = \absval{ \partial \byp_{a} } =  0$.
	Let $b = \max \discretespace$ if the maximum exists, otherwise define
	$\absval{ \partial \bxp_{b} } = \absval{ \partial \byp_{b} } =  0$. Assume $a \neq b$, i.e. $a < b$. Then, the minimum and maximum possible rank sums
	of $\bx$, over all possible values for the $d = \n - \np + \m - \mpp$ observations in $\bz \setminus \bzp$, are
	\begin{align}
		\min_{\bz \setminus \bzp \in \realR^{\dd}} \rankgiven{\bx}{\bz} 
        &= \rankgiven{\bxp}{\bzp} +  \{ (\n - \np)(\n + \np + 1) + T_1 \}/2,
		\nonumber \\
		\max_{\bz \setminus \bzp \in \realR^{\dd}} \rankgiven{\bx}{\bz} 
		&= \rankgiven{\bxp}{\bzp}  
		+ \{ \n (\n + 2\m + 1) - \np (\np +2 \mpp + 1) - T_2 \}/2,
		\nonumber 
	\end{align}
    where $T_1 = \absval{ \partial \byp_{a} } (\n - \np) + \absval{ \partial \bxp_{b} } (\m - \mpp)$
    and $T_2 = \absval{ \partial \bxp_{a} } (\m - \mpp) + \absval{ \partial \byp_{b} } (\n - \np)$.
\end{proposition}

\begin{proof}
	To start, let us denote $X \setminus X' = X^*$ and $Y \setminus Y' = Y^*$. That is,  $X^*$ is a multiset including all missing samples in $X$ and $Y^*$ is a multiset including all missing samples in $Y$. 
	
	According to the definition of rank in Definition~1, we have 
	\begin{align}  \label{supp:eqn:70}
		R_{X|Z} = \sum_{x \in X} \text{rank}_Z(x) = \sum_{x \in X'} \text{rank}_Z(x) + \sum_{x \in X^*} \text{rank}_Z(x) = R_{X'|Z} + R_{X^*|Z}. 	
	\end{align}
	Let us deal with $R_{X'|Z}$ and $R_{X^*|Z}$ separately. For the former term, we have
	\begin{align}
		R_{X'|Z} &= \sum_{x \in X'} \text{rank}_Z(x) \nonumber \\
		&= \sum_{x \in X'} (|Z_{x}| + |\partial Z_{x}|/2 + 1/2 ) \nonumber \\
		&= \sum_{x \in X'} |Z_{x}| + \sum_{x \in X'} |\partial Z_{x}|/2 + n'/2 	\label{supp:eqn:71}
	\end{align}
	For the first term of the last line, notice that for any $x \in X'$, since $Z = X \cup Y' \cup Y^*$, we have
	\begin{align*}
		|Z_{x}|  &=  | \{z \in Z: z <  x \} | \nonumber \\
		&= | \{x \in X: x <  x \} \cup \{y \in Y': y <  x \} \cup \{y \in Y^*: y <  x \} |	 \nonumber \\
		&= |X_{x}| + |Y'_{x}| + |Y^*_{x}|.
	\end{align*}
	Similarly, for any $x \in X'$, we have
	\begin{align*}
		|\partial Z_{x}| = |\partial X_{x}| + |\partial Y'_{x}| + |\partial Y^*_{x}|. 
	\end{align*} 
	Subsequently, by putting the expressions of $|Z_{x}|$ and $|\partial Z_{x}|$ back into equation \eqref{supp:eqn:71}, we have
	\begin{align*}
		\begin{split}
			R_{X'|Z} =  \underbrace{\sum_{x \in X'}  |X_{x}|}_{: = C_1}   +  \underbrace{\sum_{x \in X'} |Y'_{x}|}_{: = C_2} +  \underbrace{\sum_{x \in X'}  |Y^*_{x}|}_{: = C_3}
			+  \underbrace{\sum_{x \in X'}  |\partial X_{x}|/2}_{: =C_4} +  \underbrace{\sum_{x \in X'} |\partial Y'_{x}|/2}_{: =C_5} +  \underbrace{\sum_{x \in X'}  |\partial Y^*_{x}|/2}_{: =C_6} + n'/2.
		\end{split}
	\end{align*}
	For the same reasons as deriving the above expression for $R_{X'|Z}$, we can derive
	\begin{align*}
		\begin{split}
			R_{X^*|Z} = \underbrace{\sum_{x \in X^*}  |X_{x}|}_{: =D_1}  +  \underbrace{\sum_{x \in X^*} |Y'_{x}|}_{:=D_2} +  \underbrace{\sum_{x \in X^*}  |Y^*_{x}|}_{:=D_3}
			+  \underbrace{\sum_{x \in X^*} |\partial X_{x}|/2}_{:=D_4} +  \underbrace{\sum_{x \in X^*} |\partial Y'_{x}|/2}_{:= D_5} +  \underbrace{\sum_{x \in X^*}  |\partial Y^*_{x}|/2}_{:= D_6} + (n-n')/2.
		\end{split}
	\end{align*}
	We have shown $R_{X|Z}  = R_{X'|Z} + R_{X^*|Z} $ in equation \eqref{supp:eqn:70}, which follows
	\begin{align}
		R_{X|Z} = C_1 + C_2 + \cdots + C_6 + D_1 + D_2 + \cdots + D_6 + n/2.
	\end{align}
	
	Recall that our goal is to find the minimum possible and maximum possible values of $R_{X|Z}$ with all possible values in $X^*$ and $Y^*$. In the following, we will deal with $C_1+C_4+D_1+D_4$, $C_2 + C_5$, $C_3 + C_6$, $D_2 + D_5$, $D_3 + D_6$ separately. First, we notice that $C_1+C_4+D_1+D_4$ is a constant regardless of the values of  $X^*$ and $Y^*$. In fact,
	\begin{align*}
		C_1 + C_4 + D_1 + D_4 &= \sum_{x \in X'}  |X_{x}| + \sum_{x \in X'}  |\partial X_{x}|/2 + \sum_{x \in X^*}  |X_{x}| + \sum_{x \in X^*} |\partial X_{x}|/2 \nonumber \\
		& = \sum_{x \in X'} ( |X_{x}| +  |\partial X_{x}|/2 + 1/2) + \sum_{x \in X^*} ( |X'_{x}| +  |\partial X_{x}|/2 + 1/2) - n/2 \nonumber \\
		& = \sum_{x \in X'} \text{rank}_X(x) +  \sum_{x \in X^*} \text{rank}_X(x)  - n/2 \nonumber \\
		& = \sum_{x \in X} \text{rank}_X(x) - n/2 \nonumber \\
		& = (1+n)n/2 - n/2 = n^2 / 2.
	\end{align*}
    For $C_2 + C_5$, notice that
	\begin{align*}
		\begin{split}
			R_{X'|Z'} &= \sum_{x \in X'} \text{rank}_{Z'}(x)  = \sum_{x \in X'} (|Z'_{x}| + |\partial Z'_{x}|/2 + 1/2 )  \\
			&= \sum_{x \in X'} |Z'_{x}| + \sum_{x \in X'} |\partial Z'_{x}|/2 + n'/2  \\
			& = \sum_{x \in X'} |X'_{x}| +  \sum_{x \in X'} |Y'_{x}| + \sum_{x \in X'} |\partial X'_{x}|/2 + \sum_{x \in X'} |\partial Y'_{x}|/2  + n'/2  \\
			& = \sum_{x \in X'} |X'_{x}| + C_2 + \sum_{x \in X'} |\partial X'_{x}|/2 + C_5 + n'/2  \\
			& = \sum_{x \in X'} ( |X'_{x}| + |\partial X'_{x}|/2 + 1/2) + C_2 + C_5  \\
			& = \sum_{x \in X'} \text{rank}_{X'}(x) + C_2 + C_5  \\
			& = (1+n')n'/2 + C_2 + C_5.
		\end{split} \\
		\Rightarrow  C_2 + C_5 &= R_{X'|Z'}  -  (1+n')n'/2.
	\end{align*}
	Recall that $X'$ and $Z'$ include observed values only. Thus, $R_{X'|Z'}$ is a constant regardless of the missing values in $X^*$ and $Y^*$, which means $C_2 + C_5$ is also a constant.
	Now, adding $C_1 + C_4 + D_1 + D_4$ and $C_2 + C_5$ together, we have
	\begin{align}
		R_{X|Z} &= R_{X'|Z'} - (1+n')n'/2 + n^2/2 + C_3 + C_6 + D_2 + D_3 + D_5 + D_6 + n/2 \nonumber\\
		& = R_{X'|Z'} + (n-n')(n+n'+1)/2 + C_3 + C_6 + D_2 + D_3 + D_5 + D_6,
	\label{supp:eqn:76}
	\end{align}
	where the last equation holds because
	\begin{align*}
		- (1+n')n'/2 + n^2/2 + n/2 &= (n^2 + n - n' -(n')^2)/2\\
		& = (n^2 + nn' - nn' + n - n' -(n')^2)/2\\
		& = \{(n^2 - nn') + (nn' - (n')^2) + (n-n')\}/2\\
		& = \{n(n-n') + n'(n-n') + (n-n')\}/2\\
		& = (n-n')(n+n'+1)/2.
	\end{align*}
	Let us now consider $C_3 + C_6$, $D_2 + D_5$, $D_3 + D_6$ separately. Suppose $\text{max} \Omega = b$ is well defined. Notice that
	\begin{align}
		C_3 + C_6 &= \sum_{x \in X'}  |Y^*_{x}| + \sum_{x \in X'}  |\partial Y^*_{x}|/2 \nonumber \\
		& = \sum_{x \in X'} \left( |Y^*_{x}| +   |\partial Y^*_{x}|/2  \right) \nonumber \\
		& = \sum_{x \in X', x = b} \left( |Y^*_{x}| +   |\partial Y^*_{x}|/2  \right) +  \sum_{x \in X', x \neq b} \left( |Y^*_{x}| +   |\partial Y^*_{x}|/2  \right) \nonumber\\
		& = \sum_{x \in X', x = b} \left( |Y^*_{b}| +   |\partial Y^*_{b}|/2  \right) +  \sum_{x \in X', x \neq b} \left( |Y^*_{x}| +   |\partial Y^*_{x}|/2  \right). \label{supp:eqn:72}
	\end{align}
	Recall that $|Y^*_{b}|$ is defined as the number of elements in $Y^*$ smaller than $b$ and $|\partial Y^*_{b}|$ is defined as the number of elements in $Y^*$ equal to $b$. Since for any $y \in Y^*$, we must have $y \le b$, we can see that when
	\begin{align}
		\text{min} Y^* = \text{max} Y^*  = b, \label{supp:eqn:73}
	\end{align}
	i.e. all $y \in Y^*$ equals to $b$, $|Y^*_{b}| +   |\partial Y^*_{b}|/2$ takes its minimum value $(m-m')/2$. Further, since there are $|\partial X'_b|$ number of elements in $X'$ equalling to $b$, we have
	\begin{align*}
	 \sum_{x \in X', x = b} \left( |Y^*_{x}| +   |\partial Y^*_{x}|/2  \right) \ge |\partial X'_b| (m-m')/2,
	\end{align*}
	where the  ``='' holds if \eqref{supp:eqn:73} holds. For any $x \in X'$ and $x \neq b$. Recall that $x < b$. Then, we have  $|Y^*_{x}| +   |\partial Y^*_{x}|/2$ takes its minimum 0 when \eqref{supp:eqn:73} holds. Thus,
	\begin{align*}
		\sum_{x \in X', x \neq b} \left( |Y^*_{x}| +   |\partial Y^*_{x}|/2  \right) \ge 0,
	\end{align*} 
	where the ``='' holds when \eqref{supp:eqn:73} holds. Put the above two inequalities back into \eqref{supp:eqn:72}, we conclude
	\begin{align*}
	 C_3 + C_6 \ge |\partial X'_b| (m-m')/2,
	\end{align*}
	where ``='' holds if \eqref{supp:eqn:73} holds.	If, however, $\max \discretespace$ does not exist, let us denote $S_1 = \{s| s \in \Omega, s > \text{max} Z'\}$ as the set including all the support larger than the maximum observed value $\text{max} Z'$. Then, $S_1 \neq \emptyset$, i.e. there must exist $s \in \Omega$ such that $s > \text{max} Z'$, since if not, $\max \discretespace = \text{max} Z'$ and this is contradicted with the fact that $\max \discretespace$ does not exist. Then, we have
	\begin{align*}
		C_3 + C_6 &= \sum_{x \in X'}  |Y^*_{x}| + \sum_{x \in X'}  |\partial Y^*_{x}|/2 \ge 0
	\end{align*}
	where ``='' holds if all samples in $Y^*$ larger than $\text{max} X'$. In particular, by letting $|\partial X'_b| = 0$, we have   
	\begin{align*}
		C_3 + C_6 \ge |\partial X'_b| (m-m')/2.
	\end{align*}
	
	Suppose $\text{min} \discretespace = a$ is well defined. Then, for $D_2 + D_5$, we have
	\begin{align}
		D_2 + D_5 &= \sum_{x \in X^*} |Y'_{x}| + \sum_{x \in X^*} |\partial Y'_{x}|/2 \nonumber \\
		& = \sum_{x \in X^*} (|Y'_{x}| + |\partial Y'_{x}|/2) \nonumber \\
		& = \sum_{x \in X^*}  (|\{y \in Y': y < x \}| +  |\{y \in Y': y = x \}|/2 ) \nonumber \\
		\begin{split}
			& =  \sum_{x \in X^*} \biggl(|\{y \in Y': y = a, y < x \}|  +  |\{y \in Y': y = a, y = x \}|/2  \\
			& + |\{y \in Y': y \neq a, y < x \}|  +  |\{y \in Y': y \neq a, y = x \}|/2 \biggr ) \label{supp:eqn:74}
		\end{split}
	\end{align}
	For any $x \in X^*$, since it is defined between $[a,b]$, we have that $|\{y \in Y': y = a, y < x \}|  +  |\{y \in Y': y = a, y = x \}|/2$ takes its minimum value $|\partial Y'_a|/2$ when $x = a$. Meanwhile, since each $y \in Y$ is defined in $[a,b]$, we have $y \neq a \Rightarrow y > a$. Then, we can verify that $|\{y \in Y': y \neq a, y < x \}|  +  |\{y \in Y', y \neq a: y = x \}|/2$ takes its minimum value 0 when $x = a$. Put these results back into \eqref{supp:eqn:74} and notice that there are $n-n'$ number of elements in $X^*$. We conclude
	\begin{align*}
	 D_2 + D_5 \ge |\partial Y'_a|(n-n')/2, 
	\end{align*}
	where the ``='' holds when all $x$ in $X^*$ equals $a$, i.e.
	\begin{align}
		\text{min} X^* = \text{max} X^* = a.
		\label{supp:eqn:75}
	\end{align} 
	If, however, $\min \discretespace$ does not exist, let us denote $S_2 = \{s| s \in \Omega, s < \text{min} Z'\}$ as the set including all the support smaller than the minimum observed value $\text{min} Z'$. Then, $S_2 \neq \emptyset$, i.e. there must exist $s \in \Omega$ such that $s < \text{min} Z'$, since if not, $\min \discretespace = \text{min} Z'$ and this is contradicted with the fact that $\min \discretespace$ does not exist. Then, we have
	\begin{align*}
		D_2 + D_5 = \sum_{x \in X^*} |Y'_{x}| + \sum_{x \in X^*} |\partial Y'_{x}|/2 \ge 0
	\end{align*}
	where ``='' holds if all samples in $X^*$ smaller than $\text{min} Y'$. In particular, by letting $|\partial Y'_a| = 0$, we have   
	\begin{align*}
		D_2 + D_5 \ge |\partial Y'_a|(n-n')/2.
	\end{align*}

	Let us now consider $D_3 + D_6$. We have
	\begin{align}
		D_3 + D_6 &= \sum_{x \in X^*} |Y^*_x| + \sum_{x \in X^*} |\partial Y^*_x|/2 \nonumber \\
		&=  \sum_{x \in X^*} \left( |\{y \in Y^*: y < x\}| + |\{y \in Y^*: y = x\}|/2\right).
	\end{align}
    We notice that
	\begin{align*}
		D_3 + D_6 \ge 0,
	\end{align*}
	where the ``= '' holds when all samples in $Y^*$ larger than $\text{max} X^*$.
	
	Overall, we have shown that
	\begin{align}
		C_3 + C_6 &\ge |\partial X_b'|(m-m')/2, \label{supp:eqn:77} \\
		D_2 + D_5 &\ge |\partial Y_a'|(n-n')/2,  \label{supp:eqn:78}\\
		D_3 + D_6 & \ge 0  \label{supp:eqn:79},
	\end{align}
	where if $\text{max} \Omega = b$ is well defined, the first ``='' holds when \eqref{supp:eqn:73} holds, otherwise it holds when all samples in $Y^* > \text{max} X'$; if $\text{min} \Omega = a$ is well defined, the second ``='' holds when \eqref{supp:eqn:75} holds, otherwise it holds when all samples in $X^* < \text{min} Y'$; the last ``='' holds when all samples in $Y^*$ larger than $\text{max} X^*$. Put \eqref{supp:eqn:77}-\eqref{supp:eqn:79} back into \eqref{supp:eqn:76}, we have
	\begin{align}
		R_{X|Z} \ge R_{X'|Z'} + (n-n')(n+n'+1)/2 +  |\partial X_b'|(m-m')/2  + |\partial Y_a'|(n-n')/2 \label{supp:eqn:710}
	\end{align}    
	where the ``='' holds when all equations in \eqref{supp:eqn:77}-\eqref{supp:eqn:79} hold. In fact, the equation can always be taken regardless of whether or not $a,b$ are well defined. This argument follows by below discussions:
	\begin{itemize}
		\item 	if $\text{max} \Omega = b$ and  $\text{min} \Omega = a$ are both well defined, by letting  \eqref{supp:eqn:73} and \eqref{supp:eqn:75} both hold, we have \eqref{supp:eqn:77} and \eqref{supp:eqn:78} take equation, and \eqref{supp:eqn:79} takes equation since $\text{min} X^* = \text{max} X^* = a < b = \text{min} Y^* = \text{max} Y^*$. Thus, the equation in \eqref{supp:eqn:710} is taken.
		\item if  $\text{max} \Omega = b$  is well defined and $\text{min} \Omega = a$ does not exist, by letting \eqref{supp:eqn:73} hold and all samples in $X^* < \text{min} Y'$, we have \eqref{supp:eqn:77} and \eqref{supp:eqn:78} take equation, and \eqref{supp:eqn:79} takes equation since now $\text{max} X^* < \text{min} Y' \le b = \text{min} Y^* = \text{max} Y^*$. Thus, the equation in \eqref{supp:eqn:710} is taken.
		\item if  $\text{min} \Omega = a$  is well defined and $\text{max} \Omega = b$ does not exist, by letting \eqref{supp:eqn:75} hold and all samples in $Y^* > \text{max} X'$, we have \eqref{supp:eqn:78} and \eqref{supp:eqn:77} take equation, and \eqref{supp:eqn:79} takes equation since now $\text{min} X^* = \text{max} X^* = a \le \text{max} X' < \text{min} Y^*$. Thus, the equation in \eqref{supp:eqn:710} is taken.
		\item  if  $\text{min} \Omega = a$ and $\text{max} \Omega = b$ do not exist, by letting all samples in $X^* < \text{min} Y'$, all samples in $Y^* > \text{max} X'$, and all samples in $Y^* > \text{max} X^*$, we have  \eqref{supp:eqn:77}, \eqref{supp:eqn:78} and \eqref{supp:eqn:79} all take equation. Thus, the equation in \eqref{supp:eqn:710} is taken.
	\end{itemize}
    Hence, we conclude our result for minimum bound.
	
	To conclude maximum bound, we take the same approach as we took in the alternative proof for Proposition~\ref{prop:mainprop}. To start, applying the same argument to
	$\by$ instead of $\bx$, by symmetry would give 	\begin{align*}
		\mathop{\text{min}}\limits_{Z\setminus Z'} R_{Y|Z} =   R_{Y'|Z'} + \underbrace{(m-m')(m+m'+1)/2}_{:= E_1} + |\partial X'_a|(m-m')/2 + |\partial Y'_b| (n-n')/2,
	\end{align*}
	where $|\partial X'_a|$ and $|\partial Y'_b|$ are defined as 0 if  $\text{min} \Omega$ and $\text{max} \Omega$ do not exist, respectively.	Since 
	\begin{align*}
		R_{Z|Z} = R_{X|Z} + R_{Y|Z} = \underbrace{{(1 + n + m)(n+m)}/2}_{:=E_2}
	\end{align*}
	is a constant, therefore when $ R_{Y|Z} $ takes its minimum among all possible $Z^*$, we have $R_{X|Z}$ takes its maximum, i.e.
	\begin{align*}
		&\mathop{\text{max}}\limits_{Z\setminus Z'} R_{X|Z}  = R_{Z|Z} - \mathop{\text{min}}\limits_{Z\setminus Z'} R_{Y|Z} \nonumber \\
		& = E_2 - R_{Y'|Z'} - E_1 - |\partial X'_a|(m-m')/2 - |\partial Y'_b| (n-n')/2.
	\end{align*}
	Also, notice that 
	\begin{align*}
		&R_{Y'|Z'} + R_{X'|Z'} = \underbrace{(1 + n' + m')(n'+m')/2}_{:= E_3}   \\
		\Rightarrow &R_{Y'|Z'} = E_3 - R_{X'|Z'}
	\end{align*}
	Subsequently, we have
	\begin{align*}
		\mathop{\text{max}}\limits_{Z\setminus Z'} R_{X|Z} =  E_2 - E_3 + R_{X'|Z'} - E_1 - |\partial X'_a|(m-m')/2 - |\partial Y'_b| (n-n')/2 \nonumber \\
		= R_{X'|Z'} + (E_2 - E_3 - E_1)- |\partial X'_a|(m-m')/2 - |\partial Y'_b| (n-n')/2.
	\end{align*}
	From \eqref{supp:eqn:combine}, we have 
	\begin{align*}
		E_2 - E_3 - E_1 = \{n(n+2m+1) - n'(n'+2m'+1)\}/2.
	\end{align*} 
	Finally, we have
	\begin{align*}
		\mathop{\text{max}}\limits_{Z\setminus Z'} R_{X|Z} = R_{X'|Z'} + ( \{n(n+2m+1) - n'(n'+2m'+1)\}/2)- |\partial X'_a|(m-m')/2 - |\partial Y'_b| (n-n')/2,
	\end{align*}
	which concludes our result for the maximum bound.
\end{proof}


\subsection{Proof of Proposition~5}

\begin{proposition}
	\label{prop:mainthmties}
	Suppose that $\bx = \{\xx_1, \ldots, \xx_{\n}\}$ and
	$\by = \{\yy_1, \ldots, \yy_{\m}\}$ are samples of 
	observations, which need not necessarily be distinct, 
	from a space $\discretespace \subset \realR$.
	Suppose that $\bxp \subset \bx$ and $\byp \subset \by$ are sub-multisets
	with sizes $\np$ and $\mpp$, respectively.
	Let $a = \min \discretespace$ if the minimum exists, otherwise define
	$\absval{ \partial \bxp_{a} } = \absval{ \partial \byp_{a} } =  0$.
	Let $b = \max \discretespace$ if the maximum exists, otherwise define
	$\absval{ \partial \bxp_{b} } = \absval{ \partial \byp_{b} } =  0$.
	Then
	the Wilcoxon-Mann-Whitney statistic $\wmwstat{\bx}{\by}$ is
	bounded as follows:
	\begin{align}
		\wmwstat{\bxp}{\byp} + \Tone
		\leq
		\wmwstat{\bx}{\by}
		\leq
		\wmwstat{\bxp}{\byp} + (\m \n - \mpp \np) - \Ttwo,
		\label{eqn:wmwboundsties}
	\end{align}
	with
	$\Tone = \absval{ \partial \byp_{a} } (\n - \np)/2 
	+ \absval{ \partial \bxp_{b} } (\m - \mpp)/2$ and 
	$\Ttwo = \absval{ \partial \bxp_{a} } (\m - \mpp) 
	+ \absval{ \partial \byp_{b} } (\n - \np)$
	non-negative values
	depending on $a, b, \m, \mpp, \n, \np, \bxp$ and $\byp$.
\end{proposition}

\begin{proof}
 This proof is essentially the same as the proof of Theorem \ref{thm:mainthm}. Here we present the proof for consistency. Writing $\bz = \bx \cup \by$, Definition~1 for the Wilcoxon-Mann-Whitney statistic 
	and Equation~\eqref{eqn:prop:minR} 
	of Proposition~\ref{prop:ranksumties} gives us
	\begin{alignat}{2}
		\wmwstat{\bx}{\by}
		&{=} 
		\rankgiven{\bx}{\bz} - \n(\n+1)/2
		\nonumber \\
		&{\geq}
		\rankgiven{\bxp}{\bzp} + (\n - \np)(\n + \np + 1) /2
		- \n(\n+1)/2 + T_1
		\nonumber \\
		&= \rankgiven{\bxp}{\bzp} 
		+ (\n^2 - \np^2 + \n - \np) /2
		- (\n^2 + \n) /2 + T_1
		\nonumber \\
		&= \rankgiven{\bxp}{\bzp} 
		- \np(\np + 1) /2 + T_1
		\nonumber \\
		&{=}
		\wmwstat{\bxp}{\byp} + T_1
		\nonumber \\
		\Rightarrow
		\wmwstat{\bx}{\by}
		&\geq
		\wmwstat{\bxp}{\byp} + T_1.
		\nonumber
	\end{alignat}
	
	And then, using
	Definition~1 and Equation~\eqref{eqn:prop:maxR} 
	of Proposition~\ref{prop:mainprop}, this gives us
	\begin{alignat}{2}
		\wmwstat{\bx}{\by}
		&{=} 
		\rankgiven{\bx}{\bz} - \n(\n+1) /2
		\nonumber \\
		&{\leq}
		\rankgiven{\bx}{\bz} 
		+ \left \{\n (\n + 2\m + 1) - \np (\np +2 \mpp + 1) \right \} /2
		- \n(\n+1) /2 + T_2
		\nonumber \\
		&= \rankgiven{\bx}{\bz} 
		+ \left \{ 2\m\n + \n (\n + 1) - \np (\np +2 \mpp + 1) \right \} /2
		- \n(\n+1) /2 + T_2
		\nonumber \\
		&= \rankgiven{\bx}{\bz} 
		+ \left\{ 2\m\n - 2\mpp\np  - \np (\np + 1) \right \} /2 + T_2
		\nonumber \\
		&= \rankgiven{\bx}{\bz} - \np (\np + 1) /2
		+ (\m\n - \mpp\np) + T_2
		\nonumber \\
		&{=} 
		\wmwstat{\bxp}{\byp} + (\m\n - \mpp\np) + T_2
		\nonumber \\
		\Rightarrow
		\wmwstat{\bx}{\by} &\leq \wmwstat{\bxp}{\byp} + (\m\n - \mpp\np) + T_2,
		\nonumber
	\end{alignat}
	which proves the proposition.
\end{proof}


\subsection{Proof of Proposition~6}

\begin{proposition}  
	Suppose that $X = \{x_1,\ldots,x_n\}$ and $Y = \{y_1,\ldots,y_m\}$ are samples of real-valued observations, which need not necessarily be distinct. Suppose that $X' \subset X$ and $Y' \subset Y$ are sub-multisets with sizes $n'$ and $m'$ respectively, which are known. Define $Z = X \cup Y$ and $Z' = X' \cup Y'$ and suppose that $Z \setminus Z'$ is unknown. Suppose $X_1, Y_1$ and $X_2, Y_2$ are such that $X' \subset X_i$ and $Y' \subset Y_i$ for $i= 1,2$ and
	\begin{align*}
		W(X_1,Y_1) = \mathop{\text{min}}\limits_{Z\setminus Z'} W(X,Y),~~~ W(X_2,Y_2) = \mathop{\text{max}}\limits_{Z\setminus Z'} W(X,Y).
 	\end{align*}
	Then, among candidates for $X,Y$, it is possible $X_3, Y_3$ exist such that $X' \subset X_3$ and $Y' \subset Y_3$ and
	\begin{align*}
		W(X_1,Y_1) < W(X_3,Y_3) < W(X_2,Y_2),
	\end{align*}
    but moreover, after defining $\mu = nm/2$, the standardised statistics are such that
	\begin{align*}
		(W(X_3,Y_3)-\mu)/{\sigma(X_3,Y_3)}  \notin \left[(W(X_1,Y_1)-\mu)/{\sigma(X_1,Y_1)}, (W(X_2,Y_2)-\mu)/{\sigma(X_2,Y_2)}\right], 
	\end{align*}
	and consequently $p_3 \notin [p_\text{min}, p_{\text{max}}]$, 
    where $p_\text{min} = \text{min} \{p_1,p_2\}$ and 
    $p_\text{max} = \text{max} \{p_1,p_2\}$ 
    if $(W(X_1,Y_1)-\mu)(W(X_2,Y_2)-\mu) \ge 0$, 
    otherwise $p_\text{max} = 1$, and $p_1,p_2$ and $p_3$ are 
    the p-values for $W(X_1,Y_1), W(X_2, Y_2)$ and $W(X_3, Y_3)$, respectively.
\end{proposition}

\begin{proof}
    The condition $(W(X_1,Y_1)-\mu)(W(X_2,Y_2)-\mu) \ge 0$ is equivalent to 
    the condition that
    $(W(X_1,Y_1)- \n \m / 2)$ and $(W(X_2,Y_2)- \n\m/2)$ have the same sign, 
    which is Condition~2 in the main paper.

	We provide an example. Let us consider the case $X = \{ 1,2,3,2,2,1,1 \}$, $Y = \{3,3,3,3,3,3, y_7\}$ and the support is $\{1,2,3,4\}$, i.e. $y_7$ is a real number taking from $\{1,2,3,4\}$. Suppose $X' = X$, $Y' = Y \setminus \{y_7\}$, and denote $Z' = X' \cup Y'$. According to the definition of ranks with ties, we can verify that 
	\begin{align*}
		\text{rank}_{Z'}(1) &= |Z'_{1}| + (|\partial Z'_1| + 1)/2 = 0 + (3+1)/2 = 2,\\
		\text{rank}_{Z'}(2) &= |Z'_{2}| + (|\partial Z'_2| + 1)/2 = 3 + (3+1)/2 = 5,\\
		\text{rank}_{Z'}(3) &= |Z'_{3}| + (|\partial Z'_3| + 1)/2 = 6 + (7+1)/2 = 10.
	\end{align*}
	Subsequently,
	\begin{align*}
		W(X',Y') &= R_{X'|Z'} - n'(n'+1)/2\\
		& = 3\text{rank}_{Z'}(1) + 3\text{rank}_{Z'}(2) + \text{rank}_{Z'}(3) - 7 \cdot 8 /2 \\
		& = 6 + 15 + 10 - 28 = 3.
	\end{align*}
	Now, we are going to obtain the minimum and maximum possible $W(X,Y)$ using the result of Proposition \ref{prop:mainthmties}. Since the support $\Omega = \{1,2,3,4\}$, we have $1 = 1, b = 4$. Hence, $|\partial X'_a| = 3, |\partial Y'_a| = 0, |\partial X'_b| = 0, |\partial Y'_b| = 0$. Also, we have $n - n' = 0$ and $m -m' = 1$. Therefore, we can use  Proposition \ref{prop:mainthmties} and get
	\begin{align*}
		\mathop{\text{min}}\limits_{Z\setminus Z'} W(X,Y) &= W(X',Y') + 0 = 3,\\
		\mathop{\text{max}}\limits_{Z\setminus Z'} W(X,Y) &= W(X',Y') + (49 - 42) - 3/2 = 8.5.
	\end{align*}
	
	Let us denote $X_1 = X$, $Y_1  = Y' \cup \{4\}$ and $X_2 = X$, $Y_2 = Y' \cup \{1\}$. That is, $y_7$ takes $1$ and $4$ for $Y_1$, and $Y_2$, separately. Following the above procedures as we obtain $W(X',Y')$,  we can verify that 
	\begin{align*}
		W(X_1,Y_1) = \mathop{\text{min}}\limits_{Z\setminus Z'} W(X,Y) = 3,~W(X_2,Y_2) = \mathop{\text{max}}\limits_{Z\setminus Z'} W(X,Y) = 8.5.
	\end{align*}
	Recall that the variance of $W(X,Y)$ with ties is defined as
	\begin{align*}
		\sigma^2(X,Y) = nm(n+m+1)/12 - nm\{12(n+m)(n+m-1)\}^{-1}\sum_{i=1}^3(d_i^3 - d_i),
	\end{align*}
	and $nm = 49, n+m+1 = 15, n+m-1 = 13$. Hence, we obtain
	\begin{align*}
		\sigma^2(X_1,Y_1) &= 49\cdot 15/12 - {49}/(12 \cdot 14 \cdot 13)[(3^3 - 3) + (3^3 - 3) + (7^3 - 7)], \\ 
		&= 735/12 - 49/2184 \cdot 384\\
		&= (133770-18816)/2184 \\
		&= 114954/2184,
	\end{align*}
	and
	\begin{align*}
	   \sigma^2(X_2,Y_2) &=   49 \cdot 15/12 - \frac{49}{12 \cdot 14 \cdot 13}[(4^3 - 4) + (3^3 - 3) + (7^3 - 7)] \\
	   & =  735/12 - 49/2184 \cdot 420\\
	   & = (133770-20580)/2184\\
	   & = 113190/2184,
	\end{align*}
	respectively. For both cases, we have $\mu = nm/2 = 49/2$. Thus, it further follows
	\begin{align*}
		{(W(X_1,Y_1) - \mu)}/{\sigma(X_1,Y_1)} & = (3 - 49/2)/(114954/2184)^{1/2} \\ 
		& = -43/\{2(114954/2184)^{1/2}\} \\
		{(W(X_2,Y_2) - \mu)}/{\sigma(X_2,Y_2)} &= (8.5 - 49/2)/(735/12 - 49/2184 \cdot 384)^{1/2}\\
		& = -16/(113190/2184)^{1/2} . 
	\end{align*}
	Define $X_3 = X$, $Y_3 = Y' \cup \{3\}$. We can then verify that $W(X_3,Y_3) = 3.5$, which implies
	\begin{align*}
		\mathop{\text{min}}\limits_{Z\setminus Z'} W(X,Y) = W(X_1,Y_1) <W(X_3,Y_3) < W(X_2,Y_2) = \mathop{\text{max}}\limits_{Z\setminus Z'} W(X,Y).
	\end{align*}
	For the variance, we have
	\begin{align*}
		\sigma^2(X_3,Y_3) &= 49\cdot 15/12 - 49/(12 \cdot 14 \cdot 13)[(3^3 - 3) + (3^3 - 3) + (8^3 - 8)] \\
		&= 735/12 - 49/2184\cdot 552\\
		& = (133770-27084)/2184\\
		& = 106686/2184.
	\end{align*}
	Thus,
	\begin{align*}
		(W(X_3,Y_3) - \mu )/{\sigma(X_3,Y_3)} &= (3.5 - 49/2)/(735/12 - 49/2184\cdot 552)^{1/2}\\
		= -21/(106686/2184)^{1/2}.
	\end{align*}
	Notice that 
	\begin{align*}
		\{{(W(X_1,Y_1) - \mu)}/{\sigma(X_1,Y_1)}\}^2 & = 43^2/\{4(114954/2184)\} \\
		& = 43\cdot 43 \cdot 2184 /(4 \cdot 114954) \\
		& = 1009554/114954,\\
		\{{(W(X_2,Y_2) - \mu)}/{\sigma(X_2,Y_2)}\}^2 &= 16^2/(113190/2184)\\
		& = 16\cdot 16 \cdot 2184/113190\\
		& = 559104/113190\\
		\{{(W(X_3,Y_3) - \mu)}/{\sigma(X_3,Y_3)}\}^2 &= 21^2/(106686/2184)\\
		& = 21\cdot 21 \cdot 2184/106686\\
		& = 963144/106686.
	\end{align*}
	Now, we compare the above three numbers using the concept of cross-multiplication. Since 
	\begin{align*}
	963144 \cdot 114954 = 110717255376 > 106686 \cdot 114954 = 12263982444,
	\end{align*}
	we have
	\begin{align*}
		\{{(W(X_3,Y_3) - \mu)}/{\sigma(X_3,Y_3)}\}^2 > \{{(W(X_1,Y_1) - \mu)}/{\sigma(X_1,Y_1)}\}^2. 
	\end{align*}
	Similarly, since
	\begin{align*}
		1009554 \cdot 113190 = 114271417260 > 559104 \cdot 114954 = 64271241216,
	\end{align*}
	we have 
    \begin{align*}
		\{{(W(X_1,Y_1) - \mu)}/{\sigma(X_1,Y_1)}\}^2 > \{{(W(X_2,Y_2) - \mu)}/{\sigma(X_2,Y_2)}\}^2. 
	\end{align*}
	Thus, it follows
	\begin{align*}
		\{{(W(X_3,Y_3) - \mu)}/{\sigma(X_3,Y_3)}\}^2 &> \{{(W(X_1,Y_1) - \mu)}/{\sigma(X_1,Y_1)}\}^2 \\
		&> \{{(W(X_2,Y_2) - \mu)}/{\sigma(X_2,Y_2)}\}^2.
	\end{align*}
	Recall that $(W(X_3,Y_3)-\mu)/{\sigma(X_3,Y_3)}$, $(W(X_1,Y_1)-\mu)/{\sigma(X_1,Y_1)}$ and $(W(X_2,Y_2)-\mu)/{\sigma(X_2,Y_2)}$ are all negative real numbers. Hence,
	\begin{align*}
		{(W(X_3,Y_3) - \mu)}/{\sigma(X_3,Y_3)} &< {(W(X_1,Y_1) - \mu)}/{\sigma(X_1,Y_1)} \\
		&< {(W(X_2,Y_2) - \mu)}/{\sigma(X_2,Y_2)}.
	\end{align*}
	Therefore, 	
	\begin{align*}
		(W(X_3,Y_3)-\mu)/{\sigma(X_3,Y_3)}  \notin \left[(W(X_1,Y_1)-\mu)/{\sigma(X_1,Y_1)}, (W(X_2,Y_2)-\mu)/{\sigma(X_2,Y_2)}\right]. 
	\end{align*}
    Further, since $(W(X_3,Y_3)-\mu)/{\sigma(X_3,Y_3)}$, $(W(X_1,Y_1)-\mu)/{\sigma(X_1,Y_1)}$ and $(W(X_2,Y_2)-\mu)/{\sigma(X_2,Y_2)}$ all follow standard normal distribution and the cumulative distribution function
    $\cdfphi$ is monotonic, we have
    \begin{align*}
    	2\cdfphi((W(X_3,Y_3)-\mu)/{\sigma(X_3,Y_3)}) &\le 2\cdfphi((W(X_1,Y_1)-\mu)/{\sigma(X_1,Y_1)}) \\ &\le
    	2\cdfphi((W(X_2,Y_2)-\mu)/{\sigma(X_2,Y_2)}).
    \end{align*}
     That is, $p_3 < p_1 < p_2$, which follows $p_3 < p_{\text{min}} = \text{min} \{p_1,p_2\}$. Thus, we conclude our result.
\end{proof}


\subsection{Proof of Proposition~7}

Before proving Proposition~7, 
we introduce the following lemma.

\begin{lemma} 
    \label{lem:S1}
	Let $f(x) = x^3 - x$, where $x \in \mathbb{R}$. Suppose $a_1, \ldots, a_k$ and $b_1, \ldots, b_k$ are both non-negative integer sequence such that $a_i \le b_i$ for any $i=1,\ldots,k$. Denote
	\begin{align*}
		a_{\text{max}} = \text{max} \{a_1,\ldots,a_k\}
	\end{align*}
	and define $z$ such as
	\begin{align}
		z := a_{\text{max}} +  \sum_{i = 1}^k (b_i - a_i).
		\label{supp:eqn:l2}
	\end{align}
	Then, we have
	\begin{align}
		f(z) - f(a_{\text{max}}) \ge \sum_{i = 1}^k \left( f(b_i) - f(a_i)\right). 
		\label{supp:eqn:l3}
	\end{align}
\end{lemma}

\begin{proof}
	To start, we notice that the equation \eqref{supp:eqn:l2} can be written as
	\begin{align*}
		z &= a_{\text{max}} + b_j - a_j + \sum_{i=1, i\neq j}^k (b_i - a_i),
	\end{align*}
    where $j$ is any number in $\{1,\ldots,k\}$. Since $b_i \ge a_i$ for any $i=1,\ldots,k$, for any given $j \in \{1,\ldots,k\}$, we have
	\begin{align*}
		\sum_{i=1, i\neq j}^k (b_i - a_i) \ge 0,
	\end{align*}
    which gives us
    \begin{align*}
    	z \ge a_{\text{max}} + b_j - a_j.
    \end{align*}
    Recall that $a_{\text{max}}$ is defined as the maximum number in $\{a_1,\ldots, a_k\}$. This follows
	\begin{align}
		\begin{split}
			 z 	\ge b_j.
		\end{split}
		\label{supp:eqn:l4}
	\end{align}
	
	Subsequently, according to the definition of $f(x)$, we have
	\begin{align}
		\sum_{i = 1}^k \left( f(b_i) - f(a_i)\right) &= \sum_{i = 1}^k (b_i^3 - b_i - a_i^3 + a_i) \nonumber\\
		& = \sum_{i = 1}^k(b_i^3 - a_i^3) - \sum_{i = 1}^k(b_i - a_i). \nonumber
	\end{align}
    Using the definition of $a_{\text{max}}$ in equation \eqref{supp:eqn:l2}, we further have 
    \begin{align}
		\sum_{i = 1}^k \left( f(b_i) - f(a_i)\right) & = \sum_{i = 1}^k(b_i^3 - a_i^3)   - (z - a_{\text{max}}) \nonumber\\
		& = \sum_{i = 1}^k \{(b_i-a_i)(b_i^2 + a_ib_i + a_i^2 )\}- (z - a_{\text{max}}). \nonumber
	\end{align}
    Subsequently, the inequality \eqref{supp:eqn:l4} gives us
    \begin{align}
		\sum_{i = 1}^k \left( f(b_i) - f(a_i)\right)  & \sum_{i = 1}^k (b_i-a_i)(z^2 + za_{\text{max}} + a^2_{\text{max}})- (z - a_{\text{max}}). \nonumber
	\end{align}
    Using the definition of $a_{\text{max}}$ in equation \eqref{supp:eqn:l2} again, it follows
    \begin{align}
		\sum_{i = 1}^k \left( f(b_i) - f(a_i)\right) &= (z -a_{\text{max}})(z^2 + za_{\text{max}} + a^2_{\text{max}})- (z - a_{\text{max}}) \nonumber\\
		&= z^3 - a^3_{\text{max}} - (z - a_{\text{max}}) \nonumber\\
		&= z^3 - z - a^3_{\text{max}} + a_{\text{max}}\nonumber \\
		& = f(z) - f(a_{\text{max}}) \nonumber
	\end{align}
	Thus, we conclude our result.
\end{proof}


\begin{proposition}  \label{prop:boundsvar}
	Suppose that $X = \{x_1, \ldots, x_n\}$ and $Y = \{y_1, \ldots, y_m\}$ are samples of observations, which need not necessarily be distinct, from a space $\Omega \in \mathbb{R}$. Suppose that $X'$ and $Y'$ are sub-multisets with sizes $n'$ and $m'$, respectively, which are known and that $X' \cup Y'$ contains $e'$ distinct values with multiplicities $d_1',\ldots,d_{e'}'$ and $d_1' \le \ldots \le d_{e'}'$. Define
	\begin{align*}
		\sigma_{\text{max}}^2(X,Y) & = {mn(n+m+1)}/{12} - {mn \{12(n+m)(n+m-1)\}^{-1} \sum_{i=1}^{e'} \{ (d'_i)^3 - d'_i \} }\\
		\sigma_{\text{min}}^2(X,Y) &=  \sigma_{\text{max}}^2(X,Y) - mn \{12(n+m)(n+m-1)\}^{-1} \{d^3_{\text{max}} -d_{\text{max}} - (d'_{e'})^3 + d'_{e'}\},
	\end{align*}
	where $d_{\text{max}} = d'_{e'} + n + m - n' - m'$, then $\sigma^2(X,Y)$, the variance of Wilcoxon-Mann-Whitney statistic $W(X,Y)$, is bounded by $		\sigma_{\text{min}}^2(X,Y) \le	\sigma^2(X,Y) \le \sigma_{\text{max}}^2(X,Y).$
\end{proposition}

\begin{proof}
	To start, let us assume that $X \cup Y$ take $e$ distinct values from $\{s_1, s_2, \cdots, s_{e}\}$ and the number of values in $X \cup Y$ equals to $s_i$ is $d_i$. Without loss of generality, assume that $X' \cup Y'$ take values from $\{s_1, s_2, \cdots, s_{e'}\}$ such that the numbers of values in $X' \cup Y'$ equals to $s_i$ is $d'_i$ and $d'_1 \le \ldots \le d'_{e'}$. Define $d'_{e'+1}, \ldots, d'_{e} = 0$.  
	
    We start by proving $\sigma^2_{\max}(X,Y) \ge \sigma^2(X,Y)$. Notice that
	\begin{align*}
		\sigma^2(X,Y) = {mn(n+m+1)}/{12} - {mn \{12(n+m)(n+m-1)\}^{-1} \sum_{i=1}^{e} \{ (d_i)^3 - d_i \} },
	\end{align*}
	which follows
	\begin{align*}
		\sigma_{\text{max}}^2(X,Y)  - \sigma^2(X,Y) &= {mn \{12(n+m)(n+m-1)\}^{-1} \sum_{i=1}^{e} \{ (d_i)^3 - d_i \} }\\
		 & - {mn \{12(n+m)(n+m-1)\}^{-1} \sum_{i=1}^{e'} \{ (d'_i)^3 - d'_i \} } \\
		 & = {mn \{12(n+m)(n+m-1)\}^{-1} \sum_{i=1}^{e} [\{ (d_i)^3 - d_i \} - \{ (d'_i)^3 - d'_i \} ]}.
	\end{align*}
	Define 
	\begin{align*}
		f(x) = x^3 - x, \text{ where } x \in \mathbb{R}.
	\end{align*}
	We then have
	\begin{align*}
		\sigma_{\text{max}}^2(X,Y)  - \sigma^2(X,Y) = mn \{12(n+m)(n+m-1)\}^{-1} \sum_{i=1}^{e} (f(d_i) - f(d'_i)).
	\end{align*}
	Since $d_i$ is defined as the number of values in $X \cup Y$ taking $s_i$ while $d'_i$ is defined as the number of values in $X' \cup Y'$ taking $s_i$, we have
	\begin{align*}
		d'_i \le d_i, i = 1,2,\cdots, e. 
	\end{align*}
	Notice that $f(0) = f(1) < f(2) < \ldots$, which implies $f(d_i) - f(d'_i) \ge 0$ for any $i = 1,\ldots,e$. It follows
	\begin{align*}
		\sigma_{\text{max}}^2(X,Y)  - \sigma^2(X,Y) \ge 0.
	\end{align*}
	Thus, we have proved $	\sigma_{\text{max}}^2(X,Y) \ge 	\sigma^2(X,Y)$.
	
	We now prove $\sigma^2_{\text{min}} (X,Y) \le \sigma^2(X,Y)$. To start, notice that
	\begin{align*}
	\sigma^2(X,Y) - \sigma_{\text{min}}^2	&= {mn \{12(n+m)(n+m-1)\}^{-1} \sum_{i=1}^{e'} \{ (d'_i)^3 - d'_i \} }\\
		& + mn \{12(n+m)(n+m-1)\}^{-1} \{d^3_{\text{max}} -d_{\text{max}} - (d'_{e'})^3 + d'_{e'}\}\\
		& - {mn \{12(n+m)(n+m-1)\}^{-1} \sum_{i=1}^{e} \{ (d_i)^3 - d_i \} }\\
		& = mn \{12(n+m)(n+m-1)\}^{-1}\left[\sum_{i=1}^{e'} f(d_{i}') + f(d_{\text{max}}) -f(d_{e'}') - \sum_{i=1}^{e} f(d_i) \right].
	\end{align*}
	Since we define $d'_{e'+1}, \ldots, d'_{e} = 0$ and $f(0) = 0$, we have $\sum_{i=1}^{e'} f(d_i') = \sum_{i=1}^{e} f(d_i')$. Thus, 
	\begin{align*}
		\sigma^2(X,Y) - \sigma_{\text{min}}^2 =  mn \{12(n+m)(n+m-1)\}^{-1} \left\{f(d_{\text{max}}) - f(d'_{e'}) + \sum_{i=1}^{e} (f(d'_i) - f(d_i)) \right\}
	\end{align*}  
    From the above Lemma~\ref{lem:S1}, we have
	\begin{align*}
		f(d_{\text{max}}) - f(d'_{e'}) \ge \sum_{i=1}^{e} (f(d_i) - f(d'_i)).
	\end{align*}
	To see this, consider $d'_1, d'_2, \cdots d'_e$ to be $a_1, a_2, \cdots, a_k$ in {Lemma 1}, $d_1, d_2, \cdots d_e$ to be $b_1, b_2, \cdots, b_k$. Then, we have $a_i \le b_i$ for any $i = 1,\ldots, k$. Recall that $d'_{e'}$ is defined as the maximum over all $d'_1, d'_2, \cdots d'_e$ and $d_{\text{max}} = d'_{e'} + n +m - n' - m' =  d'_{e'} + \sum_{i = 1}^e(d_i - d'_i)$.
	
	Thus, we have shown
	\begin{align*}
		\sigma^2(X,Y) - \sigma_{\text{min}}^2  \ge 0,
	\end{align*}
	which completes our proof.
	
\end{proof}



\section{Proof of results in Section~5.3}

The proof of Corollary~3 in Section~5.3 will rely on the following proposition.

\begin{proposition}
    \label{prop:pvalties}
    Suppose that $\bx = \{\xx_1, \ldots, \xx_{\n}\}$ and
    $\by = \{\yy_1, \ldots, \yy_{\m}\}$ are samples of real-valued 
    observations, which need not necessarily be distinct.
    Suppose that $\bxp \subset \bx$ and $\byp \subset \by$ are sub-multisets
    with sizes $\np$ and $\mpp$, respectively, which are known.
    Define $\bz = \bx \cup \by$ and $\bzp=\bxp \cup \byp$
    and suppose that $\bz \setminus \bzp$ is unknown.
    Defining
    \begin{align}
        \WminXY &= \min_{\bz \setminus \bzp} \wmwstat{\bx}{\by}, \qquad
        \WmaxXY = \max_{\bz \setminus \bzp} \wmwstat{\bx}{\by},
        \nonumber
    \end{align}
    and using $\mu=\n\m/2$ and the definitions of $\sigma^2_{\min}(\bx, \by)$ 
    and $\sigma^2_{\max}(\bx, \by)$ in 
    Proposition~7, 
    define
    \begin{align}
        p_1 &= 2\cdfphi( -\absval{\WminXY - \mu}/\sigma_{\max}(\bx, \by) ),
        \quad
        p_2 = 2\cdfphi( -\absval{\WmaxXY - \mu}/\sigma_{\max}(\bx, \by) ),
        \nonumber \\
        p_3 &= 2\cdfphi( -\absval{\WmaxXY - \mu}/\sigma_{\min}(\bx, \by) ),
        \quad
        p_4 = 2\cdfphi( -\absval{\WminXY - \mu}/\sigma_{\min}(\bx, \by) ).
        \nonumber
    \end{align}
    Then, denoting the $p$-value of $W(X,Y)$ by $p$, we have the bounds:
    \begin{alignat}{2}
        &p_3 \leq p \leq p_1, \quad &&\textrm{if } \WminXY - \mu \geq 0 
        \textrm{ and } \WmaxXY - \mu \geq 0,
        \nonumber \\
        &p_4 \leq p \leq p_2, \quad &&\textrm{if } \WminXY - \mu < 0 
        \textrm{ and } \WmaxXY - \mu < 0,
        \nonumber \\
        &\min\{p_3, p_4\} \leq p \leq 1, \quad &&\textrm{otherwise}.
        \nonumber 
    \end{alignat}
\end{proposition}

\begin{proof}
    While $\wmwstat{\bx}{\by}$ and $\sigma^2(\bx, \by)$ are not known, 
    the bounds $\WminXY$, 
    $\WmaxXY$, $\sigma^2_{\min}(\bx, \by)$ and $\sigma^2_{\max}(\bx, \by)$
    are available such that
    \begin{align}
        \WminXY &\leq \wmwstat{\bx}{\by} \leq \WmaxXY, 
        \nonumber \\
        \sigma_{\min}(\bx, \by) &\leq \sigma(\bx, \by)  
        \leq \sigma_{\max}(\bx, \by).
        \nonumber 
    \end{align}
    The are now four cases to consider, depending on the signs of
    $\WminXY - \mu$ and $\WmaxXY - \mu$.
    To ease of notation, and since the proof only concerns the
    sets $\bx$ and $\by$, for the remainder of the proof we write
    $\sigma_{\min}$, $\sigma$ and $\sigma_{\max}$ for 
    $\sigma_{\min}(\bx, \by)$, $\sigma(\bx, \by)$ 
    and $\sigma_{\max}(\bx, \by)$, respectively.
    \begin{itemize}
        \item \textbf{Case I:} $\WminXY - \mu < 0$ and $\WmaxXY - \mu < 0$
    \end{itemize}
    Since $\wmwstat{\bx}{\by} \leq \WmaxXY$, 
    \begin{align}
        &\wmwstat{\bx}{\by} - \mu \leq \WmaxXY - \mu 
        \nonumber \\
        \Rightarrow
        &\sigma (\wmwstat{\bx}{\by} - \mu) 
        \leq \sigma (\WmaxXY - \mu ).
        \label{ineq:caseone_one}
    \end{align}
    Furthermore, this case implies
    $\wmwstat{\bx}{\by} - \mu \leq \WmaxXY - \mu < 0$, i.e. that 
    $\wmwstat{\bx}{\by} - \mu$ is also negative. Now, since 
    $\sigma_{\max}\geq \sigma$ and $\wmwstat{\bx}{\by} - \mu$ is negative, 
    \begin{align}
        &\sigma_{\max}(\wmwstat{\bx}{\by} - \mu) 
        \leq \sigma(\wmwstat{\bx}{\by}- \mu )
        \nonumber \\
        \Rightarrow
        &\sigma_{\max}(\wmwstat{\bx}{\by} - \mu) 
        \leq \sigma(\WmaxXY- \mu ),
        \nonumber
    \end{align}
    where the last line uses Inequality~\eqref{ineq:caseone_one}.
    Dividing both sides by $\sigma_{\max} \sigma$,
    this then implies 
    \begin{align}
        & (\wmwstat{\bx}{\by} - \mu)/\sigma
        \leq  (\WmaxXY- \mu )/ \sigma_{\max},
        \nonumber
    \end{align}
    providing an upper bound for
    $(\wmwstat{\bx}{\by} - \mu)/\sigma(\bx, \by)$.

    A lower bound is obtained similarly;
    since
    $\sigma \geq \sigma_{\min}$
    and $\WminXY - \mu$ is negative,
    \begin{align}
        \sigma(\wmwstat{\bx}{\by} - \mu) 
        \leq  \sigma_{\min}(\wmwstat{\bx}{\by} - \mu).
        \label{ineq:caseone_two}
    \end{align}
    Furthermore, since $\WminXY - \mu \leq \wmwstat{\bx}{\by} - \mu$, 
    \begin{align}
        &\sigma(\WminXY - \mu) 
        \leq  \sigma(\wmwstat{\bx}{\by} - \mu)
        \nonumber \\
        \Rightarrow
        &\sigma(\WminXY - \mu) 
        \leq  \sigma_{\min}(\wmwstat{\bx}{\by} - \mu),
        \nonumber 
    \end{align}
    where the last line us Inequality~\eqref{ineq:caseone_two}, and therefore,
    after dividing both sides by $\sigma$,
    \begin{align}
        &(\WminXY - \mu) / \sigma_{\min} 
        \leq (\wmwstat{\bx}{\by} - \mu) / \sigma,
        \nonumber 
    \end{align}
    which combined with the upperbound above, provides
    \begin{align}
        & (\WminXY - \mu) / \sigma_{\min} 
        \leq (\wmwstat{\bx}{\by} - \mu) / \sigma
        \leq (\WmaxXY - \mu) / \sigma_{\max} .
        \nonumber
    \end{align}
    For ease of notation, let us denote
    $\qmin = (\WminXY - \mu) / \sigma_{\min}$, 
    $q = (\wmwstat{\bx}{\by} - \mu) / \sigma$, 
    and
    $\qmax = (\WmaxXY - \mu) / \sigma_{\max}$.
    This implies $\qmin \leq q \leq \qmax < 0$ 
    (the last inequality is from the Case I condition).
    Therefore, since the cumulative distribution function
    $\cdfphi$ is monotonic, 
    \begin{align}
        &\qmin \leq q \leq \qmax 
        \nonumber \\
        \Rightarrow 
        & -\absval{\qmin} \leq -\absval{q} \leq -\absval{\qmax}
        \nonumber \\
        \Rightarrow 
        &2\cdfphi( -\absval{\qmin}) \leq 2\cdfphi(-\absval{q}) 
        \leq 2\cdfphi( -\absval{\qmax})
        \nonumber \\
        \Rightarrow 
        &p_4 \leq 2\cdfphi(-\absval{q}) 
        \leq p_2
        \nonumber
    \end{align}
    Next we see that, 
    since $-\absval{q} \leq 0 \Rightarrow \cdfphi(-\absval{q}) \leq 0.5$,
    \begin{align}
        &2\cdfphi(-\absval{q}) \leq 1
        \nonumber \\
        \Rightarrow
        &1-2\cdfphi(-\absval{q}) \geq 0,
        \nonumber 
    \end{align}
    and therefore, by definition of the $p$-value of $\wmwstat{\bx}{\by}$,
    \begin{align}
        p &= 
        1 - \absval{1-2\cdfphi(-\absval{q})} 
        = 1 - \{ 1-2\cdfphi(-\absval{q}) \}
        = 1 - 1 + 2\cdfphi(-\absval{q}) \}
        =2\cdfphi(-\absval{q}) \},
        \nonumber 
    \end{align}
    and so $p_4 \leq p \leq p_2$, which proves the result for Case I.

    \begin{itemize}
        \item \textbf{Case II:} $\WminXY - \mu \geq 0$ and $\WmaxXY - \mu \geq 0$
    \end{itemize}
    Since by definition $\WminXY \leq  \wmwstat{\bx}{\by} \leq \WmaxXY$, this case
    then implies
    $0 \leq \WminXY - \mu \leq  \wmwstat{\bx}{\by} - \mu \leq \WmaxXY - \mu$.
    By definition, 
    $\sigma_{\min} \leq \sigma \leq \sigma_{\max}.$
    Therefore, 
    \begin{align}
        {\sigma}(\WminXY - \mu)
        &\leq {\sigma_{\max}}(\WminXY - \mu)
        \nonumber \\
        \Rightarrow
        (\WminXY - \mu)/{\sigma_{\max}}
        &\leq (\WminXY - \mu)/{\sigma}
        \nonumber \\
        &\leq (\wmwstat{\bx}{\by} - \mu)/{\sigma}
        \nonumber \\
        &\leq (\wmwstat{\bx}{\by} - \mu)/{\sigma_{\min}}
        \nonumber \\
        &\leq (\WmaxXY - \mu)/{\sigma_{\min}}, 
        \nonumber
    \end{align}
    or, more specifically, after taking absolute values of both sides and 
    multiplying through by $-1$,
    \begin{align}
        -\absval{\WmaxXY - \mu}/{\sigma_{\min}}
        \leq -\absval{\wmwstat{\bx}{\by} - \mu}/{\sigma(\bx, \by)}
        \leq -\absval{\WminXY - \mu}/{\sigma_{\max}} .
        \nonumber
    \end{align}
    Since the cumulative distribution function $\cdfphi$ is nondecreasing, 
    $a \leq b$ implies $2\cdfphi(a) \leq 2\cdfphi(b)$, and so applying
    the function $2\cdfphi$ to the values in the inequality above, 
    we have $p_3 \leq p \leq p_1$.
    \begin{itemize}
        \item \textbf{Case III:} $\WminXY - \mu \geq 0$ and $\WmaxXY - \mu < 0$
    \end{itemize}
    If this were true, then 
    \begin{align}
        \WmaxXY - \mu < 0 \leq \WminXY - \mu
        \quad
        \Rightarrow 
        \quad
        \WmaxXY  < \WminXY ,
        \nonumber 
    \end{align}
    which is a contradiction, since by definition $\WmaxXY  \geq \WminXY$. 
    Therefore, this case is not possible.

    \begin{itemize}
        \item \textbf{Case IV:} $\WminXY - \mu < 0$ and $\WmaxXY - \mu \geq 0$
    \end{itemize}
    In this case, $\wmwstat{\bx}{\by} - \mu$ can be either positive or negative, so 
    we consider these two subcases.

    \begin{itemize}
        \item \textbf{Case IVa:} $\WminXY - \mu < 0$ and $\WmaxXY - \mu \geq 0$ 
            and $\wmwstat{\bx}{\by} - \mu < 0$
    \end{itemize}
    Since $\wmwstat{\bx}{\by} - \mu < 0$ and $\sigma \geq \sigma_{\min}$, then
    \begin{align}
        \sigma(\wmwstat{\bx}{\by} - \mu) 
        \leq \sigma_{\min} (\wmwstat{\bx}{\by} - \mu),
        \nonumber
    \end{align}
    or, equivalently, 
    \begin{align}
        \sigma_{\min} (\wmwstat{\bx}{\by} - \mu)
        \geq 
        \sigma(\wmwstat{\bx}{\by} - \mu). 
        \nonumber
    \end{align}
    Then, since $\wmwstat{\bx}{\by} - \mu \geq \WminXY - \mu$, this leads to 
    \begin{align}
        \sigma_{\min} (\wmwstat{\bx}{\by} - \mu)
        \geq \sigma(\wmwstat{\bx}{\by} - \mu) 
        \geq \sigma(\WminXY - \mu) .
        \nonumber
    \end{align}
    From the above, ignoring the middle term we have
    \begin{align}
        \sigma(\WminXY - \mu) 
        \leq
        \sigma_{\min} (\wmwstat{\bx}{\by} - \mu),
        \nonumber
    \end{align}
    which implies
    \begin{align}
        (\WminXY - \mu) / \sigma_{\min} 
        \leq  (\wmwstat{\bx}{\by} - \mu) / \sigma,
        \nonumber
    \end{align}
    and since these terms are negative this implies
    \begin{align}
        -\absval{\WminXY - \mu} / \sigma_{\min} 
        \leq  -\absval{\wmwstat{\bx}{\by} - \mu} / \sigma,
        \nonumber
    \end{align}
    which implies that $p_4 \leq p$. 
    Next we consider the other subcase, $\wmwstat{\bx}{\by} - \mu \geq 0$.
    \begin{itemize}
        \item \textbf{Case IVb:} $\WminXY - \mu < 0$ and $\WmaxXY - \mu \geq 0$ 
            and $\wmwstat{\bx}{\by} - \mu \geq 0$
    \end{itemize}
    Since $0 \leq \wmwstat{\bx}{\by} - \mu \leq \WmaxXY - \mu$, 
    and $\sigma_{\min} \leq \sigma$, then 
    \begin{align}
        \sigma_{\min} (\wmwstat{\bx}{\by} - \mu) 
        \leq \sigma(\WmaxXY - \mu), 
        \nonumber
    \end{align}
    which implies
    \begin{align}
        (\wmwstat{\bx}{\by} - \mu) / \sigma
        \leq (\WmaxXY - \mu) /\sigma_{\min}, 
        \nonumber
    \end{align}
    which, since these terms are positive, implies
    \begin{align}
        -\absval{\WmaxXY - \mu} /\sigma_{\min}(\bx, \by)
        \leq -\absval{\wmwstat{\bx}{\by} - \mu} / \sigma(\bx, \by),
        \nonumber
    \end{align}
    which implies $p_3 \leq p$. Cases IVa and IVb therefore together imply
    $\min \{p_3, p_4\} \leq p$ (either $p_3 \leq p$ or $p_4 \leq p$, depending 
    on the sign of $\wmwstat{\bx}{\by} - \mu$, and so their minimum is guaranteed
    to be a lower bound for $p$). 
    Since $p \leq 1$ by definition, Case IV 
    therefore implies $\min \{p_3, p_4\} \leq p \leq 1$. This proves the result.
\end{proof}


\subsection{Proof of Corollary~3}

\begin{corollary}
    \label{cor:tiesresult}
    Suppose that $\bx = \{\xx_1, \ldots, \xx_{\n}\}$ and
    $\by = \{\yy_1, \ldots, \yy_{\m}\}$ are samples of 
    observations, which need not necessarily be distinct,
    from a space $\discretespace \subset \realR$.
    Suppose that $\bxp \subset \bx$ and $\byp \subset \by$ are sub-multisets
    with sizes $\np$ and $\mpp$, respectively, which are known, 
    and compute $\partial \bxp_{a}$, $\partial \bxp_{b}$,
    $\partial \byp_{a}$ and $\partial \byp_{b}$ from 
    Proposition~\ref{prop:mainthmties} and 
    $\sigma_{\max}^2(\bx, \by)$ from Proposition~7,
    and define
    \begin{align}
        &\WminXY =
        \wmwstat{\bxp}{\byp} + \absval{ \partial \byp_{a} } (\n - \np)/2 
        + \absval{ \partial \bxp_{b} } (\m - \mpp)/2,
        \quad \mu = \n\m/2,
        \nonumber \\
        &\WmaxXY =
        \wmwstat{\bxp}{\byp} + (\m \n - \mpp \np) - 
        \absval{ \partial \bxp_{a} } (\m - \mpp) 
        + \absval{ \partial \byp_{b} } (\n - \np),
        \nonumber \\
        &p_1 = 2\cdfphi( -\absval{\WminXY - \mu}/\sigma_{\max}(\bx, \by) ),
        \quad
        p_2 = 2\cdfphi( -\absval{\WmaxXY - \mu}/\sigma_{\max}(\bx, \by) ).
        \nonumber
    \end{align}
    where $\cdfphi$ is the cumulative distribution function of the standard 
    normal distribution.
    Then given a significance threshold $\alpha \in (0, 1)$ and 
    defining $p = \max\{p_1, p_2\}$, if $p < \alpha$
    and the two terms $(\WminXY - \mu)$ and $(\WminXY - \mu)$ 
    have the same sign, then the data 
    will yield a significant result, regardless of the values in 
    $(\bx \cup \by) \setminus (\bxp \cup \byp)$.
\end{corollary}

\begin{proof}
    Proposition~5 in the main paper gives
    \begin{align}
        \WminXY &=
        \wmwstat{\bxp}{\byp} + \absval{ \partial \byp_{a} } (\n - \np)/2 
        + \absval{ \partial \bxp_{b} } (\m - \mpp)/2,
        \nonumber
    \end{align}
    and
    \begin{align}
        \WmaxXY &=
        \wmwstat{\bxp}{\byp} + (\m \n - \mpp \np) - 
        \absval{ \partial \bxp_{a} } (\m - \mpp) 
        + \absval{ \partial \byp_{b} } (\n - \np).
        \nonumber 
    \end{align}
    Proposition~\ref{prop:pvalties} shows that any $p$-value
    for $\wmwstat{\bx}{\by}$ must be bounded by 
    both $p_1$ and $p_2$ if 
    the two terms $(\WminXY - \mu)$ and $(\WminXY - \mu)$ 
    have the same sign; in the case that those two terms do not have
    the same sign, then the upper bound is $1$.

    Therefore, if $(\WminXY - \mu)$ and $(\WminXY - \mu)$ have the same
    sign, if $p=\max \{p_1, p_2 \}$ and $p < \alpha$, we can declare a
    significant result, regardless of the values of the missing data, 
    which proves the corollary.
\end{proof}


\section{Additional experiments}

\subsection{Additional experiment for data missing completely at random; similar to Figure~1}

\begin{figure}
    \includegraphics[width=\textwidth]{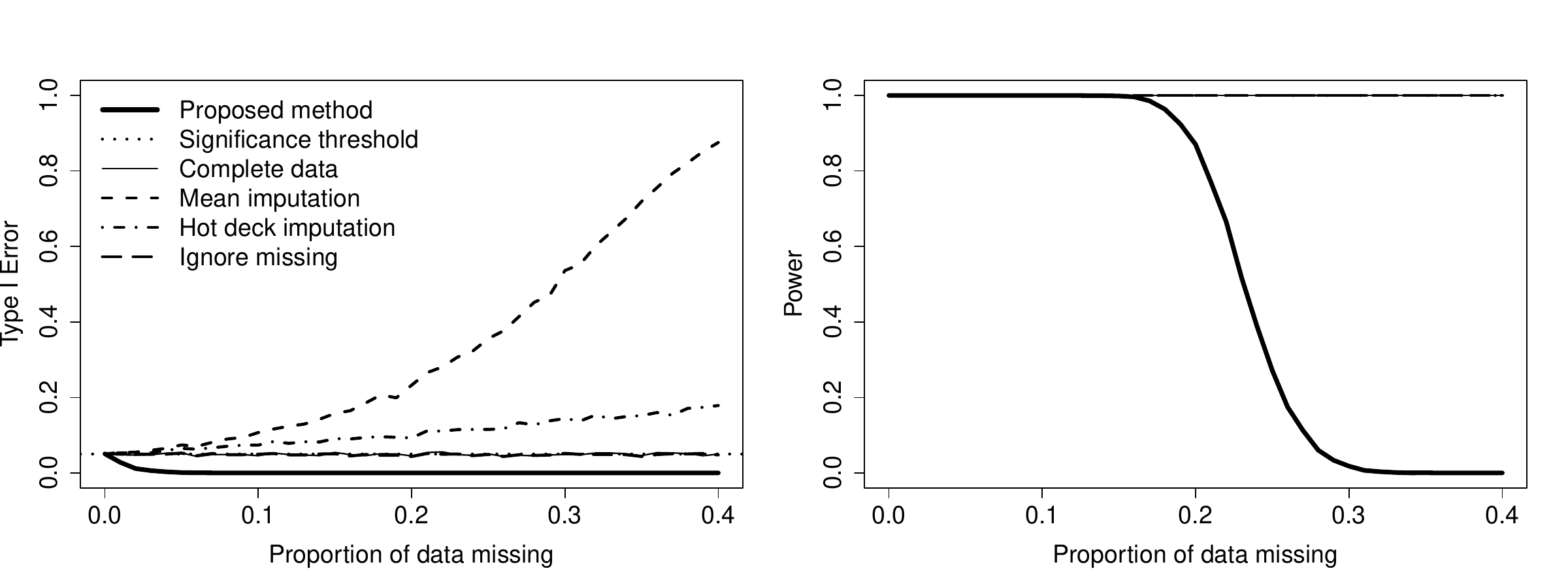}
    \caption{The Type I error and statistical power of the proposed method and 
    the standard Wilcoxon-Mann-Whitney test after the missing data has either known or
    has been imputed or ignored as the proportion of missing data increases.
    The data is missing completely at random.
    (Left) Type I error : $\mathrm{N}(0,1)$ vs $\mathrm{N}(0,1)$; 
    (Right) Power: $\mathrm{N}(0,1)$ vs $\mathrm{N}(3,1)$. For both figures, 
    a significance threshold of $\alpha=0.05$ has been used and the total
    sample sizes are $\n=100$, $\m=100$, and $5000$ trials were used..}
\label{fig:mcar2}
\end{figure}

Figure~\ref{fig:mcar2} is a similar case to Figure~1 in the main paper, 
but shows the increased power when the first sample consists of observations
from $\mathrm{N}(0,1)$, while the second sample consists of observations 
from $\mathrm{N}(3,1)$. For both samples, the data is missing completely at random.
The proposed method still had good power when over $20\%$ of the data is missing.

\subsection{Additional experiments for data missing not at random; similar to Figure~2}

\begin{figure}
    \includegraphics[width=\textwidth]{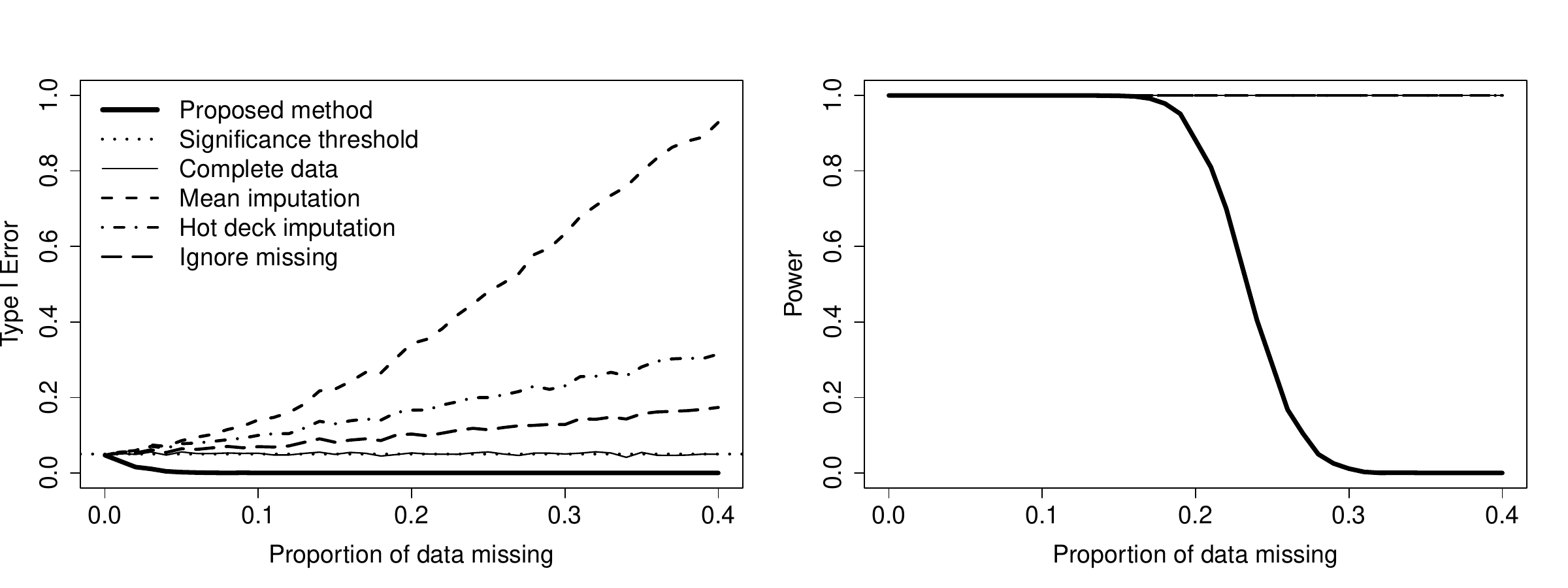}
    \caption{The Type I error and statistical power of the proposed method and 
    the standard Wilcoxon-Mann-Whitney test after the missing data has either known or
    has been imputed or ignored as the proportion of missing data increases.
    The data is missing not at random, according to the mechanism in 
    Equation~(4), where only observations greater than $0$
    are possibly missing from both samples.
    (Left) Type I error : $\mathrm{N}(0,1)$ vs $\mathrm{N}(0,1)$; 
    (Right) Power: $\mathrm{N}(0,1)$ vs $\bm{\mathrm{N}(3,1)}$. For both figures, 
    a significance threshold of $\alpha=0.05$ has been used and the total
    sample sizes are $\n=100$, $\m=100$, and $5000$ trials were used.}
\label{fig:mnar2}
\end{figure}

Figure~\ref{fig:mnar2} is a similar case to Figure~2 in the main paper, 
but shows the increased power when the first sample consists of observations
from $\mathrm{N}(0,1)$, while the second sample consists of observations 
from $\mathrm{N}(3,1)$. For both samples, the data is missing \textbf{not at random}, 
with only values greater than $0$ being possibly missing.
Again, the proposed method still had good power
when over $20\%$ of the data is missing.

\begin{figure}
    \includegraphics[width=\textwidth]{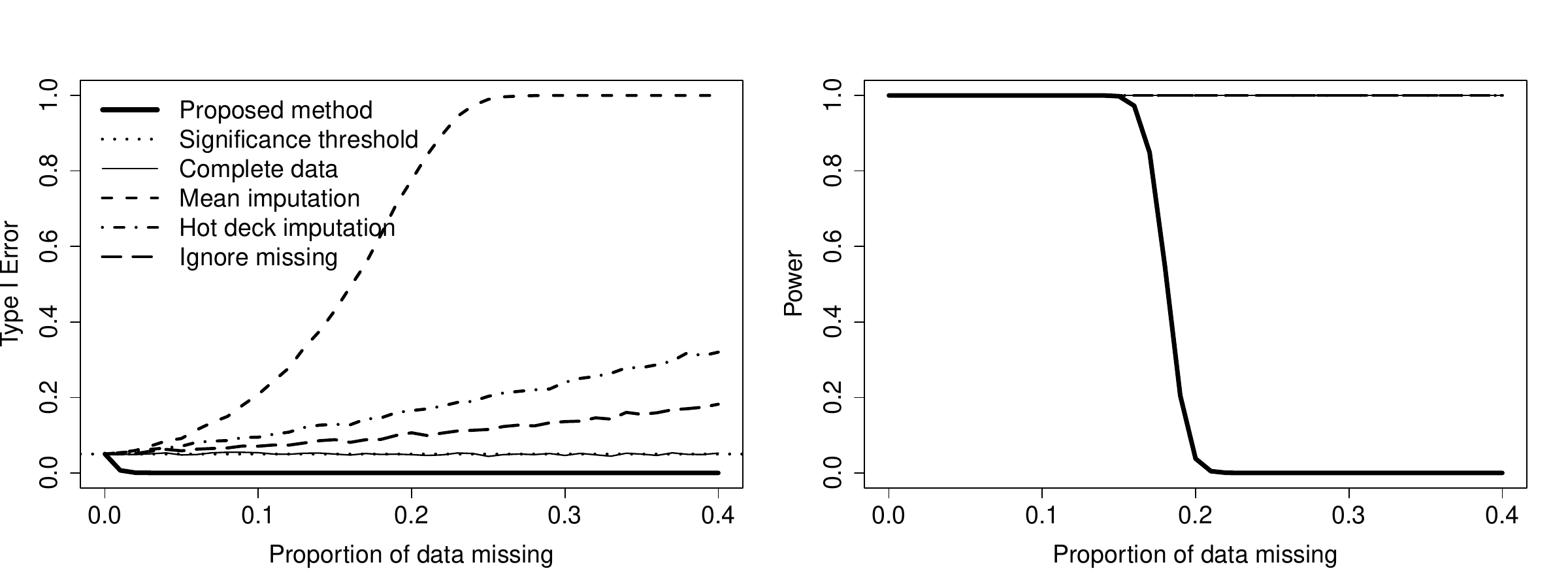}
    \caption{The Type I error and statistical power of the proposed method and 
    the standard Wilcoxon-Mann-Whitney test after the missing data has either known or
    has been imputed or ignored as the proportion of missing data increases.
    The data is missing not at random, according to the mechanism in 
    Equation~(4), where only observations greater than $0$
    are possibly missing from both samples.
    (Left) Type I error : $\mathrm{N}(0,1)$ vs $\mathrm{N}(0,1)$; 
    (Right) Power: $\mathrm{N}(0,1)$ vs $\mathrm{N}(1,1)$. For both figures, 
    a significance threshold of $\alpha=0.05$ has been used and the total
    \textbf{sample sizes} are $\n=1000$, $\m=1000$, and $5000$ trials were used.}
\label{fig:mnar3}
\end{figure}

Figure~\ref{fig:mnar3} shows results under the same conditions as
and Figure~2, but with the sample sizes increased to $n=m=1000$.
The power figure shows data from $\bx$ sampled from $\mathrm{N}(0, 1)$
and data from $\by$ sampled from $\mathrm{N}(1, 1)$.
Again, the proposed method still had good power
when over $20\%$ of the data is missing.

\begin{figure}
    \includegraphics[width=\textwidth]{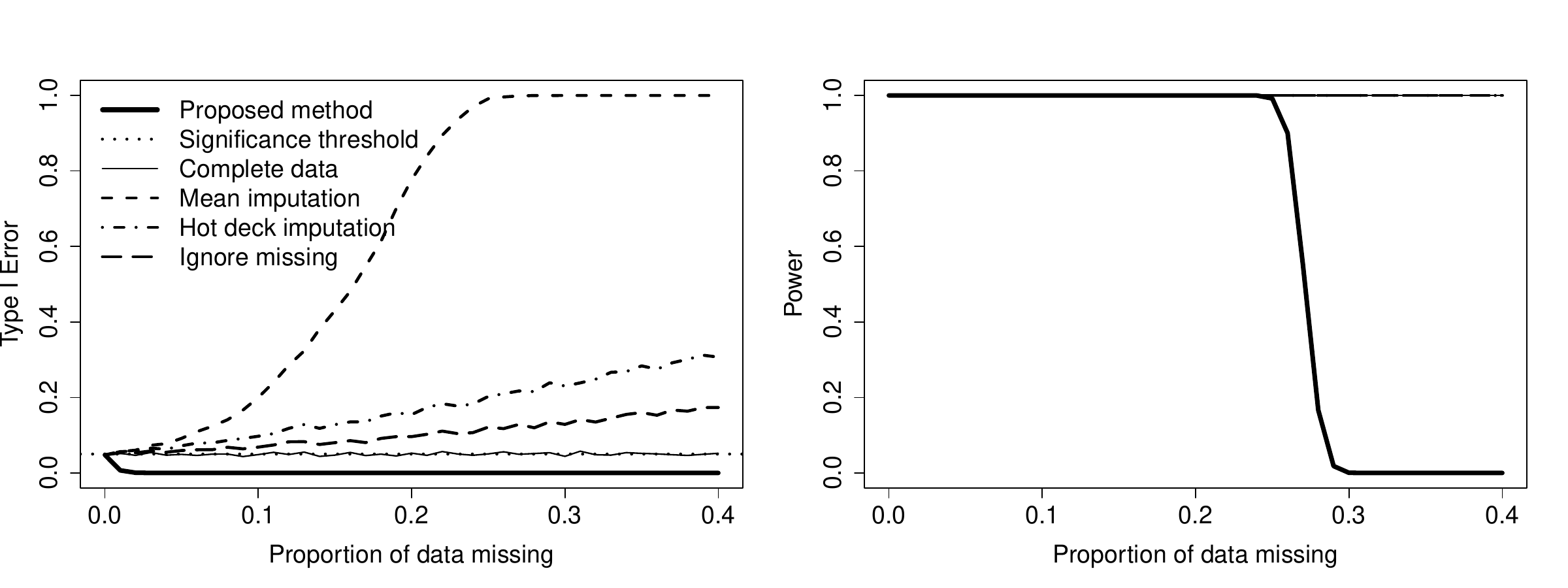}
    \caption{The Type I error and statistical power of the proposed method and 
    the standard Wilcoxon-Mann-Whitney test after the missing data has either known or
    has been imputed or ignored as the proportion of missing data increases.
    The data is missing not at random, according to the mechanism in 
    Equation~(4), where only observations greater than $0$
    are possibly missing.
    (Left) Type I error : $\mathrm{N}(0,1)$ vs $\mathrm{N}(0,1)$; 
    (Right) Power: $\mathrm{N}(0,1)$ vs $\bm{\mathrm{N}(3,1)}$. For both figures, 
    a significance threshold of $\alpha=0.05$ has been used and the total
    \textbf{sample sizes} are $\n=1000$, $\m=1000$, and $5000$ trials were used.}
\label{fig:mnar4}
\end{figure}

Figure~\ref{fig:mnar4} shows results under the same conditions as
and Figure~\ref{fig:mnar3}, with the sample sizes  $n=m=1000$, 
but also 
with the power figure showing data from $\bx$ sampled from $\mathrm{N}(0, 1)$
and data from $\by$ sampled from $\mathrm{N}(3, 1)$.
In this case, the proposed method has good power almost until
$30\%$ of the data is missing.

\subsection{Additional experiments for data with ties; similar to Figure~3}

\begin{figure}
    \includegraphics[width=\textwidth]{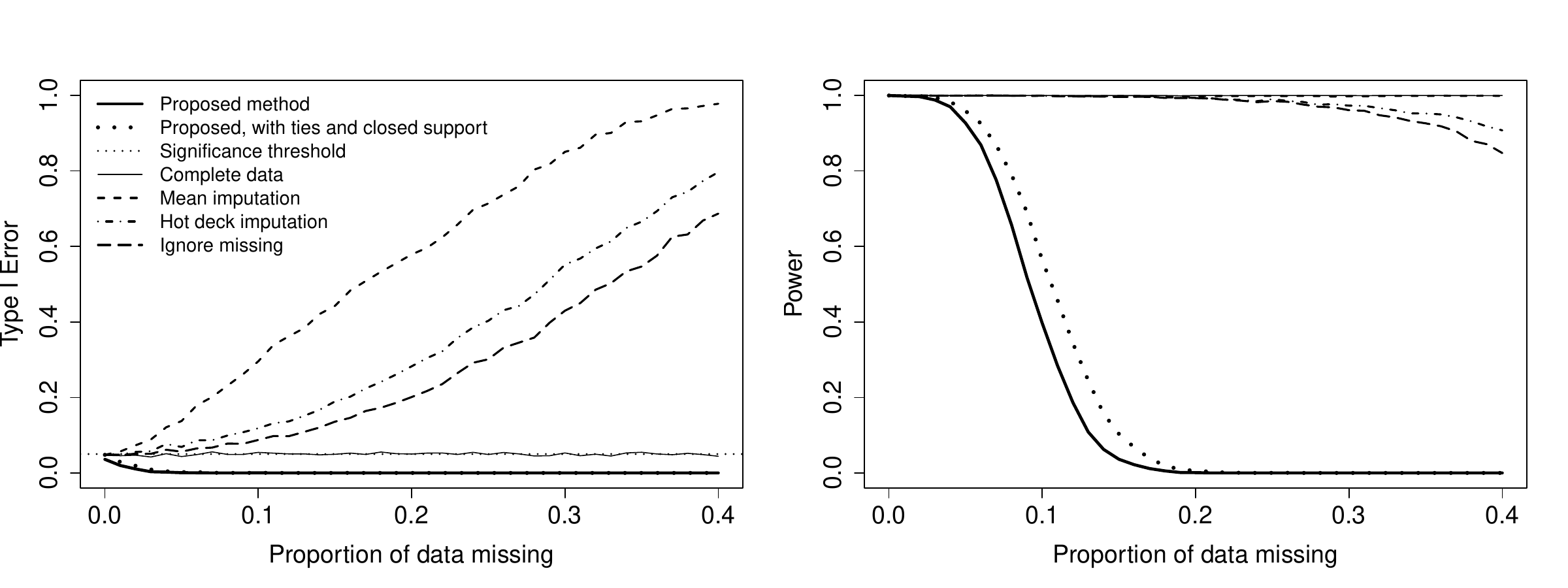}
    \caption{The Type I error and statistical power of the proposed method and 
    the standard Wilcoxon-Mann-Whitney test after the missing data is either known or
    has been imputed or ignored as the proportion of missing data increases.
    The data for the first sample $\bx$ is missing completely at random but the
    data for the second sample $\by$ is missing not at random, with
    only observations greater than $0$
    are possibly missing.
    (Left) Type I error : $\mathrm{Pois}(1)$ vs $\mathrm{Pois}(1)$; 
    (Right) Power: $\mathrm{Pois}(1)$ vs $\bm{\mathrm{Pois}(2)}$. For both figures, 
    a significance threshold of $\alpha=0.05$ has been used and the total
    sample sizes are $\n=100$, $\m=100$, and $5000$ trials were used.}
\label{fig:mnar2ties}
\end{figure}

Figure~\ref{fig:mnar2ties} is similar to Figure~3 in the main paper, 
but shows decreased power when the second sample consists
of observations sampled from a $\mathrm{Pois}(2)$ distribution;
recall that in Figure~3 in the main paper the second sample
of observations were sampled from a $\mathrm{Pois}(3)$ distribution.

\begin{figure}
    \includegraphics[width=\textwidth]{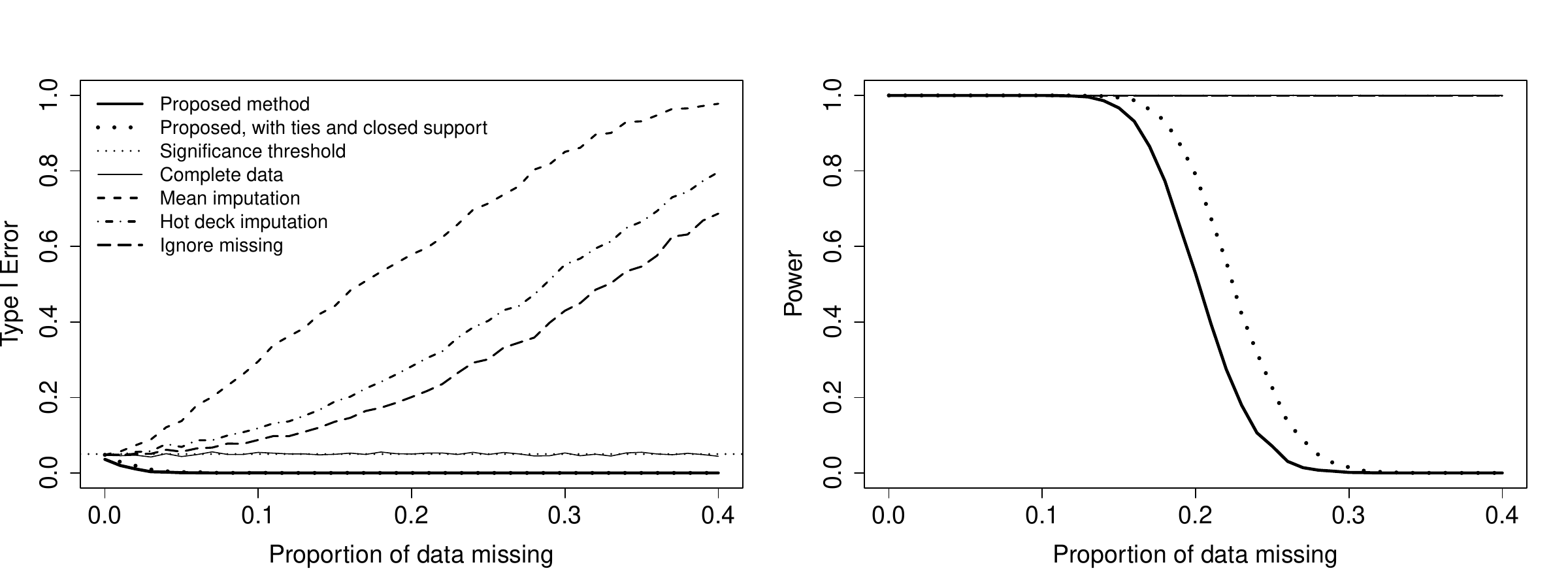}
    \caption{The Type I error and statistical power of the proposed method and 
    the standard Wilcoxon-Mann-Whitney test after the missing data is either known or
    has been imputed or ignored as the proportion of missing data increases.
    The data for the first sample $\bx$ is missing completely at random but the
    data for the second sample $\by$ is missing not at random, with only 
    observations greater than $0$
    are possibly missing.
    (Left) Type I error : $\mathrm{Pois}(1)$ vs $\mathrm{Pois}(1)$; 
    (Right) Power: $\mathrm{Pois}(1)$ vs $\bm{\mathrm{Pois}(4)}$. For both figures, 
    a significance threshold of $\alpha=0.05$ has been used and the total
    sample sizes are $\n=100$, $\m=100$, and $5000$ trials were used.}
\label{fig:mnar3ties}
\end{figure}

In Figure~\ref{fig:mnar3ties}, on the other hand, the second sample
of observations are sampled from a  $\mathrm{Pois}(4)$ distribution.
In this case, the proposed method (which takes ties and closed support
into account; thickly-dotted line) has good power when up to $20\%$
of the data is missing.

\begin{figure}
    \includegraphics[width=\textwidth]{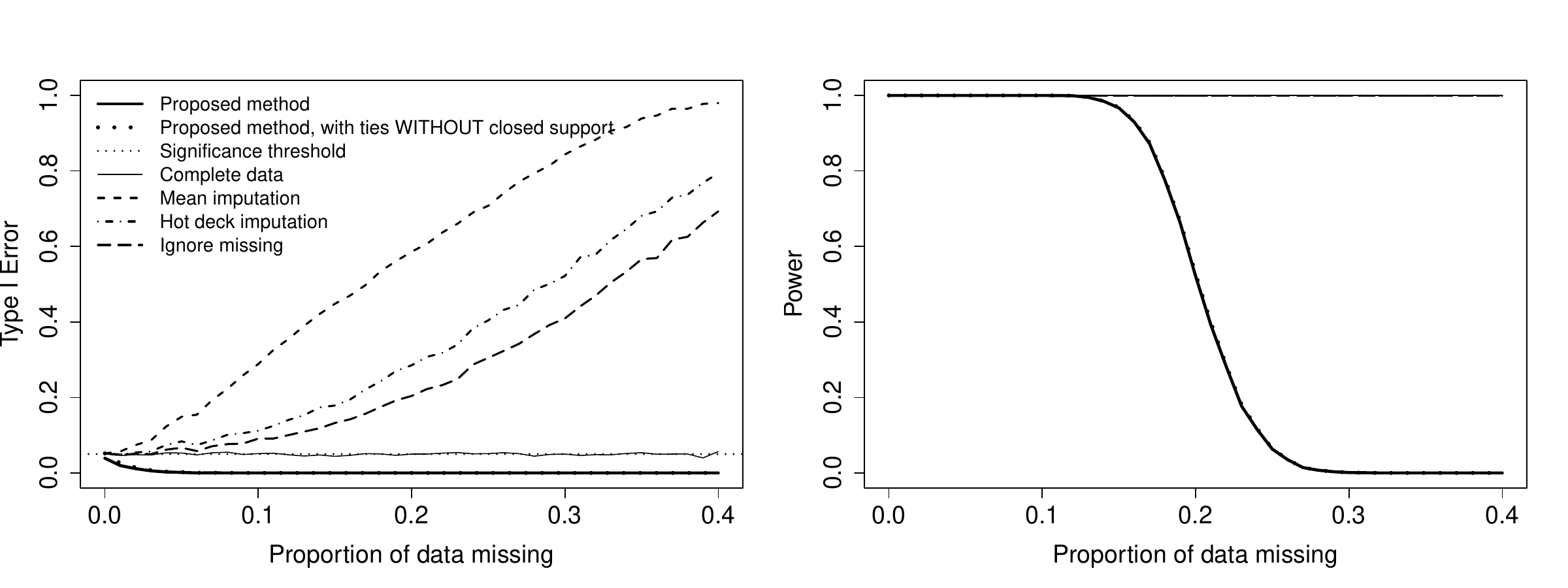}
    \caption{The Type I error and statistical power of the proposed method and 
    the standard Wilcoxon-Mann-Whitney test after the missing data is either known or
    has been imputed or ignored as the proportion of missing data increases.
    The data for the first sample $\bx$ is missing completely at random but the
    data for the second sample $\by$ is missing not at random, with only 
    observations greater than $0$
    are possibly missing. \textbf{We also plot the modified method, but without taking
    closed support into account.}
    (Left) Type I error : $\mathrm{Pois}(1)$ vs $\mathrm{Pois}(1)$; 
    (Right) Power: $\mathrm{Pois}(1)$ vs $\bm{\mathrm{Pois}(4)}$. For both figures, 
    a significance threshold of $\alpha=0.05$ has been used and the total
    sample sizes are $\n=100$, $\m=100$, and $5000$ trials were used.}
\label{fig:mnar4ties}
\end{figure}

In Figure~\ref{fig:mnar4ties}, the data for both the Type I error
and power plots is the same as for Figure~\ref{fig:mnar3ties}, 
but we show the proposed method taking ties into account, but not
taking closed support into account (thickly-dotted line).
In this case, the performance is almost exactly the same as for the 
proposed method (which does not take ties or closed support into account).
This shows the importance of taking the closed support into account, 
as illustrated in Figure~3 and Figures~\ref{fig:mnar2ties} 
and \ref{fig:mnar3ties},
since Poisson-distributed observations have a lower bound of $0$.

\begin{figure}
    \includegraphics[width=\textwidth]{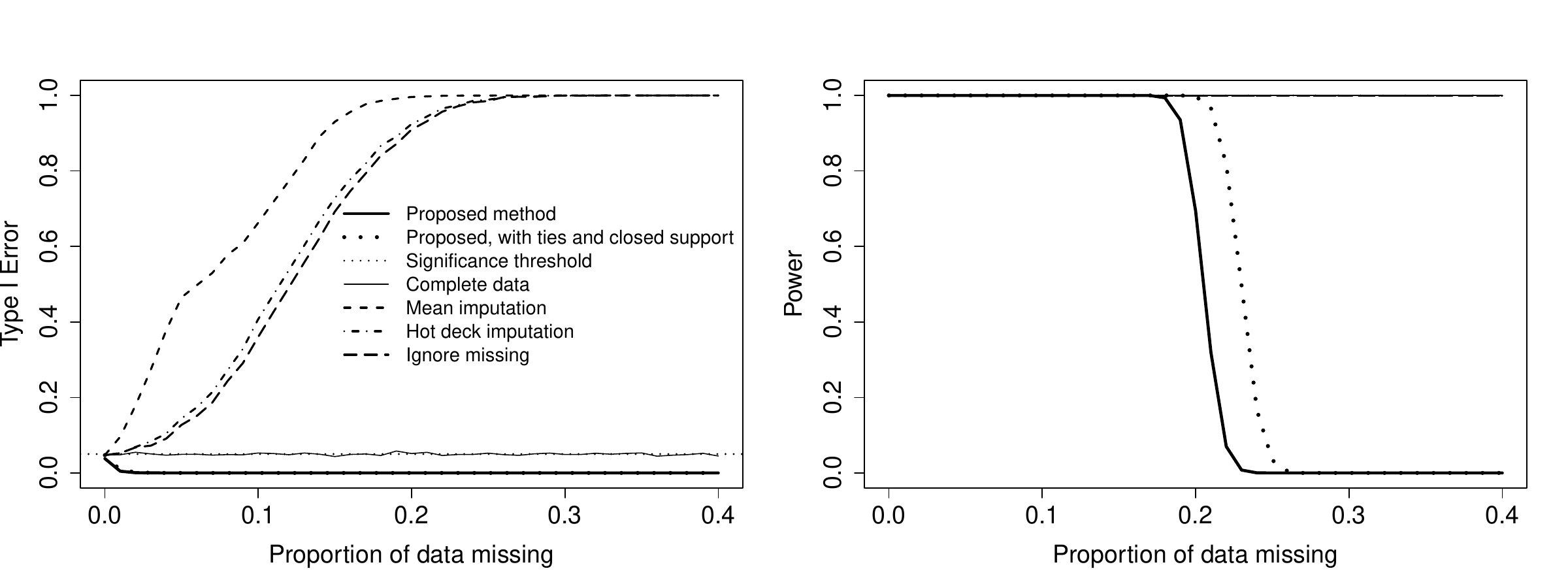}
    \caption{The Type I error and statistical power of the proposed method and 
    the standard Wilcoxon-Mann-Whitney test after the missing data is either known or
    has been imputed or ignored as the proportion of missing data increases.
    The data for the first sample $\bx$ is missing completely at random but the
    data for the second sample $\by$ is missing not at random, with only 
    observations greater than $0$
    are possibly missing.
    (Left) Type I error : $\mathrm{Pois}(1)$ vs $\mathrm{Pois}(1)$; 
    (Right) Power: $\mathrm{Pois}(1)$ vs $\mathrm{Pois}(3)$. For both figures, 
    a significance threshold of $\alpha=0.05$ has been used and the total
    \textbf{sample sizes} are $\n=1000$, $\m=1000$, and $5000$ trials were used.}
\label{fig:mnar5ties}
\end{figure}

Figure~\ref{fig:mnar5ties}, the data is the same as for Figure~3, 
but now with the sample sizes increased to $n=m=1000$.
We see an increase in power in the proposed method(s), as well as 
an increase in the Type I error in the methods which use imputation.

\begin{figure}
    \includegraphics[width=\textwidth]{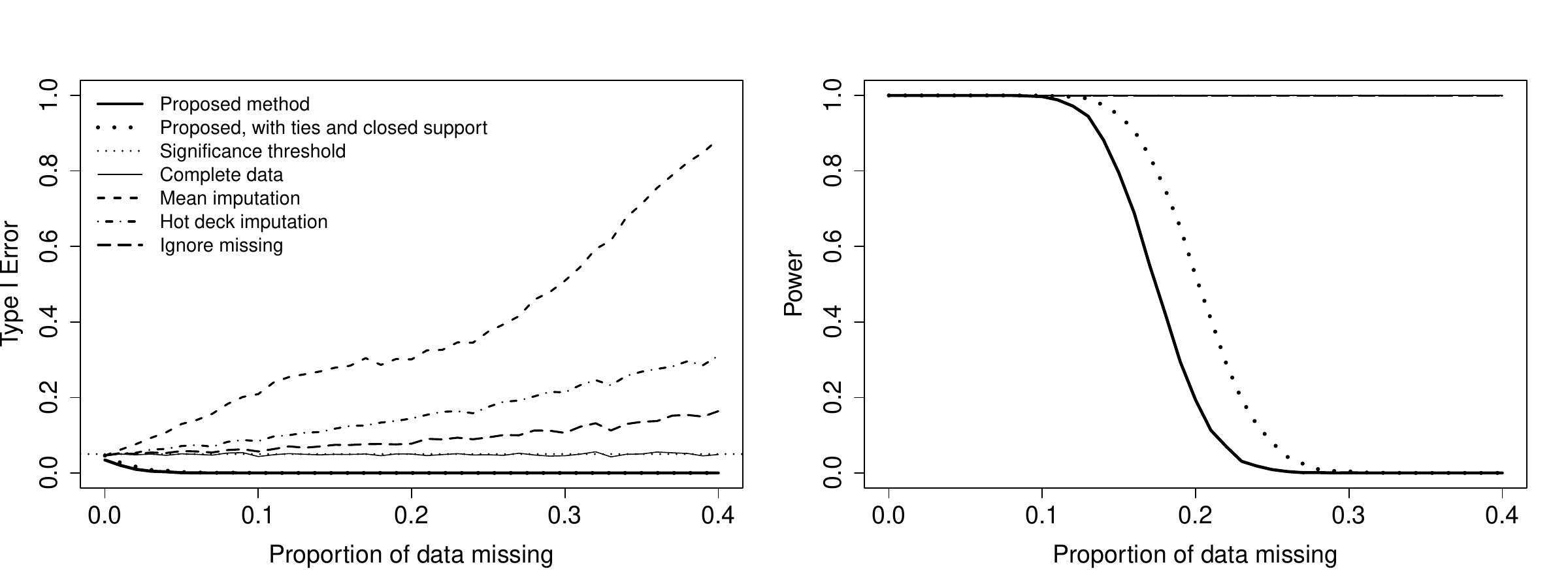}
    \caption{The Type I error and statistical power of the proposed method and 
    the standard Wilcoxon-Mann-Whitney test after the missing data is either known or
    has been imputed or ignored as the proportion of missing data increases.
    The data from \textbf{both samples are missing not at random}, with only 
    observations greater than $0$ are possibly missing.
    (Left) Type I error : $\mathrm{Pois}(1)$ vs $\mathrm{Pois}(1)$; 
    (Right) Power: $\mathrm{Pois}(1)$ vs $\mathrm{Pois}(3)$. For both figures, 
    a significance threshold of $\alpha=0.05$ has been used and the total
    sample sizes are $\n=100$, $\m=100$, and $5000$ trials were used.}
\label{fig:mnar6ties}
\end{figure}

Finally, Figure~\ref{fig:mnar6ties} shows the same data as for Figure~3, 
but now with both samples containing data missing not at random, with only
observations larger than $0$ being possibly missing.
Interestingly, there is a slight reduction in Type I error for the 
imputation methods, although they still fail to control the Type I error.

\clearpage

\section{Additional tables for clinical trials data analysis}

\begin{table}
    \def~{\hphantom{0}}
    \caption{Results from statistical analysis, with $p$-values}{%
        \scalebox{0.8}{
\begin{tabular}{lccccccc}
\\
                                     &   1.25 mg  & 2.5 mg    & 5 mg      & 7.5 mg       & 10 mg                 & 15 mg                 & 20 mg                 \\ 
    Ignoring missing                 & 0.196      & 0.208     & 0.264     & 0.002        & $3.9 \times 10^{-4}$  & $7.4 \times 10^{-6}$  & $6.5 \times 10^{-8}$  \\
   Ignoring missing, Holm-corrected & 0.587      & 0.587     & 0.587     & 0.007        & 0.002                 & $4.4 \times 10^{-5}$  & $4.6 \times 10^{-7}$  \\
    Proposed method                  & 1.000      & 1.000     & 1.000     & 0.112        & 0.072                 & 0.011                 & $1.4 \times 10^{-4}$  \\
    Proposed, Holm-corrected  & 1.000      & 1.000     & 1.000     & 0.450        & 0.358                 & 0.064                 & $9.4 \times 10^{-4}$  \\
\end{tabular}}
}
\label{tab:UACRpval2}
\end{table}

\begin{table}
    \def~{\hphantom{0}}
    \caption{Results from statistical analysis, with two-sided $p$-values}{%
        \scalebox{0.8}{
\begin{tabular}{lccccccc}
\\
                                         &   1.25 mg  & 2.5 mg    & 5 mg      & 7.5 mg       & 10 mg                 & 15 mg                 & 20 mg                 \\ 
    Ignoring missing two-sided & 0.392      & 0.415     & 0.529     & 0.003        & $7.8 \times 10^{-4}$  & $1.5 \times 10^{-5}$  & $1.3 \times 10^{-7}$  \\
    Ignoring missing two-sided, Holm-corrected & 1.000      & 1.000     & 1.000     & 0.013        & $3.9 \times 10^{-3}$  & $8.9 \times 10^{-5}$  & $9.1 \times 10^{-7}$  \\
    Proposed two-sided         & 1.00       & 1.00      & 1.00      & 0.225        & 0.143                 & 0.021                 & $2.7 \times 10^{-4}$  \\
    Proposed two-sided, Holm-corrected    & 1.00       & 1.00      & 1.00      & 0.900        & 0.716                 & 0.128                 & $1.8 \times 10^{-3}$  \\
\end{tabular}}
}
\label{tab:UACRpval3}
\end{table}

Table~\ref{tab:UACRpval2} is the same as Table 3, but also including the 
Holm-corrected $p$-values for the case when the missing data are ignored.
Table~\ref{tab:UACRpval3} shows the $p$-values when the two-sided version
of our proposed method is used.

\clearpage
\bibliographystyle{plainnat}
\bibliography{refs}


\end{document}